\documentclass[twocolumn]{aastex631}

\graphicspath{{./}{figures/}}
\usepackage[toc,page]{appendix}

\usepackage{float}
\usepackage{amsmath}
\usepackage{afterpage}
\usepackage{enumitem}
\usepackage{pifont}

\usepackage{statmath}

\usepackage[caption=false]{subfig}
\usepackage{appendix}
\usepackage{amssymb}
\usepackage{array}

\def\kms{${\text{km s}^{-1}}$} 

\def\mjy{mJy~beam$^{-1}$}
\def\jy{Jy~beam$^{-1}$}

\def\farcs{\ensuremath{\overset{\prime\prime}{.}}}

\def\arcm{\ensuremath{^{\prime}}}
\def\arcs{\ensuremath{^{\prime\prime}}}
\def\msun{\ensuremath{M_\odot}}
\def\lsun{\ensuremath{L_\odot}}

\begin{document}
\title{Searching for Molecular Jets from High-Mass Protostars}
\shorttitle{Molecular Jets from High-Mass Protostars}

\author[0000-0003-0090-9137]{Tatiana M. Rodr\'iguez}
\affiliation{Physics Department, New Mexico Tech, 801 Leroy Pl., Socorro, NM 87801, USA.}

\author{Peter Hofner}
\affiliation{Physics Department, New Mexico Tech, 801 Leroy Pl., Socorro, NM 87801, USA.}
\affiliation{Adjunct Astronomer at the National Radio Astronomy Observatory, 1003 Lopezville Road, Socorro, NM 87801, USA.}

\author[0000-0001-8745-2613]{Isaac Edelman}
\affiliation{Physics Department, New Mexico Tech, 801 Leroy Pl., Socorro, NM 87801, USA.}

\author[0000-0001-6755-9106]{Esteban D. Araya}
\affiliation{Physics Department, Western Illinois University, 1 University Circle, Macomb, IL 61455, USA.}
\affiliation{Physics Department, New Mexico Tech, 801 Leroy Pl., Socorro, NM 87801, USA.}

\author[0000-0001-8596-1756]{Viviana Rosero}
\affiliation{National Radio Astronomy Observatory, 1003 Lopezville Rd., Socorro, NM 87801, USA.}

\begin{abstract}
We report Very Large Array (VLA) observations in the Q~band toward 10 ionized jet candidates to search for SiO emission, a well-known shocked gas tracer.
We detected 7~mm continuum counterparts toward 90\% of the jet candidates. In most cases, the jet candidate is located toward the center of the 7~mm core, and the high masses ($\approx 100\,$\msun) and densities ($\approx 10^7\, \text{cm}^{-3}$) of the cores suggest that the central objects are very young high-mass protostars. 
We detected SiO $J=1-0$ emission associated with 6 target sources. In all cases, the morphology and spectrum of the emission is consistent with what is expected for molecular jets along an outflow axis, thus confirming the jet nature of 60\% of our sample.
Our data suggest a positive correlation between the SiO luminosity $L_{SiO}$, and both the bolometric luminosity $L_{Bol}$ and the radio luminosity $S_\nu d^2$ of the driving sources.
\end{abstract}

\section{Introduction}
\label{sec:intro}

Many questions remain unanswered regarding the origin of high-mass ($M_*\gtrsim 8M_\odot$) stars. Large scale molecular outflows are observed ubiquitously in high-mass star-forming (HMSF) regions \citep[e.g.,][]{Zhang01,Wu05}, which argues in favor of a formation scenario similar to that of lower mass stars, i.e., via disk accretion \citep[e.g.,][]{Cesaroni17,Oliva20,Williams22}. In this scenario, the emerging protostar accretes material from its surroundings while ejecting material along its poles. These bipolar jets play a key role in the formation process, getting rid of excess angular momentum and allowing accretion to proceed. Furthermore, because of the injection of turbulence in the surrounding medium, outflows play an important role in the future star formation in the region \citep[e.g.][]{Tanaka17,Grudic22}.
Massive young stellar objects (MYSOs) reach the conditions necessary for nuclear burning while still deeply embedded and actively accreting, making the detection of au-scale ionized jets an observational challenge. This has hindered efforts to address fundamental questions regarding the nature of the mass flows, as the number of known jets driven by MYSOs has only recently increased from a handful, thanks to high resolution and sensitivity surveys \citep[e.g.,][]{Purser21,Kavak21}. Increasing this still small number is an important task. 

A survey that significantly contributed to the study of jets from MYSOs was published by \cite{Rosero16,Rosero2019}. The authors conducted deep ($3\sim10\,\mu$\jy\ rms), sub-arcsecond resolution (0\farcs4) VLA observations at 1.3 and 6~cm toward HMSF regions. Their sample was carefully chosen to target MYSOs in the earliest stages of star formation, i.e., prior to the formation of hyper-compact (HC) H~II regions. They observed 18 cold molecular clumps (CMC), 15 cold molecular clumps with mid-infrared association (CMC-IR), and 25 hot molecular cores (HMC), and detected a total of 70 sources of radio emission, with a detection rate of 6\%, 53\%, and 100\% for CMC, CMC-IR, and HMC, respectively. 
While several of these sources were shown to be ionized jets, about 30 of them were classified as jet candidates based on two key characteristics: they are unresolved at 0\farcs4 and have a rising spectral index $\alpha$ (i.e., $0.1<\alpha<1.5$, with $S_\nu \propto \nu^\alpha$).  
A rising spectral index at cm wavelengths is expected from thermal emission from a partially optically thick ionized jet, as described by \cite{Reynolds86}. However, as \cite{Rosero2019} show in their Figure 2, a spherical ionized region of constant density can also account for their measured $\alpha$ values. Confirmation of the jet nature of the candidates in the \cite{Rosero16,Rosero2019} study is therefore needed.

One way to distinguish between the models of the origin of the radio continuum put forward by \cite{Rosero2019}, is to determine whether the continuum sources display a morphology consistent with ionized jets at higher angular resolution, i.e., elongated in the direction of the large scale molecular flows, which are present toward most of the sources. 
Alternatively, one could use molecular jet observations to differentiate between the models. In the earliest stages of formation, when the protostar is still deeply embedded, we expect that a molecular flow would be associated with an ionized jet. Thus, the detection of a molecular jet should allow us to differentiate whether the radio continuum emission arises from a mass ejection phenomenon or from an extremely compact, constant density, ionized region.
In this work, we conduct SiO(1$-$0) observations toward a sub-sample of jet candidates from the \cite{Rosero16,Rosero2019} study to search for molecular jets. 

The abundance of the SiO molecule is highly enriched in shocked gas regions, making it an ideal probe for our science goal. SiO in star forming regions can be attributed to a variety of processes. \cite{Schilke97} showed that the production of SiO in the gas phase can be linked to the sputtering of Si-bearing dust grains in C-type shocks in the jet working surface, and \cite{Anderl13} explored the role of grain-grain collisions in C-type shocks in the SiO abundance enhancement. Recently, it has been shown that SiO in the gas phase could originate at the base of jets (i.e., within the dust sublimation radius) due to evaporation \citep[e.g.,][]{Lee2017,Lee20,Podio21,Dutta2022}.

We selected 9 regions from the \cite{Rosero16,Rosero2019} survey that host a jet candidate and where molecular outflow tracers have been detected. These regions are: IRAS sources 18345$-$0641, 18440$-$0148, 18517$+$0437, 18553$+$0414, 19012$+$0536, 20293$+$3952, 20343$+$4129, G53.11$+$00.05 mm2, and G53.25$+$00.04 mm2. 
Additionally, we included IRAS 19266$+$1745 in our sample. This region contains 3 sources with a negative spectral index ($\alpha <-0.5$). The nature of the continuum emission in this region is addressed in Section~\ref{19266}.

\medskip

This paper is organized as follows: in Section~\ref{sec:observations} we describe details of the observations, as well as the data calibration and imaging process.
In Section~\ref{sec:results}, we present our observational results, which we discuss in Section~\ref{sec:discussion}. Lastly, Section~\ref{sec:conclusions} contains a summary of this work and our conclusions.

Regarding the format of the paper, to improve the readability, we include in the main body of the text the data of only one source (IRAS 18517$+$0437), serving as an example.
Analogous images and comments for all regions can be found in Appendix~\ref{sec:appendix}.

\section{Observations}
\label{sec:observations}
We obtained Q-band ($\lambda=7\,$mm) observations of our target sources with NRAO's\footnote{The National Radio Astronomy Observatory is a facility of the National Science Foundation operated under cooperative agreement by Associated Universities, Inc.} Karl G. Jansky Very Large Array (VLA) in the D configuration on March 28, 2021. The phase center for each target region is given in columns 2 and 3 of Table~\ref{tab:sources}. We adopted distances and luminosities from \cite{Rosero2019}, unless indicated otherwise; these are presented in columns 4 and 5. Additionally, in column 6, we list the classification adopted by these authors of the evolutionary stage of the source, i.e., CMC, CMC-IR, or HMC (see \citealt{Rosero16} Section 2.1 for more details).

The main target of our observations was the SiO $J=1-0$ line ($\nu_o=43.42376\,$GHz). We used two 8-bit samplers of the WIDAR correlator with dual polarization. To fully exploit the capabilities of the receiver, 1~GHz subbands were centered at 43.25, and also at 48.57 GHz, thus also including observations of the CS $J=1-0$ emission line ($\nu_o=48.99095\,$GHz). 
To cover these two molecular lines, we set up 64~MHz wide spectral windows (SPWs), with $1024\times50\,$kHz channels. We also 
configured 10 broad band SPWs to measure the Q-band continuum emission, each with $128\times1\,$MHz channels. 

Flux density and bandpass calibration was based on observations of the quasars 3C286 and J1733$-$1304, respectively. The complex gain calibrators used are presented in Table~\ref{tab:calibrators}.
Typical time on source was approximately 13.5 minutes.

\begin{deluxetable*}{lcccccccc}
\tablenum{1}
\label{tab:sources}
\tabletypesize{\scriptsize}
\tablecaption{Observed Sources and Detection Summary}
\tablecolumns{7}
\tablewidth{0pt}
\tablehead{
\colhead{Region} & \colhead{R.A.} & \colhead{Dec} & \colhead{Distance} & \colhead{L} & \colhead{Type} & \twocolhead{Detection}\\
\colhead{} & \colhead{[h m s]} & \colhead{[$^{\circ}$ \arcm\ \arcs]} & \colhead{[kpc]} & \colhead{[log$_{10}$ L$_\odot$]} & & 7 mm & SiO & CS
}
\startdata
18345$-$0641 & 18 37 16.8 & $-$ 06 38 30 & 9.5$^a$ & 5.3$^*$ & HMC & y & y & y\\
18440$-$0148 & 18 46 36.6 & $-$ 01 45 21 &  5.2$^b$ & 3.5$^*$ & HMC & y & n & y\\
18517$+$0437 & 18 54 14.2 & $+$ 04 41 40 &  1.9 & 3.8$^{c,*}$ & HMC & y & y & y\\
18553$+$0414 & 18 57 53.6 & $+$ 04 18 16 & 12.3 & 4.8 & HMC & y & y & y \\
19012$+$0536 & 19 03 45.3 & $+$ 05 40 42 & 4.2 & 4.0 & HMC & y & y & y\\
19266$+$1745 & 19 28 55.6 & $+$ 17 52 00 &  9.5 & 4.4 & HMC & y & n  & y\\
G53.11$+$00.05 mm2 & 19 29 20.6 & $+$ 17 57 18 &  1.9 & 1.9 & CMC & y & y & y\\
G53.25$+$00.04 mm2 & 19 29 33.5 & $+$ 18 00 54 &  2.0 & 2.1 & CMC-IR & n & n & n\\
20293$+$3952 & 20 31 12.9 & $+$ 40 03 22 & 1.3/2.0$^{d}$ & 3.1/3.5$^{d}$ & HMC & y & y & y\\
20343$+$4129 & 20 36 07.5 & $+$ 41 40 09 & 1.4 & 3.0 & HMC & y & n  & y\\
\enddata
\tablenotetext{}{The right ascension and declination values are given in the J2000 epoch.}
\tablenotetext{a}{From \cite{Szymczak07}.}
\tablenotetext{b}{From \cite{Urquhart18}.}
\tablenotetext{c}{From \cite{Lu14}.}
\tablenotetext{d}{Near/far.}
\tablenotetext{*}{Corrected for the adopted distance.}
\end{deluxetable*}

\begin{deluxetable}{lcl}
\tablenum{2}
\label{tab:calibrators}
\tabletypesize{\scriptsize}
\tablecaption{Calibrators List}
\tablecolumns{6}
\tablewidth{0pt}
\tablehead{
\colhead{Calibrator} & \colhead{A.P.$^{a}$} & \colhead{Sources Calibrated}
}
\startdata
J2007$+$4029 & B & 20293$+$3952, 20343$+$4129 \\
J1925$+$2106 & A & 19266$+$1745, G53.11$+$00.05, G53.25$+$00.04 \\
J1832$-$1035 & C & 18345$-$0641 \\
J1851$+$0035 & C & 18440$-$0148, 18517$+$0437, 19012$+$0536, \\
 & & 18553$+$0414\\
\enddata
\tablenotetext{a}{Astrometry precision (A.P.) classifications A, B, and C correspond to positional accuracies of $<$2 mas, 2-10 mas, and 0.01-0.15\arcs, respectively. From the VLA calibrator list: \url{https://science.nrao.edu/facilities/vla/observing/callist}.}
\end{deluxetable}

The data were processed through NRAO's Common Astronomy Software Applications (CASA, \citealt{casa}) Calibration Pipeline, using version 6.1.2. For each source we combined all 128$\,$MHz SPWs to create a continuum map corresponding to a central frequency of 45.84 GHz. The typical synthesized beam size and rms for naturally weighted continuum maps are 2\farcs0$\,\times\,$1\farcs6 and $91\,\mu\text{Jy beam}^{-1}$, respectively.
Regarding the molecular emission, the typical beam size, rms, and channel width of the SiO(1$-$0) velocity cubes are 2\farcs0$\times$1\farcs6, 4.5 mJy~beam$^{-1}$, and 1.5~\kms, respectively. 
A slight taper was applied to the CS(1$-$0) data to account for the extended nature of the emission. The typical beam size, rms, and channel width of the CS(1$-$0) velocity cubes are 2\farcs4$\times$2\farcs0, 11 mJy~beam$^{-1}$, and 1.5~\kms, respectively.
Columns 7, 8, and 9 of Table~\ref{tab:sources} show whether continuum, SiO, and CS emission was detected (y) or not (n) in each region.

\section{Results}
\label{sec:results}

\subsection{7 mm Continuum}
\label{sec:results-cont}
We detected 7~mm continuum sources (hereafter referred to as cores) in all the observed regions (referred to with the first 5 digits of their name), except for G53.25. Some regions contain only one core, while others host several. We were able to identify a total of 23 discrete 7~mm cores. 
The cores in each region are denominated C$_i$, with $i=1,2,...$, increasing with RA. 
We were able to associate 16 cores with radio continuum sources. 
We note that not all the sources in the sample were detected at 6~cm in the \cite{Rosero16,Rosero2019} survey, hence we will refer to the 1.3~cm emission only when making comparisons with radio continuum.
In Table~\ref{tab:7mm-cores} we list the regions observed, the 7~mm cores detected, and their centimeter counterpart identified by \cite{Rosero16} using the designation given by these authors. 

\begin{figure}[htb]
    \centering
    \includegraphics[width=0.49\textwidth]{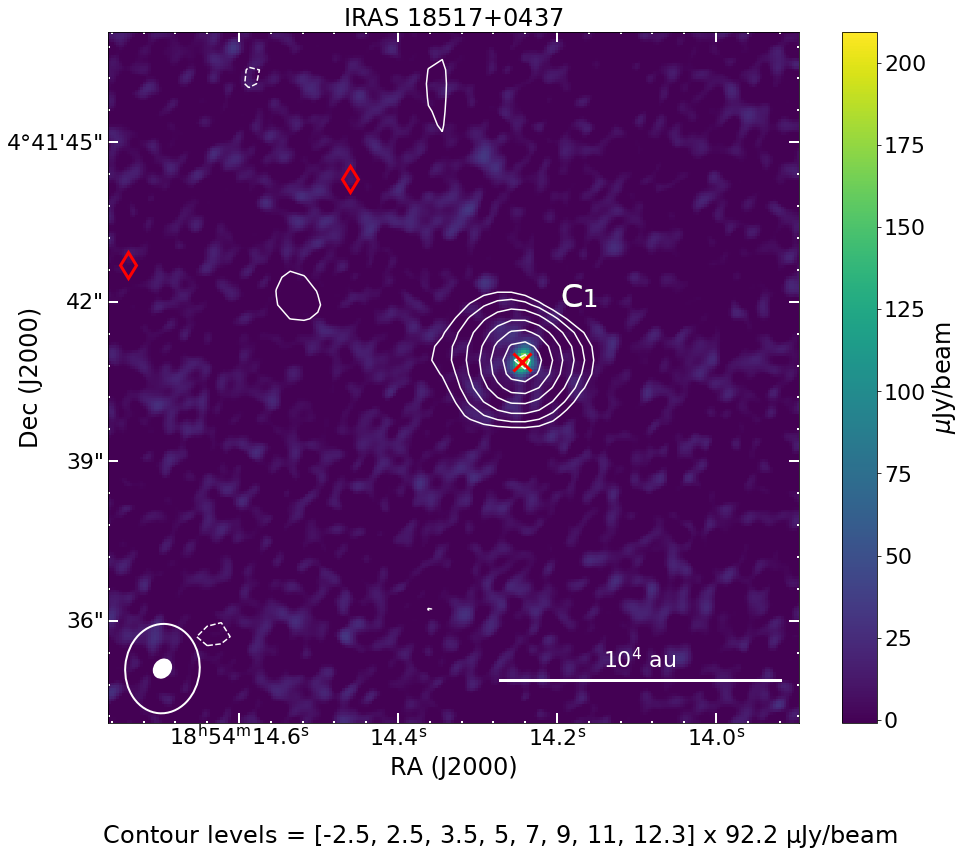}
    \caption{The white contours show the 7~mm continuum emission toward 18517 C$_1$, and the 1.3~cm emission from \cite{Rosero16} is shown in color. The filled and empty white ellipses in the bottom left represent the 1.3~cm and 7~mm synthesized beam sizes, respectively. We note that the centimeter and millimeter emission peaks are coincident. The red diamond and $\times$ show the position of the methanol 44  and 25~GHz emission from \cite{Rodriguez-Garza17} and Sanchez-Tovar, E. et al. (submitted), respectively.
    }
    \label{fig:cont-mm-cm}
\end{figure}

In Figure~\ref{fig:cont-mm-cm}, we show the 7~mm core 18517 C$_1$ in contours overlaid on the 1.3~cm emission.
The peak emission at 1.3~cm, as exemplified in Fig.~\ref{fig:cont-mm-cm}, is essentially coincident with that of the 7~mm continuum in 10 cores (18345 C$_1$, 18517 C$_1$, 18553 C$_1$, 19012 C$_1$, 19012 C$_2$, 19266 C$_1$, 19266 C$_3$, 19266 C$_5$, G53.11 C$_1$, and 20343 C$_3$), while for the 6 other cores, the 1.3~cm source is located near the edge of the 7~mm core (20293 C$_1$, C$_3$, and C$_4$, 20343 C$_1$, 18440 C$_1$, and 19266 C$_2$). We discuss this position offset in Section~\ref{sec:discussion-cont}.

We carried out a 2-D Gaussian fit to all cores using the CASA task \texttt{imfit}. Based on the results of this fit, we indicate in column 4 of Table~\ref{tab:7mm-cores} whether the core is resolved (R), or unresolved/resolved in at least one direction (U). 
The deconvolved core size and integrated flux density $S_\nu$ are given in columns 5 and 6, respectively. We note that the size and PA reported for U sources are those of the synthesized beam. 
Additionally, we classified as extended (E) two sources that present a structure several times the beam size, 19012~C$_1$ (Fig.~\ref{app:19012-cont}) and 20343~C$_3$ (Fig.~\ref{app:20343-cont}). Both present extended emission also at 1.3 and 6~cm. Based on their morphology and spectral energy distribution (SED), these are most likely H~II regions. 
Since the goal of our work is the study of jet candidates, we will exclude these two cores from the following analysis and discussion.

\begin{figure}[h!]
    \centering
    \includegraphics[width=0.45\textwidth]{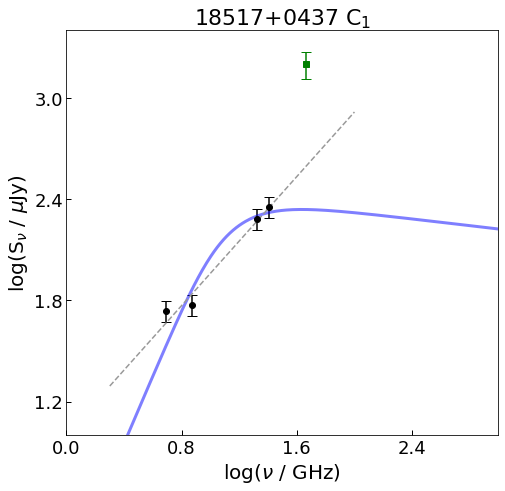}
    \caption{SED of 18517 C$_1$. The black dots represent the 1.3 and 6~cm emission from \cite{Rosero16}, while the green square shows the 7~mm flux density. The error bar of the 7~mm flux density represents the Gaussian fit error plus 10\% to account for calibration uncertainties. The solid, blue line and dotted, gray line represent the spherical ionized gas model and power-law fit from \cite{Rosero2019}, respectively.
    }
    \label{fig:SED}
\end{figure}

We plotted the SED (including the 7~mm emission) of the cores with non-extended radio continuum counterpart.
In Figure~\ref{fig:SED}, we show the resulting SED of 18517 C$_1$. 
The blue solid and the gray dashed lines show the ionized gas sphere model and power-law fit from \cite{Rosero2019}, respectively.
As in the case for 18517, for most sources the flux density rises significantly at 7~mm, which indicates that thermal emission from dust begins to dominate at this wavelength. 
In the case of 18440~C$_1$ (Fig.~\ref{app:18440-cont}, right), 19012~C$_2$ (Fig.~\ref{app:19012-cont}, first right panel), 20293~C$_3$ (Fig.~\ref{app:20293-cont}, second right panel) and C$_4$ (Fig.~\ref{app:20293-cont}, third right panel), the 7~mm flux is consistent within errors with one of the ionized gas models. This suggests that no dust emission was detected toward these cores.

\begin{deluxetable*}{lcccrc}
\tablenum{3}
\label{tab:7mm-cores}
\tabletypesize{\scriptsize}
\tablecaption{Measured Parameters of the 7 mm Cores}
\tablecolumns{6}
\tablewidth{0pt}
\tablehead{
\colhead{Region} & \colhead{Core} & \colhead{1.3\,cm\,Source$^a$} & \colhead{Type$^b$} & \colhead{Size$^c$} & \colhead{$S_\nu$}  \\
\colhead{} & \colhead{} & \colhead{} & \colhead{} & \colhead{[\arcs\ $\times$ \arcs,$^\circ$]} & \colhead{[mJy]} 
}
\startdata
18345$-$0641 & C$_1$  & A & U & 2.1 $\times$ 1.6, $-$9  & 0.63 $\pm$0.10\\
18440$-$0148 & C$_1$ & A & U & 2.0 $\times$ 1.6, $-$3 & 0.43 $\pm$0.07 \\
             & C$_2$ & - & R & 2.1 $\times$ 1.4, 0 & 0.77 $\pm$0.05 \\
18517$+$0437 & C$_1$ & A & R & 1.0 $\times$ 0.4, 85 & 1.59 $\pm$0.19 \\
18553$+$0414 & C$_1$ & A  & R & 1.5 $\times$ 1.1, 125 & 1.39 $\pm$0.15 \\
19012$+$0536 & C$_1$ & G39.389$-$0.143 & E & 3.4 $\times$ 3.2, 9 & 1.86 $\pm$0.15 \\
            & C$_2$  & A & R& 1.7 $\times$ 0.7, 68 & 1.94 $\pm$0.18 \\
19266$+$1745 & C$_1$  & A & R  & 1.9 $\times$ 0.5, 93 & 1.10 $\pm$0.06\\
            & C$_2$ & C & U & 1.9 $\times$ 1.6, $-$40 & 0.36 $\pm$0.04 \\
            & C$_3$ & B & R & 1.5 $\times$ 0.5, 110 & 0.48 $\pm$0.02\\
            & C$_4$ & - & U & 1.9 $\times$ 1.6, $-$40 & 0.35 $\pm$0.01 \\
            & C$_5$ & G53.037$+$0.115 & U & 1.9 $\times$ 1.6, $-$40 & 0.39 $\pm$0.01\\
            & C$_6$ & - & U & 1.9 $\times$ 1.6, $-$40 & 0.36 $\pm$0.01\\
G53.11$+$00.05mm2 & C$_1$ & A & U & 1.9 $\times$ 1.6, $-$37 & 0.59 $\pm$0.08 \\
20293$+$3952 & C$_1$ & E & R & 1.6 $\times$ 1.5, 24 & 0.72 $\pm$0.06 \\
             & C$_2$ & - & U & 2.2 $\times$ 1.6, $-$79 & 0.59 $\pm$0.03 \\
             & C$_3$ & C$^d$ & R & 1.5 $\times$ 1.3, 23 & 1.78 $\pm$0.11\\
             & C$_4$ & G78.976$+$0.358 & U & 2.2 $\times$ 1.6, $-$79 & 1.75 $\pm$0.11 \\
20343$+$4129 & C$_1$ & B & U & 2.2 $\times$ 1.6, $-$80 & 0.39 $\pm$0.01 \\
             & C$_2$ & - & R & 1.2 $\times$ 1.0, 86 & 0.49 $\pm$0.02 \\
             & C$_3$ & A & E & 2.7 $\times$ 1.4, $-$55 & 1.23 $\pm$0.20 \\
             & C$_4$ & - & R & 1.2 $\times$ 0.8, 49 & 0.54 $\pm$0.04 \\
             & C$_5$ & - & R & 1.4 $\times$ 1.0, 46 & 0.66 $\pm$0.05 \\
\enddata
\tablenotetext{a}{Nomenclature from \cite{Rosero16}.}
\tablenotetext{b}{Unresolved (U), resolved (R), or extended (E).}
\tablenotetext{c}{For U sources, the size listed is that of the 7~mm synthesized beam.}
\tablenotetext{d}{Includes radio source 20293$+$3952 B \citep[see][Figure 2]{Rosero16}.}
\end{deluxetable*}

\begin{deluxetable*}{lccccr}
\tablenum{4}
\label{tab:core-mass}
\tabletypesize{\scriptsize}
\tablecaption{Derived Parameters of the 7 mm Cores}
\tablecolumns{8}
\tablewidth{0pt}
\tablehead{
\colhead{Source}  & \twocolhead{Geom. Mean FWHM} & \colhead{M$_d$} & \colhead{n(H$_2$)} & \colhead{N(H$_2$)} \\
\colhead{}  &  \colhead{[\arcs]}&  \colhead{[$10^3$ au]} & \colhead{[M$_\odot$]} & \colhead{[$10^7\,\text{cm}^{-3}$]} & \colhead{[$10^{24}\,\text{cm}^{-2}$]}
}
\startdata
18345 C$_1$  & 1.9 & 17.8  & 83 $\pm$22 & 0.3 $\pm$0.1 & 0.6 $\pm$0.2\\
18440 C$_1$ & 1.8 & 9.3  & $-^a$ & $-^a$ & $-^a$  \\
18440 C$_2$  & 1.7 & 8.8  & 113 $\pm$17 & 0.9 $\pm$0.1 & 1.4 $\pm$0.2 \\
18517 C$_1$  & 0.7 & 1.3  & 14 $\pm$3 & 171 $\pm$41 & 21.5 $\pm$5.2 \\
18553 C$_1$  & 1.3 & 15.9  & 627 $\pm$155 & 3.7 $\pm$0.9 & 6.0 $\pm$1.5 \\
19012 C$_2$  & 1.1 & 4.7  & $-^a$ & $-^a$ & $-^a$ \\
19266 C$_1$  & 1.7 & 16.0  & 429 $\pm$80 & 2.5 $\pm$0.5 & 4.0 $\pm$0.7 \\
19266 C$_2$ & 1.7 & 16.6  & 92 $\pm$24 & 0.5 $\pm$0.1 & 0.8 $\pm$0.2 \\
19266 C$_3$  & 0.8 & 7.8  & 189 $\pm$34 & 9.6 $\pm$1.7 & 7.5 $\pm$1.3 \\
19266 C$_4$ & 1.7 & 16.6  & 141 $\pm$25 & 0.7 $\pm$0.1 & 1.2 $\pm$0.2 \\
19266 C$_5$ & 1.7 & 16.6 & 152 $\pm$27 & 0.8 $\pm$0.1 & 1.3 $\pm$0.2 \\
19266 C$_6$ & 1.7 & 16.6  & 142 $\pm$25 & 0.7 $\pm$0.1 & 1.2 $\pm$0.2 \\
G53.11 C$_1$& 1.7 & 3.3  & 9 $\pm$1 & 5.9 $\pm$1.1 & 2.0 $\pm$0.4 \\
20293 C$_1$  & 1.5 & 2.0/3.0$^b$  & 3/8$^b$ $\pm$1/2 & 11.0/7.2$^b$ $\pm$2.6/1.7 & 2.2 $\pm$0.5 \\
20293 C$_2$ & 1.9 & 2.4/3.7$^b$  & 4/10$^b$ $\pm$1/2 & 7.1/4.6$^b$ $\pm$1.3/0.8 & 1.7 $\pm$0.3 \\
20293 C$_3$ & 1.4  & 1.8/2.8$^b$  & $-^a$ & $-^a$ & $-^a$  \\
20293 C$_4$ & 1.9 & 2.4/3.7$^b$  & $-^a$  & $-^a$  & $-^a$  \\
20343 C$_1$ & 1.9 & 2.6  & 3 $\pm$0.6 & 4.4 $\pm$0.8 & 1.2 $\pm$0.2 \\
20343 C$_2$  & 1.1 & 1.5  & 4 $\pm$1 & 30.1$\pm$5.5 & 4.5 $\pm$0.8 \\
20343 C$_4$  & 0.9 & 1.3  & 4 $\pm$1 & 57.5 $\pm$10.9 & 7.6 $\pm$1.4 \\
20343 C$_5$  & 1.2 & 1.6  & 5 $\pm$1 & 25.2 $\pm$4.8 & 4.1 $\pm$0.8\\
\enddata
\tablenotetext{a}{No dust emission was detected.}
\tablenotetext{b}{Near/far.}
\end{deluxetable*}

Assuming that the $7\,$mm emission arises from optically thin dust, we calculated the mass of each core using the equation 
\begin{equation}
M_d=\frac{d^2S_{\nu}R_g}{B_{\nu}(T_d)\kappa_{\nu}}.
\end{equation}
In this expression, $d$ is the distance to the source, $R_g$ is the gas-to-dust ratio, $B_{\nu}(T_d)$ is the Planck function for a dust temperature $T_d$, and $\kappa_{\nu}$ is the dust opacity (see \citealt{Hildebrand83}).
Most of these quantities are unknown, hence we made the following assumptions.
First, based on \cite{Lu14} and \cite{Rathborne2010}, we took 30~K as the dust temperature for all cores.
Second, to obtain the dust opacity, we interpolated the value of $\kappa_{1.3mm}$ for cores with thin ice mantles and a density of $10^6\,\text{cm}^{-3}$ from \cite{Ossenkopf94}, using the power law relation $\kappa_\nu = \kappa_{\nu_0} (\frac{\nu}{\nu_0})^\beta$. We took an intermediate value for the dust emissivity index $\beta$ of 1.5, which results in $\kappa_{7mm}=0.08\,\text{cm}^2\,\text{g}^{-1}$.
Third, $R_g$ was estimated using equation 2 in \cite{Giannetti17}, which depends on the galactocentric distance of the sources. The $R_g$ values obtained range between $\sim65$ and 135.
Finally, we extrapolated the ionized gas power-law fit to 7~mm and subtracted the resulting value from the total integrated flux for those cores with centimeter counterpart and rising spectral index, i.e., 18345~C$_1$, 18517~C$_1$, 18553~C$_1$, G53.11 C$_1$, 20293~C$_1$, and 20343~C$_1$. This was done to account for the contribution from ionized gas and hence the estimated mass of these cores is a lower limit. 

From the derived masses, we obtained the density $n(H_2)$ and column density $N(H_2)$ of the emission. For the calculation of these quantities, we assumed the cores to have a spherical geometry with a diameter equal to their geometric mean FWHM. 
The geometric mean FWHM of each core in units of arcsec and au, as well as the obtained masses, densities, and column densities are presented in Table~\ref{tab:core-mass}. We note that for unresolved sources the densities obtained are lower limits.


\subsection{SiO $J=1-0$}
\label{subsec:SiO}
\begin{figure}
    \centering
    \includegraphics[width=0.49\textwidth]{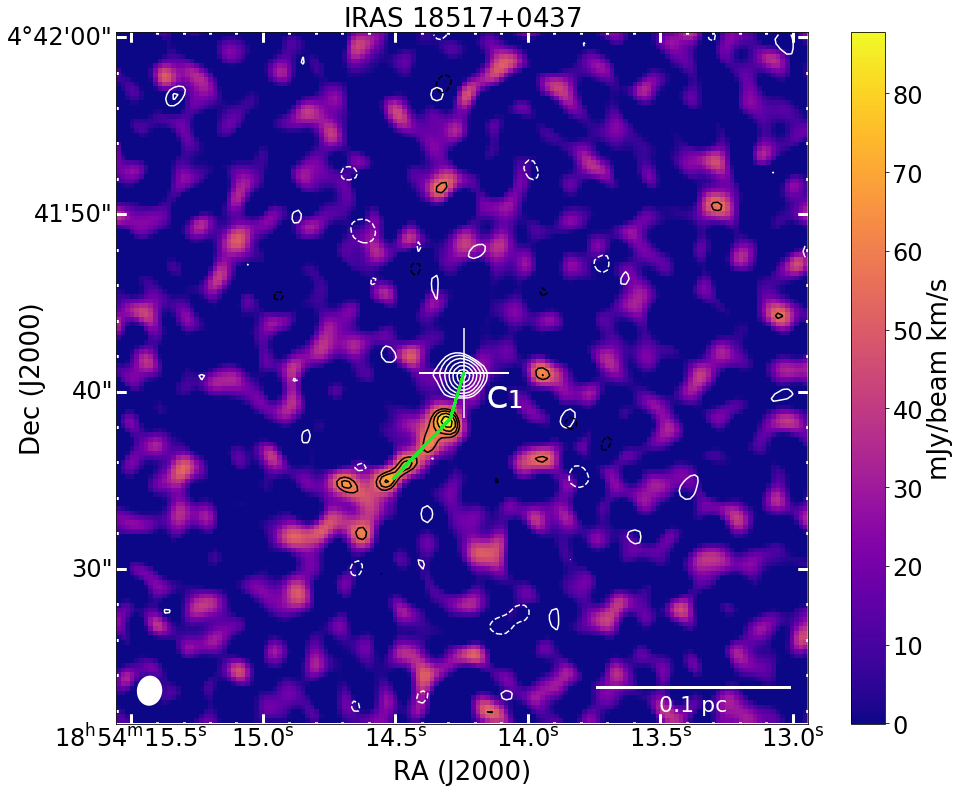}
    \caption{SiO integrated intensity in 18517 is shown in color and black contours. The latter represent $[-3, 3, 3.5, 4, 4.5]\,\times\sigma_{SiO}$, with $\sigma_{SiO}=18.1\,$\mjy~\kms. White contours show the 7~mm emission, and are the same as in Figure~\ref{fig:cont-mm-cm}. The white cross marks the position of the radio continuum source.
    The filled, white ellipse in the bottom left represents the synthesized beam size. The SiO(1$-$0) emission is highly collimated, and is clearly associated with the jet candidate embedded in the 7~mm core. The emission is monopolar, and extends to about 0.08~pc from the $7\,$mm core. The green line was drawn to guide the reader on the possible direction of the emission.
    }
    \label{fig:SiO-mom0}
\end{figure}

We detected SiO emission in 6 of the 10 regions observed: 18345, 18517, 18553, 19012, G53.11, and 20293.
Figure~\ref{fig:SiO-mom0} shows the SiO~(1$-$0) integrated intensity (moment-0) map in 18517.
We find a number of shared characteristics in the morphology of the SiO emission. First, in all cases the SiO emission appears to originate in close proximity to the radio continuum source. Second, the emission appears in elongated structures, and is highly asymmetric or monopolar. In 18517 (Fig.~\ref{app:18517-mol}), 19012 (Fig.~\ref{app:19012-mol}), and 20293 (Fig.~\ref{app:20293-mol}), the SiO emission appears quite collimated and elongated on scales $\gtrsim0.1\,$pc. On the other hand, in 18345 (Fig.~\ref{app:18345-mol}), 18553 (Fig.~\ref{app:18553-mol}), and G53.11 (Fig.~\ref{app:G53-mol}), we observe a few clumps of strong emission at distances $\gtrsim0.1\,$pc from the core. In all cases the SiO emission seems to be clumpy. 

\begin{figure}
    \centering
    \includegraphics[width=.4\textwidth]{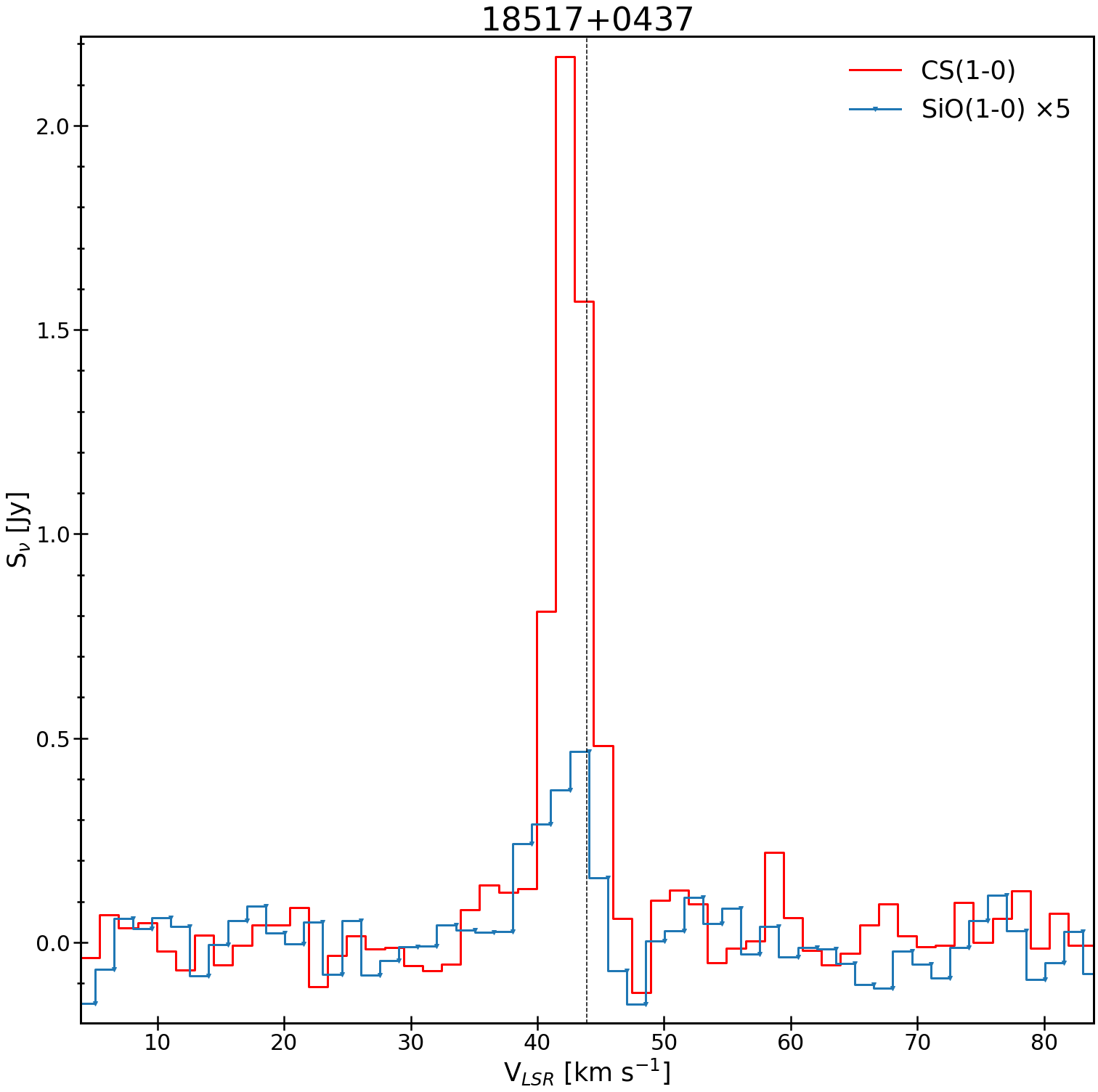}
    \caption{SiO$\,\times 5$ and CS~$J=1-0$ emission in 18517 are shown in red and blue, respectively.
    The SiO and CS spectra were obtained integrating over all the emission that appears to be spatially connected to C$_1$ and enclosed in the $3\sigma$ and 35\% contour levels from Fig.~\ref{fig:SiO-mom0} and \ref{fig:CS-mom0}, respectively. The SiO emission is highly asymmetric, and shows a clear line wing toward blue-shifted velocities. The CS line is single peaked and symmetric. The vertical dotted line marks the systemic velocity.
    } 
    \label{fig:spectra}
\end{figure}

In Figure~\ref{fig:spectra}, we show the spectrum of the SiO emission multiplied by 5 in 18517 in blue, obtained integrating over all the emission that seems to be associated with C$_1$ within the $3\sigma$ contour level, as shown in Figure~\ref{fig:SiO-mom0}. 
The vertical dotted line in Figure~\ref{fig:spectra} marks the systemic velocity, taken from \cite{Bronfman96}, derived from single dish CS(2$-$1) observations.
We took the systemic velocities of 18345, 19012, and 20293 from \cite{Bronfman96} as well, while the measurements of NH$_3\,(J,K)=(1,1)$ from \cite{Wienen15} and C$^{18}$O(1$-$0) from \cite{Zhang17} were used to obtain the systemic velocity of 18553 and G53.11, respectively.
The peak flux density $S_\nu^{peak}$ and rms of the SiO line emission measured from the integrated spectrum in each region are listed in columns 2 and 3 of Table~\ref{tab:SiO-vel}. 
Due to the asymmetry observed in most cases, we characterize the line width $\Delta v$ with the full width at zero power (FWZP), which ranges between $\sim7.5\,$\kms\ (18517, Fig.~\ref{fig:spectra}) and 24~\kms\ (20293, Fig.~\ref{app:20293-mol}).
We also found that usually the SiO line peaks blue-shifted from the adopted systemic velocity. The observed velocity offsets $v_{of\!f}$ range between 1.6~\kms\ (18345, Fig.~\ref{app:18345-mol}) and almost 4~\kms\ (20293, Fig.~\ref{app:20293-mol}). 
In Table~\ref{tab:SiO-vel} we list $v_{of\!f}$ and $\Delta v$ of the line measured in each region in columns 4 and 5, respectively.

\begin{deluxetable}
{l c c c c c c}
\tablenum{5}
\label{tab:SiO-vel}
\tabletypesize{\scriptsize}
\tablecaption{Properties of the SiO Flows}
\tablecolumns{7}
\tablewidth{0pt}
\tablehead{
\colhead{Region} & \colhead{$S_\nu^{peak}$} & \colhead{rms} & \colhead{$v_{of\!f}^{a}$} & \colhead{$\Delta v$} & \colhead{$\int S_\nu\, dv$} & \colhead{L$_{SiO}$} \\
\colhead{}       & \colhead{[mJy]}       & \colhead{[mJy]}       & \colhead{[\kms]}    & \colhead{[\kms]}     & \colhead{[Jy \kms]}       & \colhead{[$10^{-7} L_\odot$]}
}
\startdata
18345  & 113.6 & 18.2   & $-$1.6     & 21.0  & 0.95  & 26.6 \\
18517  & 93.3  & 13.1   & $-$0.0     &  7.5  & 0.46  & 0.5 \\
18553  & 53.5  & 10.8   & $-$2.8     &  9.0  &  0.31  & 14.7  \\
19012  & 203.4 & 24.6   & $-$2.7     & 22.5  &  2.08  & 11.4 \\
G53.11  & 21.2 & 4.3    & $-$0.0     & 12.0  &  0.12  &  0.1\\
20293 & 191.1  & 19.0   & $-$3.8     & 24.0  &  1.85  & 1.0/2.3$^{b}$\\
\enddata
\tablenotetext{a}{Velocity offset between the peak emission of the SiO spectra and the adopted systemic velocity.}
\tablenotetext{b}{Near/far}
\end{deluxetable}

To estimate the energetics of the emission, we calculated for each region the SiO(1$-$0) luminosity using the expression 
\begin{equation}
    L_{\text{SiO}} = 4\pi\, d^2\, \int S_\nu \, dv,
\end{equation}
where $d$ and $\int S_\nu\, dv$ are the distance and integrated flux of the emission, respectively. 
In columns 6 and 7 of Table~\ref{tab:SiO-vel}, we list the measured $\int S_\nu\ dv$ values and implied luminosities. The obtained SiO(1$-$0) luminosities range between $10^{-8}\sim10^{-6}\,$\lsun.

\subsection{CS $J=1-0$}
\label{subsec:CS}
\begin{figure}
    \centering
    \includegraphics[width=0.49\textwidth]{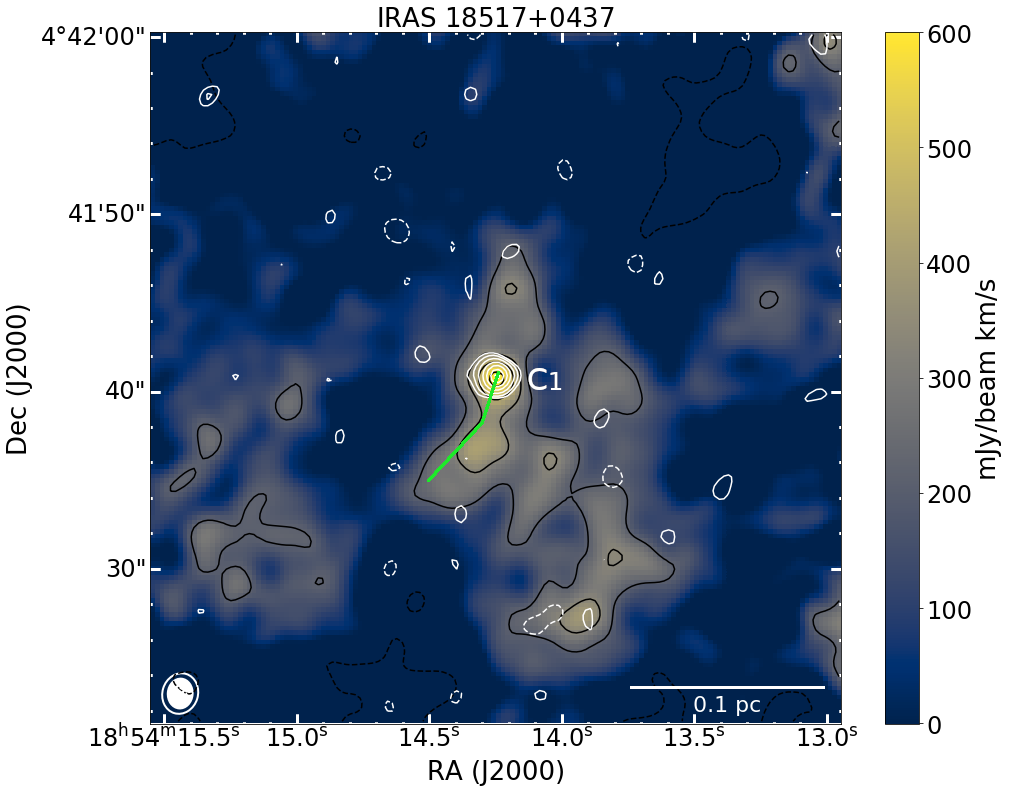}
    \caption{CS(1$-$0) integrated intensity map toward 18517 in color and black contours. The contour levels represent $-$35\%, 35\%, 55\%, 75\%, and 95\% of the peak CS(1$-$0) emission, with the peak emission and rms of the map being 600 and 147~\mjy~\kms, respectively. The white contours show the 7~mm emission and are the same as in Fig.~\ref{fig:cont-mm-cm}. The outlined and filled white ellipses in the bottom left represent the CS and 7~mm continuum synthesized beam sizes, respectively. The green line was drawn to guide the reader on the possible direction of the SiO emission and is the same as in Figure~\ref{fig:SiO-mom0}.
    The CS emission presents an elongated morphology in the North-South direction, although slightly asymmetric, more intense to the South. Its peak coincides with the $7\,$mm continuum peak.    }
    \label{fig:CS-mom0}
\end{figure}

We detected CS(1$-$0) emission in 90\% of the regions observed, which is all except G53.25.
In Figure~\ref{fig:CS-mom0}, we show the CS moment-0 map toward the 18517 region.
While in some cases the CS emission peak is essentially coincident with the $7\,$mm core (e.g., 18553, Fig.~\ref{app:18553-mol}), in other cases it is found at a considerable distance from the jet candidate (e.g., G53.11, Fig.~\ref{app:G53-mol}), and in other regions we are not able to associate it with a singular core (e.g., 18440, Fig.~\ref{app:18440-mol}).
Additionally, and unlike the SiO flows, we find significant morphology variations in the different regions. The CS emission is usually clumpy, quite extended in some cases (e.g., 20343, Fig.~\ref{app:20343-mol}) and elongated or filamentary in others (e.g., 20293, Fig.~\ref{app:20293-mol}).
 
In Figure~\ref{fig:spectra}, we show in red the spectrum of the CS(1$-$0) emission in 18517, obtained integrating over all the emission that appears to be connected to C$_1$ and within the 35\% contour level shown in Figure~\ref{fig:CS-mom0}. We find that most CS spectra show single peak and symmetric lines, with the exception of 20293 (Fig.~\ref{app:20293-mol}). 
We note, however, that due to our lack of sufficiently small spacings, not all CS emission was recovered by our observations, therefore, additional observations with more compact arrays are needed to reliably map the CS distribution in our sample.

\section{Discussion}
\label{sec:discussion}

\subsection{Continuum emission}
\label{sec:discussion-cont}
The association of the radio continuum sources with dust cores traced by 1.2~mm emission was introduced by \cite{Rosero16}, who based their analysis on available single-dish data. The higher resolution of our 7~mm data now allow us to clearly connect the dust cores and ionized gas and thus investigate where star formation occurs. The cores traced by our observations have typical sizes of $\approx 10,000\,$au, and 
most of the radio continuum sources in our sample are located toward the center of the 7~mm cores (see Section~\ref{sec:results-cont} and Figures \ref{app:18345-cont}, \ref{app:18517-cont}, \ref{app:18553-cont}, \ref{app:19012-cont}, \ref{app:19266-cont}, \ref{app:G53-cont}, and \ref{app:20343-cont}). 
This confirms that the cores are associated with YSOs, and indicates these YSOs are deeply embedded, which, in turn, suggests they are in an early evolutionary stage. Additionally, the large masses ($\approx 100\,M_\odot$) and high densities ($\approx 10^7\,$cm$^{-3}$) estimated for most cores confirm that these are high-mass objects.
Therefore, our observations reveal a scenario similar to that predicted by the core accretion model \citep[e.g.,][]{mckee03}, with an accreting, emerging protostar in the center of a dust core, which propels an ionized jet. 

However, not in all cases do the 1.3~cm and 7~mm peaks exactly coincide (e.g., \ref{app:18440-cont}, left).
We observe slight offsets between the radio continuum sources and the $7\,$mm core peak across the sample that range between $500\sim4,300\,$au, with a mean separation of about 2,000~au.
We note that all the cores where a position offset is observed are weak detections (i.e., peak $\lesssim5\sigma$) and in all cases the offset is within the 7~mm beam size. It is then possible that this position discrepancy is an effect of low signal-to-noise ratio in our observations and we caution about an overinterpretation of this feature. 
Nonetheless, position discrepancies were also noted by \cite{Rosero16} in several cases when associating the radio continuum sources with the single-dish 1.2~mm cores.
The offset reported by these authors range between 4,000 and 10,000~au (median values for CMC-IR and HMC).
Additionally, \cite{Liu21} found 6~cm and 1.3~mm continuum peak offsets in a number of their sources as well, with typical separation distances of approximately 2,500~au.
If such offsets are indeed real, they could be interpreted as the result of a jet-cloud collision. This would indicate that the observed radio continuum emission is a jet knot located at a certain distance from the driving source. Since only one of the offset sources has been associated with an SiO outflow (20293 C$_1$, see Fig.~\ref{app:20293-mol}), we are not able to extend this interpretation to all our case studies.

\subsection{SiO emission}
\label{sec:discussion-sio}

Shocks are thought to be the main source of gaseous SiO. 
Both shocks associated with cloud-cloud collisions and with ionized jets can provide the conditions to highly enrich the SiO abundance.
In the former, the SiO emission is observed as large cloud scale flows with relatively small ($<2\sim5\,$\kms) line widths \citep[e.g.,][]{Jimenez-serra10,Cosentino18,Cosentino20}. Our observations, on the other hand, present projected sizes and line widths of $<1\,$pc and $\gtrsim 10\,$\kms\ (see Table~\ref{tab:SiO-vel}), respectively. Hence, we discard this scenario as the origin of the SiO emission detected.

As mentioned before, we observe clumpy and collimated SiO structures, which are a common trait of jet-driven molecular flows. 
In favor of this interpretation, molecular outflows have been previously detected in 18345, 18517, 18553, 19012, and 20293 using probes such as CO and HCO$^+$ (see Appendix~\ref{sec:appendix} for details). In all cases where an image of the outflow is available, we find that the direction of the SiO emission is consistent with the outflow axis.
Additionally, morphological and line asymmetries are usual in these kind of objects, and monopolar SiO flows are rather common in both low- and high-mass protostars as well \citep[e.g.,][]{Zapata06,Nony20,Jhan22}. It is worth noting that the nature of this monopolarity is not well understood; while some authors argue that this feature is intrinsic \citep[e.g.,][]{Codella14}, others propose that it can be explained by asymmetries in the ambient gas \citep[e.g.,][]{FernandezLopez13}.
In conclusion, it is our interpretation that the observed SiO emission is driven by ionized jets.

The next logical question is whether these SiO flows are associated with jet candidates.
We are able to morphologically connect all the observed SiO flows with a 7~mm core associated with a jet candidate, thus we propose that these are the driving sources. In fact, according to \cite{Liu21}, sources with luminosities $\gtrsim10^2\,$\lsun\ can drive strong enough jets to sputter the surrounding dust grains, allowing the detection of SiO. Based on the estimations made by \cite{Rosero2019} using Hi-GAL data, all the sources in our sample associated with an SiO outflow have luminosities $>10^2\,$\lsun, which supports our interpretation.

\subsubsection{Comments on Detection Rate}
The SiO detection rate is generally greater than 50\% in the early stages of high-mass star formation, and it decreases for more evolved objects \citep[see Section 5.1 of][for a more detailed discussion]{Liu21}.
In our sample, which consists mostly of sources in the HMC evolutionary phase, we detected SiO jets in 6 of the 10 regions observed, which translates to a detection rate of 60\%. 
Regarding the other 40\% of our sample, either the SiO emission is lacking in those regions or the sensitivity of our observations is insufficient to detect it. 
If the non-detection is due to the latter, then the SiO line intensity in those regions is weaker than $\sim15\,$\mjy, which is our $3\sigma$ limit. 
The regions where we did detect SiO emission have distances that range between $\sim2$ and 12.3 kpc, thus a correlation with distance is not immediately clear. Note that the outflows associated with more distant sources are orders of magnitude more luminous than those driven by sources located within 2~kpc from the Sun. Of the non-detected regions, 18440 and 19266 have distances larger than 5~kpc, while G53.25 and 20343 are within 2~kpc.

Most previous surveys have used higher $J$ transitions of the SiO molecule \citep[e.g.,][]{Gibb07,Csengeri16,Li19,Liu21}, which are expected to be brighter, and hence easier to detect. Nonetheless, our detection rate is similar to what most other studies have found, so it appears that VLA SiO $J=1-0$ observations are useful to trace jets in high-mass star forming regions.

\subsubsection{Energetic and spatial correlations} 

\begin{figure}
    \centering
    \includegraphics[width=0.4\textwidth]{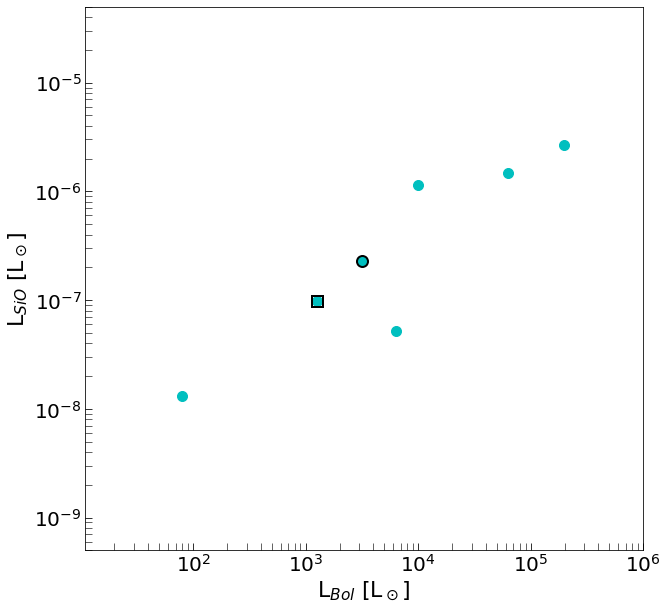}
    \includegraphics[width=0.4\textwidth]{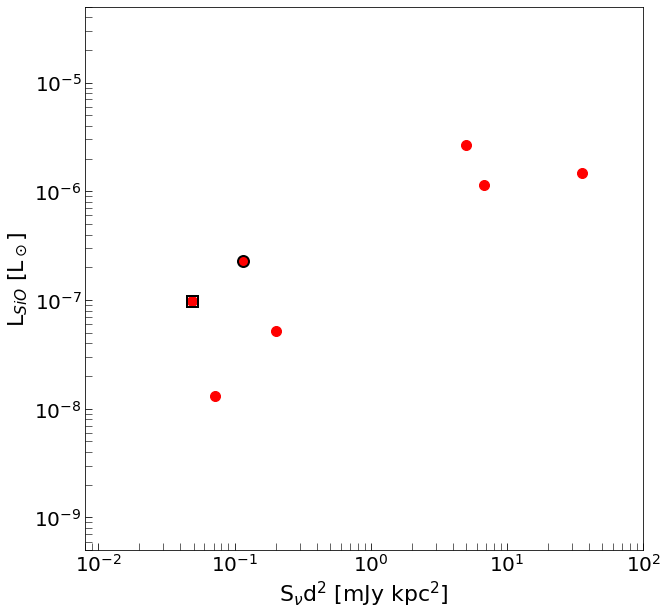}
    \caption{SiO luminosity $L_{SiO}$ versus bolometric luminosity $L_{Bol}$ (top) and 1.3~cm radio luminosity $S_{\nu}d^2$ (bottom) plots. The outlined square and circle show the luminosity values for 20293~C$_1$ if the near or far distance was adopted, respectively. 
    Although the scatter in our data is significant, our observations suggest a positive correlation between $L_{SiO}$ and both $L_{Bol}$ and $S_{\nu}d^2$. 
    }
    \label{fig:luminosity-plots}
\end{figure}

\cite{Liu21} observed the SiO$(5-4)$ emission toward a set of infrared dark clouds associated with MYSOs with a resolution comparable to ours ($\sim1$\arcs). 
The SiO luminosities we measured are similar to those they report after considering the expected intensity ratio between the $J=1-0$ and $J=5-4$ lines, assuming LTE conditions with $T=30\,$K. 
Additionally, we found a positive correlation between the SiO luminosity $L_{SiO}$ and the bolometric luminosity $L_{Bol}$ of the sources, as shown in the top panel of Figure~\ref{fig:luminosity-plots}. This has also been seen in previous studies \citep[e.g.,][]{Liu21,Codella99,Liu22} and, although the scatter in our observations is significant, supports the idea that more luminous objects drive stronger SiO jets.

A positive correlation was also observed between $L_{SiO}$ and the radio luminosity $S_{1.3cm}d^2$, which we present in the bottom panel of Figure~\ref{fig:luminosity-plots}. This points to a connection between the outflowing ionized and molecular gas, and suggests that more radio-luminous objects drive more luminous flows, similar to what Figure~9 of \cite{Rosero2019} shows. 

We found no clear correlation between $L_{SiO}$ and $L_{Bol}/M_d$, often used as an indicator of the evolutionary stage of a YSO. 
It is possible, though, that this lack of correlation is impacted by the fact that the sources targeted in this work are all extremely young and expected to be of similar age.

Regarding other outflow tracers, we find that the SiO flows seem to be oriented in the same direction as the molecular outflows traced by other molecules. Unfortunately, the available data are highly inhomogeneous (see Appendix~\ref{sec:appendix} for details). This hampered our attempt to search for correlations between the CO (or HCO$^+$) and SiO flows, from both a morphological and an energetic point of view. 

\begin{figure}
    \centering
    \includegraphics[width=0.49\textwidth]{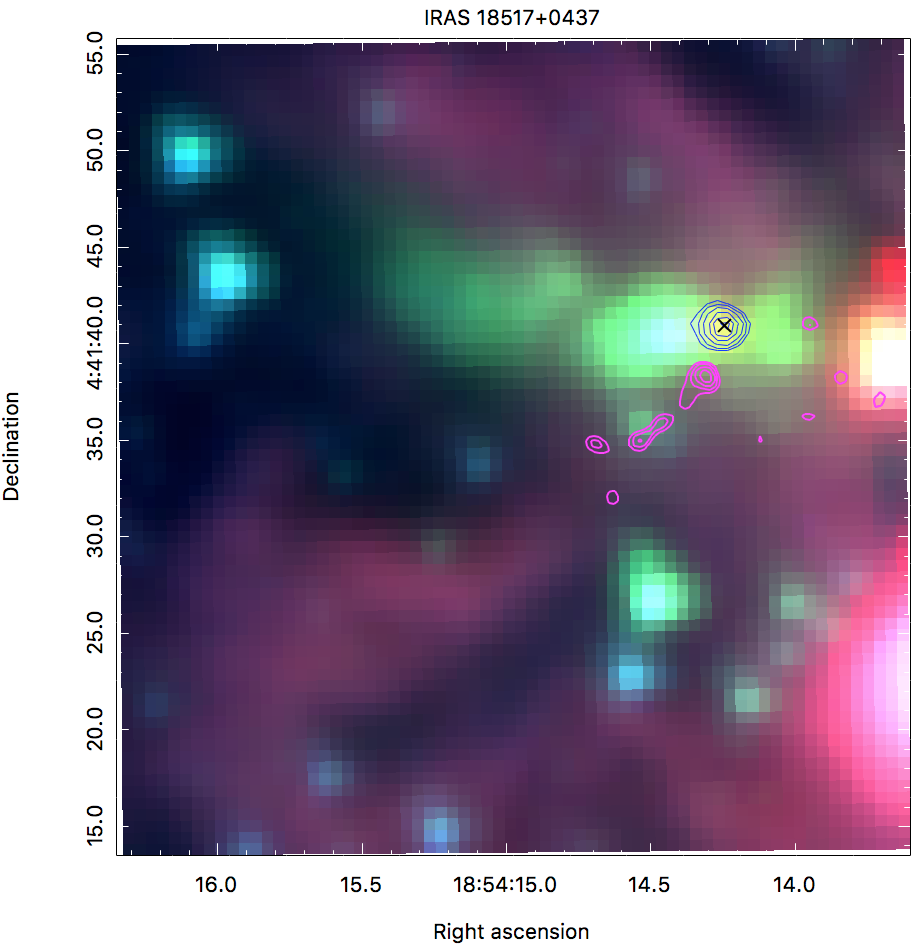}
    \caption{Spitzer IRAC color composite of 18517, showing the 3.6 (blue), 4.5 (green), and 8.0~$\mu$m (red) emission. The blue and magenta contours show the 7~mm continuum and SiO integrated intensity and are the same as in Figures~\ref{fig:SiO-mom0} and \ref{fig:cont-mm-cm}, respectively. 
    The black $\times$ marks the position of the radio continuum source. We note that the green band emission appears to be elongated at an angle of about 45$^\circ$ from the SiO outflow axis.
    }
    \label{fig:green-band}
\end{figure}

We also searched the literature and archival data for spatial correlations between the SiO and the shocked gas tracer 2.122~$\mu$m H$_2$, as well as the 4.5~$\mu$m (also known as green band) excess from the Spitzer Space Telescope\footnote{The Spitzer Space Telescope was operated by the Jet Propulsion Laboratory, California Institute of Technology under a contract with NASA.} as provided by \cite{IRSA22}.
We found H$_2$ features in 18345 \citep{Varricatt10,Varricatt13} and 20293 \citep{Beuther04}, where only the former appears to be coincident with an SiO knot. We only found significant Spitzer 4.5~$\mu$m extended excess emission in 18517~(Fig.~\ref{fig:green-band}) and 19012~(Fig.~\ref{app:19012-IR}). In both cases, the green band emission seems to be associated with the jet candidate, but elongated in a direction quasi perpendicular to the SiO flow.

\section{Summary and Conclusions}
\label{sec:conclusions}
The results of this work can be summarized as follows:
\begin{enumerate}
    \item We detected 7~mm continuum emission toward 90\% of the jet candidates, and identified a total of 23 individual cores. 
    
    \item We found that, in most cases, the radio continuum source is essentially coincident with the 7~mm peak, suggesting these are deeply embedded objects. Additionally, the large masses and densities of the cores  support the idea that these are MYSOs in the earliest stages of formation.
    
    \item We detected SiO(1$-$0) flows in 6 of the targeted regions. In all cases, the flows appear to be associated with a jet candidate, thus confirming the jet nature of 60\% of our sample. 
    
    \item Although based on a small sample with substantial scatter, our data suggest a positive correlation between the SiO luminosity ($L_{SiO}$) and both the bolometric ($L_{Bol}$) and radio luminosity ($S_\nu d^2$) of the driving sources.
    
\end{enumerate}

This is, to the best of our knowledge, the first search for molecular jets carried out with the VLA using the SiO $J=1-0$ line. 
Our results demonstrate both the suitability of this transition to trace molecular flows from MYSOs, and the capability of the VLA to conduct this kind of work at relatively high frequency and with only a few minutes of time on-source. 
Our study adds to the growing list of studies with high detection rate of SiO in the earliest stages of star formation \citep[e.g.,][]{Lopez-Sepulcre11,Scengeri16,Liu21,Liu22}.
\\
\\
\small{
\textit{Acknowledgments.} We wish to thank the anonymous referee for comments and suggestions that helped improve this work. P. H. and E. D. A. acknowledge support from NSF grants AST–1814011, and AST–1814063, respectively.
}

\bibliography{main}{}
\bibliographystyle{aasjournal}

\begin{appendix}

\section{Comments on Individual Sources}
\label{sec:appendix}

\renewcommand{\thefigure}{\Alph{section}\arabic{section}.\arabic{figure}}
\setcounter{figure}{0}

\subsection{IRAS 18345$-$0641}
\begin{figure}[H]
    \centering
    \includegraphics[width=0.53\textwidth]{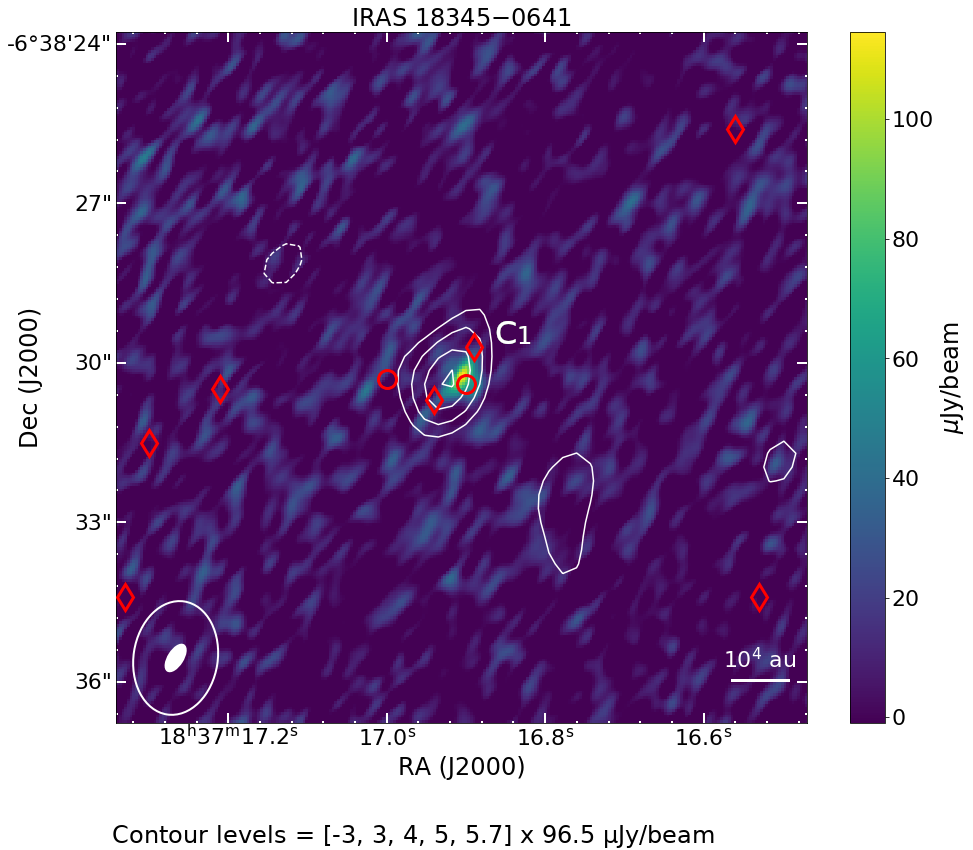}
    \includegraphics[width=0.45\textwidth]{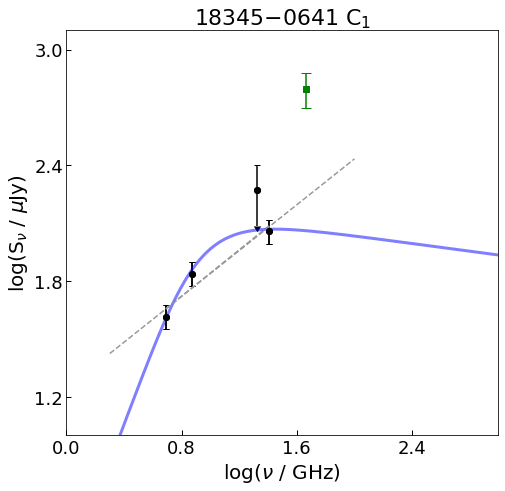}
    \caption{\textit{Left}: the 7~mm continuum emission toward 18345 is shown in white contours overlaid on the 1.3~cm color image from \cite{Rosero16}.
    The outlined and filled white ellipses in the bottom left represent the 7~mm and 1.3~cm synthesized beam sizes, respectively. The red circles and diamonds show the 6.7 and 44 GHz methanol masers from \cite{Beuther02c} and \cite{Rodriguez-Garza17}, respectively.
    \textit{Right}: SED of C$_1$. The black dots show the radio continuum data from \cite{Rosero16}, while the green square represents the 7~mm integrated flux obtained from the 2D Gaussian fit detailed in section~\ref{sec:results-cont}.
    The dashed-gray and continuous-blue lines show the power law fit and ionized gas sphere model from \cite{Rosero2019}, respectively. 
    }
    \label{app:18345-cont}
\end{figure}

\begin{figure}[H]
    \centering
    \includegraphics[width=0.49\textwidth]{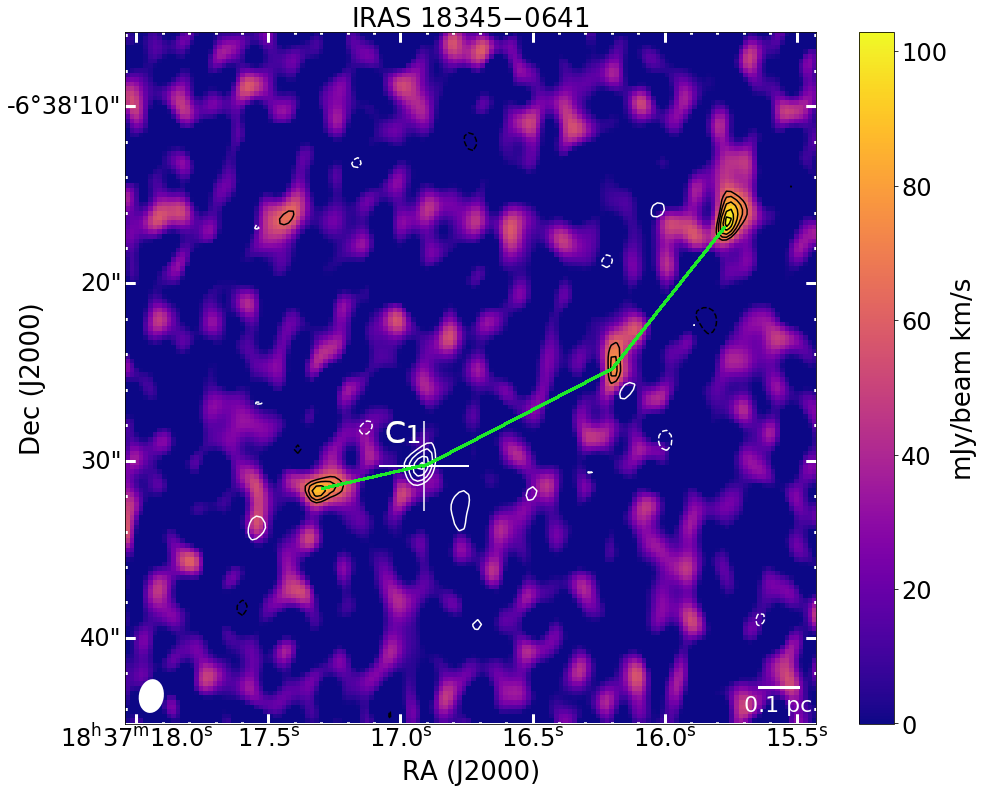}
    \includegraphics[width=0.5\textwidth]{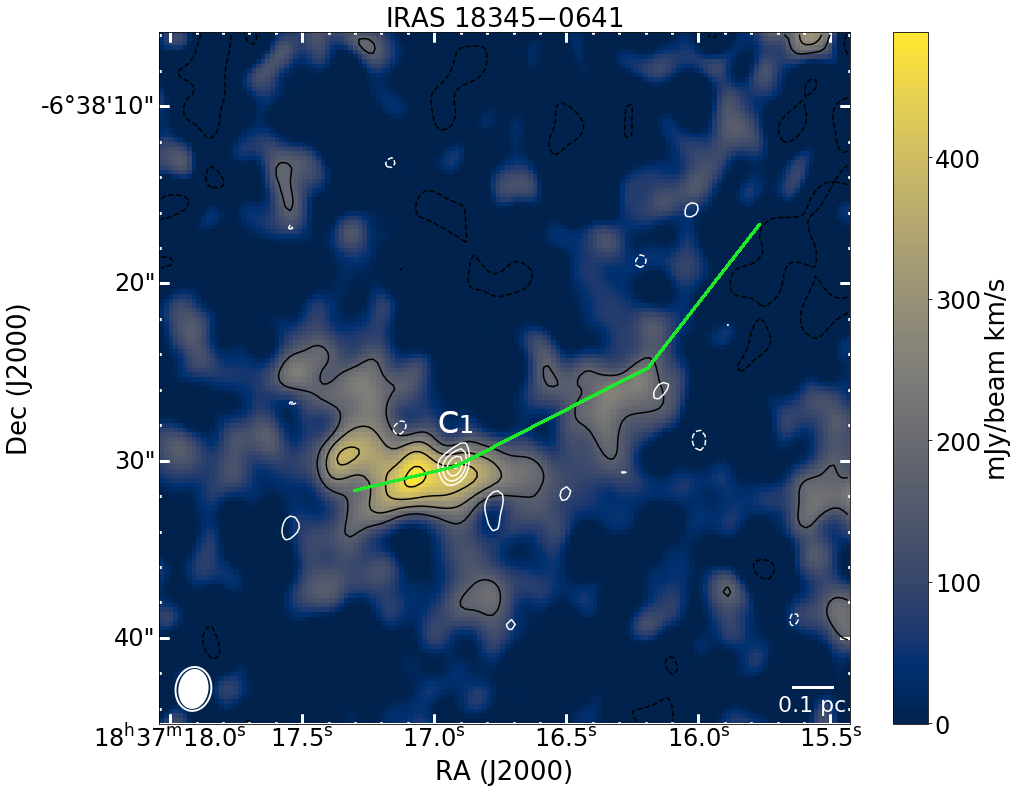}\\
    \includegraphics[width=0.43\textwidth]{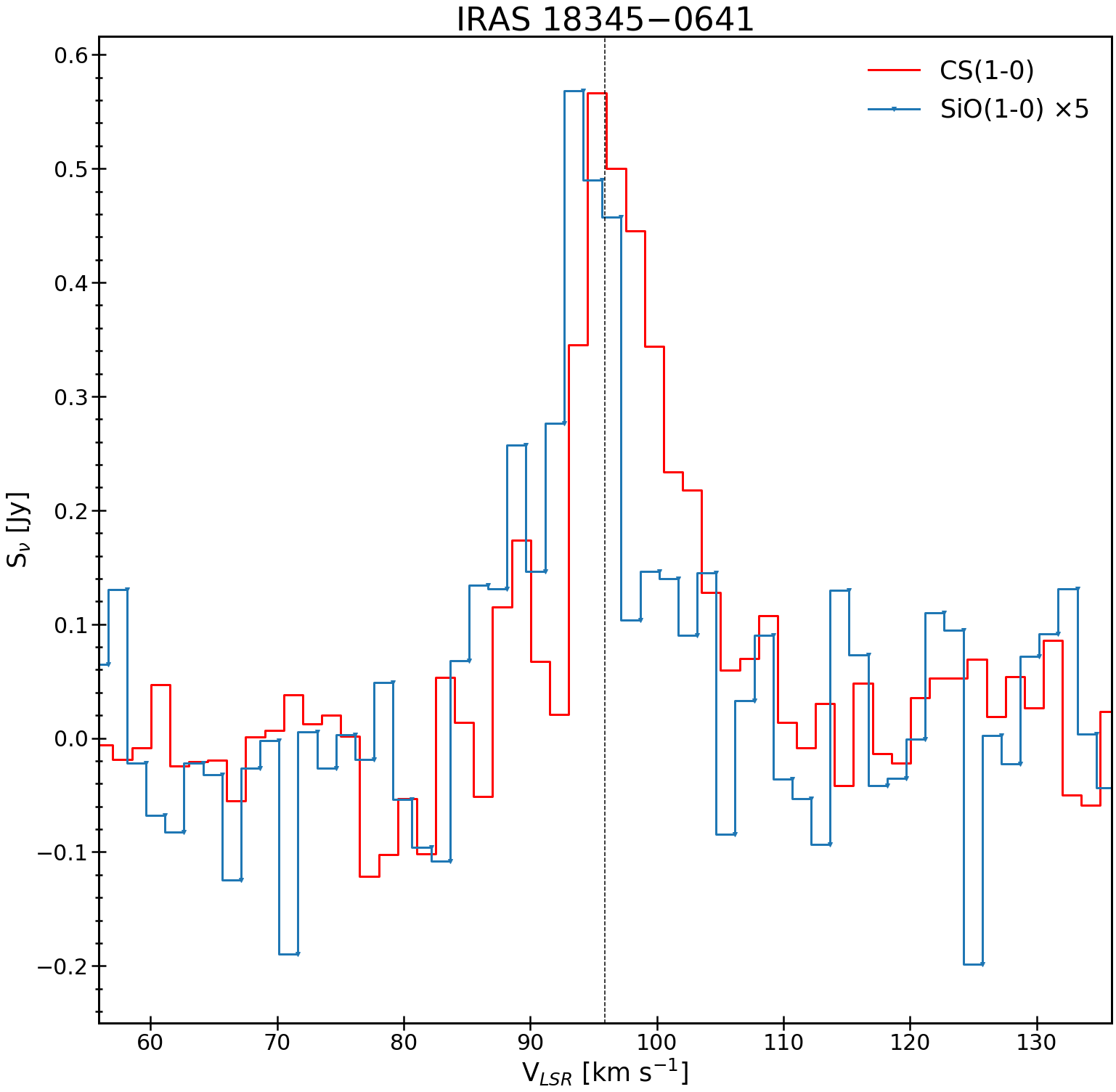}
    \caption{\textit{Top left:} SiO(1$-$0) integrated intensity in 18345 is shown in color and black contours. The latter represent [$-$3, 3, 3.5, 4, 4.7]$\,\times\sigma_{SiO}$ levels, with $\sigma_{SiO}=20.8\,$mJy~beam$^{-1}\,$\kms. White contours are the same as in Fig.~\ref{app:18345-cont} left, and show the 7~mm continuum emission. The white cross marks the position of the radio continuum source. The filled, white ellipse in the bottom left represent the SiO and 7~mm continuum synthesized beam size.
    \textit{Top right:} CS(1$-$0) integrated intensity in 18345 in color and black contours. The black contours represent $-$35\%, 35\%, 55\%, 75\%, and 95\% of the peak CS(1$-$0) emission, with the peak emission and rms of the map being 490 and 108~\mjy~\kms, respectively. White contours show the 7~mm continuum emission and are the same as in Fig.~\ref{app:18345-cont} left, while the outlined and filled white ellipses in the bottom left represent the CS and 7~mm continuum synthesized beam size, respectively.
    In both top panels the field of view is the same, and the green lines were drawn to guide the reader on the possible flow directions.
    \textit{Bottom:} spectra of the SiO ($\times5$) and CS $J=1-0$ emission in 18345 in blue and red, respectively. The spectra were obtained integrating over all the emission that appears to be connected to C$_1$. The vertical dotted line marks the systemic velocity.
    }
    \label{app:18345-mol}
\end{figure}

We detected one 7~mm core in this region, C$_1$, shown in the left panel of Figure~\ref{app:18345-cont}. The radio continuum and 7~mm peaks are essentially coincident, and the SED is shown in the right panel.  
We detected both SiO(1$-$0) and CS(1$-$0) emission associated with C$_1$.
We identified three SiO knots, one to the East and two to the North-West of the core (see Fig~\ref{app:18345-mol}, top left).
The CS emission has a smooth appearance (see Fig.~\ref{app:18345-mol}, top right) with the bulk of the emission extending to the East of the core, and the peak located near the Eastern SiO knot. In the bottom panel we show the spectra of the SiO ($\times5$) and CS line emission in blue and red, respectively. 
These were obtained integrating over the three SiO knots shown in the top left panel, and over the CS emission that appears to be connected with C$_1$ and enclosed in the 35\% contour level shown in the top right panel.

NIR observations show a faint H$_2$ $2.12\,\mu$m feature to the East of the core \citep{Varricatt10,Varricatt13}, which position is consistent with the Eastern SiO knot as well. 
Single dish \citep{Beuther02, Varricatt13,Szymczak07} and interferometric \citep{Morgan21} observations reveal the presence of an outflow in this region, elongated in the East-West direction. 
H$_2$O and OH masers were detected in single dish observations \citep{Sridharan02}, and subsequent interferometric observations detected $6.7$ and $44\,$GHz methanol maser emission \citep[][]{Beuther02c,Rodriguez-Garza17}. Additionally, a VLBI 6.7 GHz methanol maser ring centered on C$_1$ was reported by \cite{Bartkiewicz09}.

\newpage

\subsection{IRAS 18440$-$0148}
\renewcommand{\thefigure}{\Alph{section}\arabic{subsection}.\arabic{figure}}
\setcounter{figure}{0}

\begin{figure}[h]
    \centering
    \includegraphics[width=0.53\textwidth]{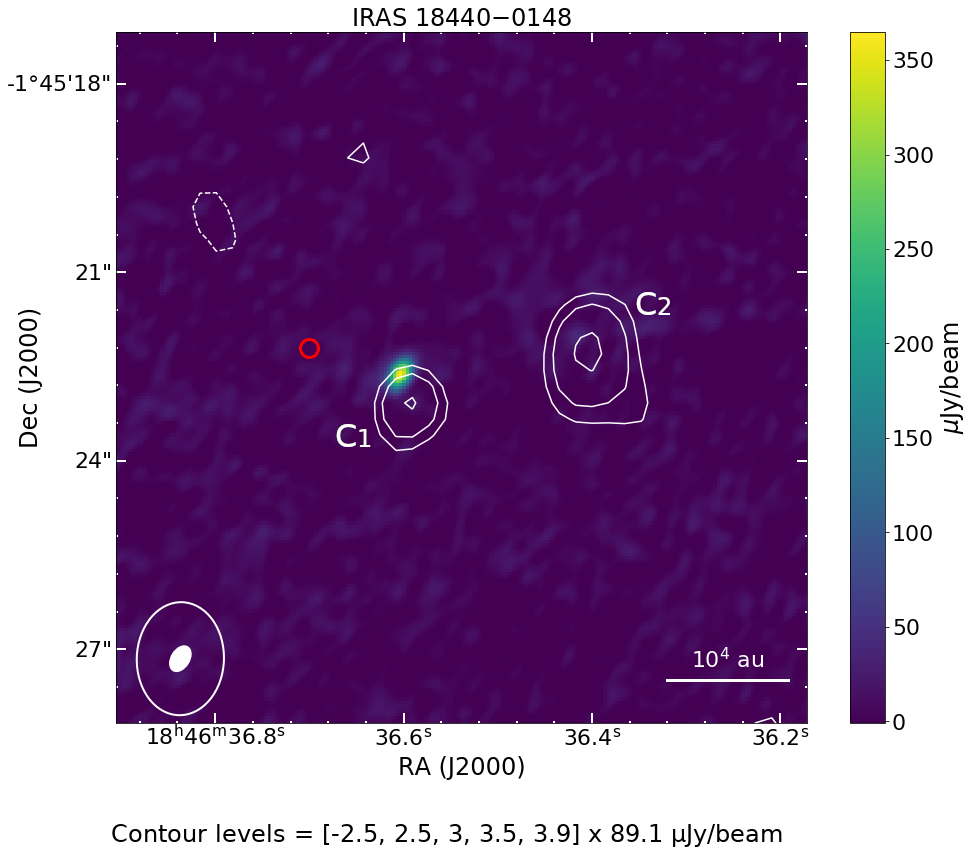}
    \includegraphics[width=0.45\textwidth]{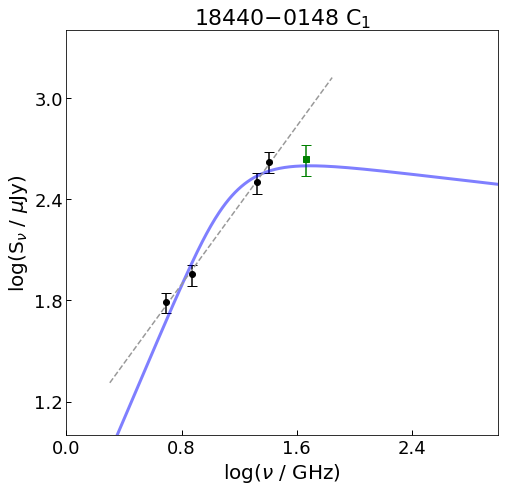}
    \caption{Same as Fig.~\ref{app:18345-cont}, but for 18440. The red circle shows the position of the 6.7 GHz methanol maser from \cite{Beuther02c}.
    }
    \label{app:18440-cont}
\end{figure}

\begin{figure}[H]
    \centering
    \includegraphics[width=0.57\textwidth]{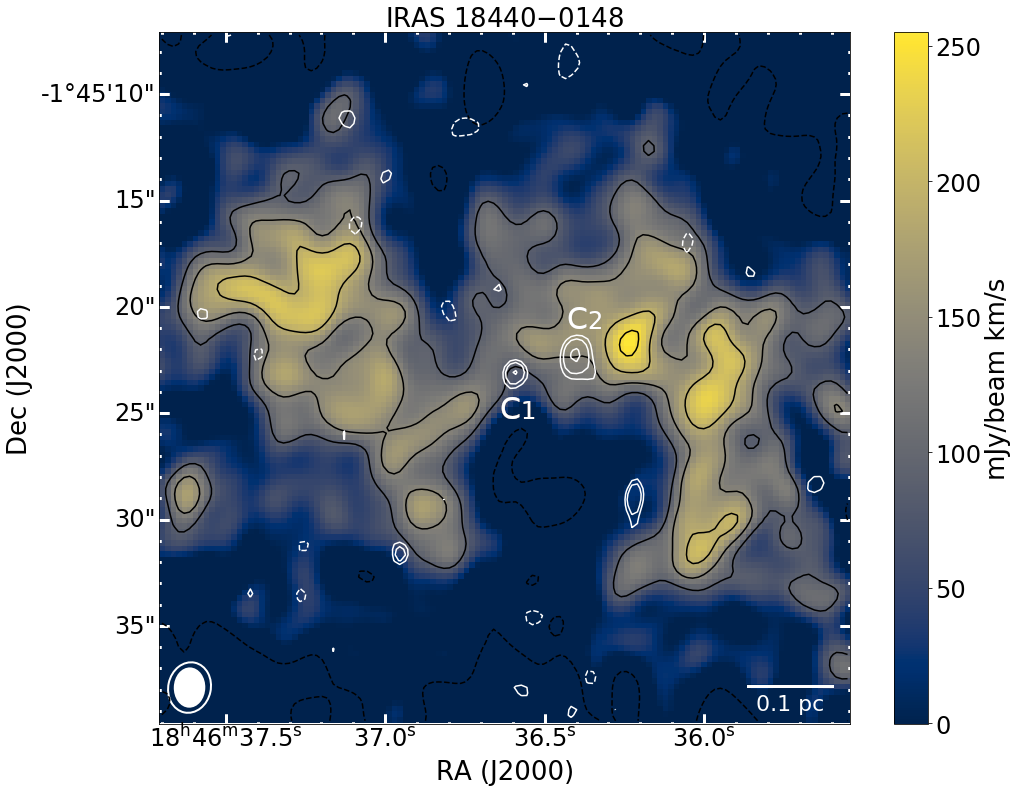}
    \includegraphics[width=0.42\textwidth]{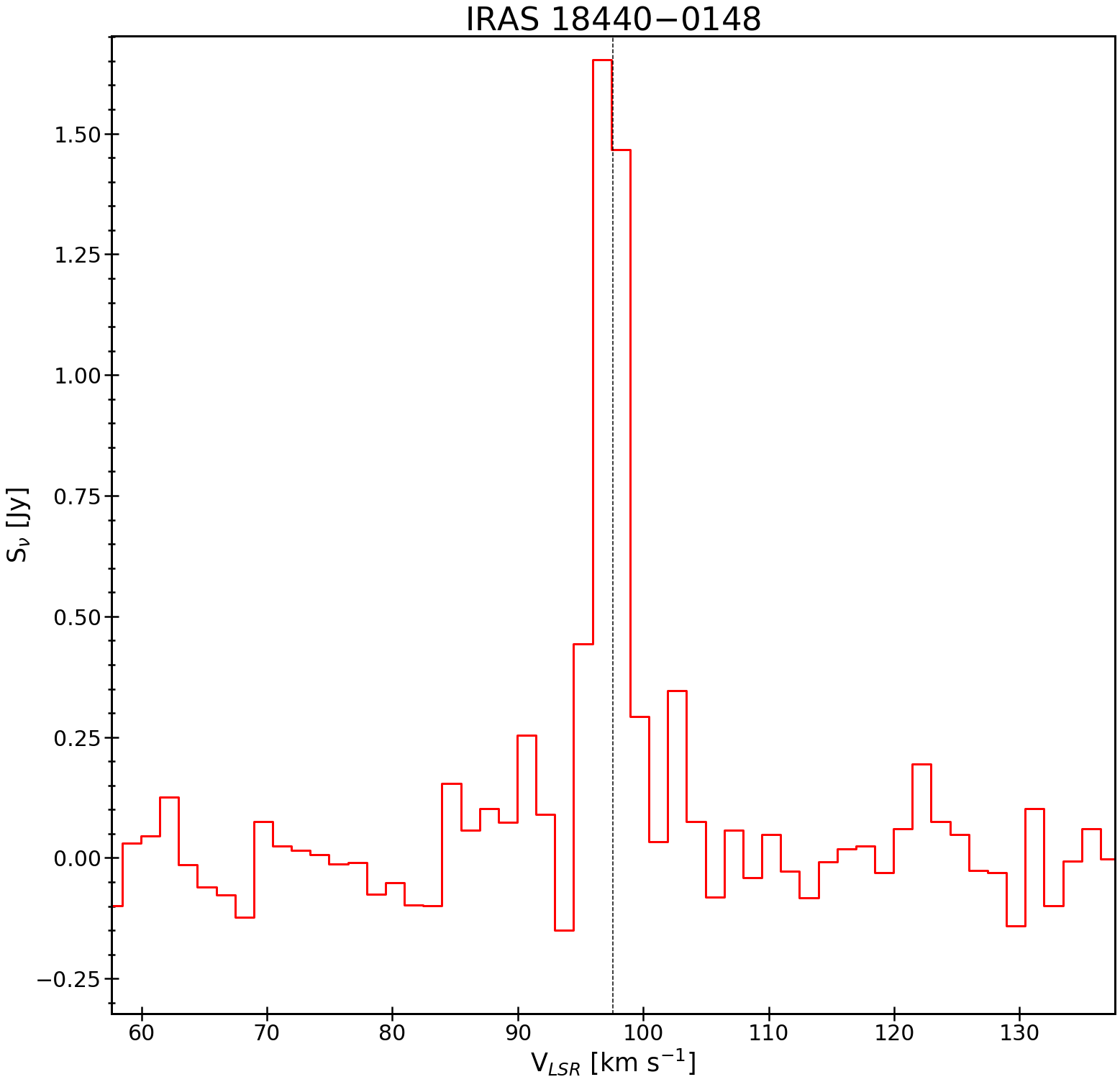}
    \caption{\textit{Left:} CS(1$-$0) integrated intensity in 18440 is shown in color and black contours. The latter represent $-$35\%, 35\%, 55\%, 75\%, and 95\% of the peak CS(1$-$0) emission, with the peak emission and rms of the map being 255 and 92~\mjy~\kms, respectively. White contours show the 7~mm continuum emission, and are the same as Fig.~\ref{app:18440-cont} (left). The outlined and filled white ellipses in the bottom left represent the CS and 7~mm continuum synthesized beam size, respectively.
    \textit{Right:} spectrum of the CS(1$-$0) line emission in 18440, obtained integrating over all the emission enclosed by the 35\% contour level from the left panel.
    The vertical dotted line marks the systemic velocity.
    }
    \label{app:18440-mol}
\end{figure}

We detected two 7~mm cores in this region, C$_1$ and C$_2$, shown in the left panel of Figure~\ref{app:18440-cont}. 
The C$_1$ core is coincident with the ionized source within the 7~mm synthesized beam, while C$_2$ does not have a radio continuum counterpart. 
In the right panel, we show the SED of C$_1$. The 7~mm flux is consistent with the ionized gas sphere model, which indicates that the emission is entirely ionized and there is no dust contribution detected at 7~mm toward C$_1$.
We did not detect SiO(1$-$0) emission in this region. If this is due to a lack of sensitivity, it would imply emission weaker than $\sim20\,$\mjy. 
We detected CS(1$-$0) emission, which integrated intensity map is shown in the left panel of Figure~\ref{app:18440-mol}. The emission is rather extended, and is found both East and West from the cores. In the right panel, we show the spectrum of the CS line emission, obtained  integrating over all the emission enclosed by the 35\% contour level from the left panel. 
\cite{Sridharan02} and \cite{Yang18} detected CO(2$-$1) and $^{13}$CO(3$-$2) line wings in their single-dish observations.
Additionally, \cite{Beuther02c} detected 6.7 GHz methanol maser emission in this region using ATCA.

\newpage

\subsection{IRAS 18517$+$0437}
\renewcommand{\thefigure}{\Alph{section}\arabic{subsection}.\arabic{figure}}
\setcounter{figure}{0}
We detected one 7~mm core, C$_1$, shown in the left panel of Figure~\ref{app:18517-cont}. The radio continuum source is located in the center of the core. In the right panel, we present the SED of C$_1$, where we see the flux rising at 7~mm, as expected from emission dominated by dust. 
We detected a highly collimated, monopolar SiO(1$-$0) flow in this region, located to the South-East of the core. The SiO integrated intensity is shown in the left panel of Figure~\ref{app:18517-mol}.
We also detected CS(1$-$0) emission associated to the core, shown in the right panel. The CS peak coincides with the continuum peak within a North-South filament. 
In the left panel of Figure~\ref{app:18517-IR} we show the spectra of the SiO ($\times5$) and CS line emission associated with C$_1$. The lines present different characteristics: the monopolar SiO emission shows a blue wing, and the CS line presents a more symmetric profile. 
In the right panel we show a Spitzer IRAC color composite image of the region. There is green band excess consistent with an extended green object (EGO) in a direction about 45$^\circ$ from the SiO flow. 

\cite{Sridharan02} reported the detection of CO(2$-$1) line wings in this region using single-dish observations. Subsequently, \cite{Lopez-Sepulcre10} detected HCN(1$-$0) and C$\,^{18}$O(2$-$1) in 18517, as well as a North-South HCO$^+$(1$-$0) outflow, using the IRAM 30~m telescope. 
Additionally, water and methanol masers were detected in single-dish \citep[e.g.,][]{Sridharan02,Tan20} and radio interferometric \citep[e.g.,][]{Surcis15,Rodriguez-Garza17} observations. In a more recent study, Sanchez-Tovar, E. et al. (submitted) detected CH$_3$OH 25 GHz emission toward C$_1$.

\begin{figure}[H]
    \centering
    \includegraphics[width=0.53\textwidth]{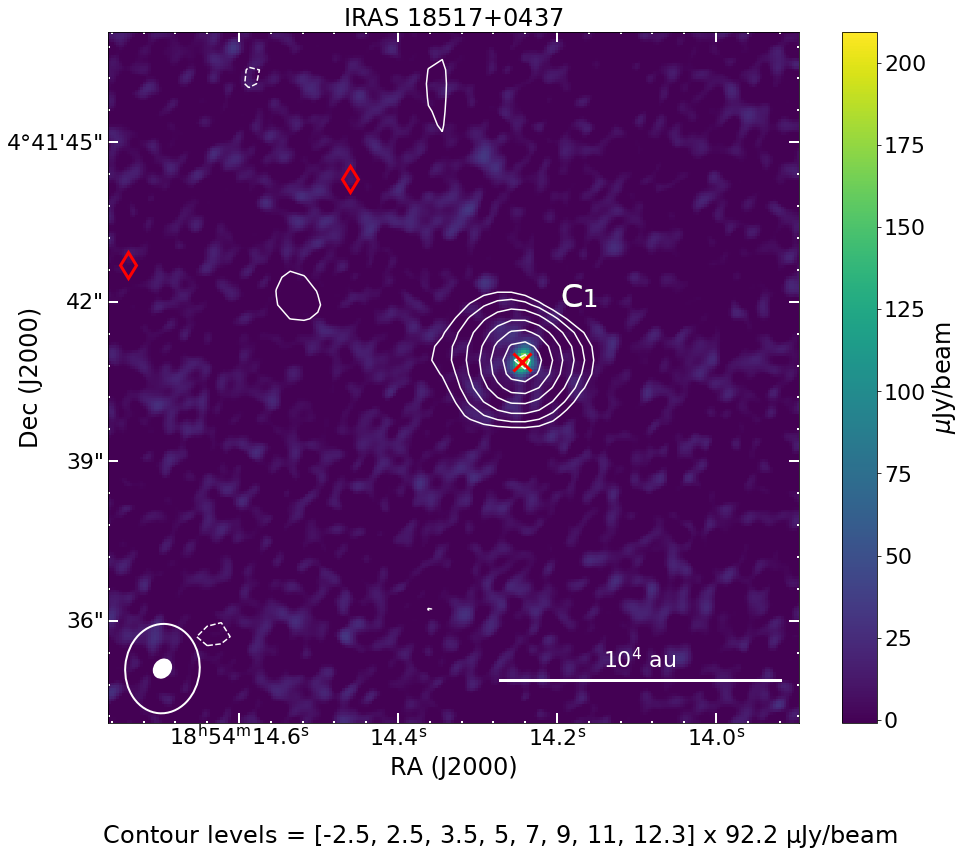}
    \includegraphics[width=0.45\textwidth]{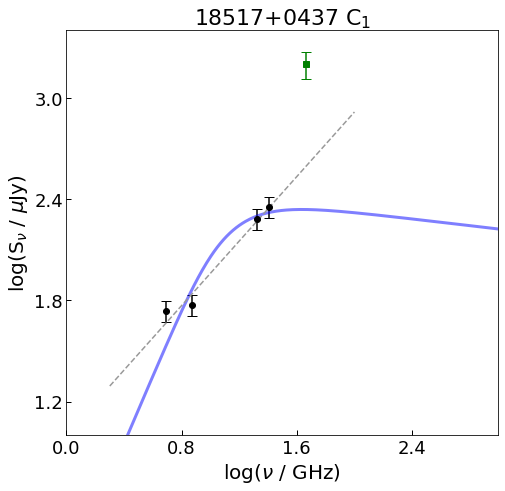}
    \caption{Same as Fig.~\ref{app:18345-cont}, but for 18517. The red cross and diamonds in the left panel mark the position of the 25 and 44 GHz methanol emission from Sanchez-Tovar, E. et al. (submitted) and \cite{Rodriguez-Garza17}.
    }
    \label{app:18517-cont}
\end{figure}

\begin{figure}[H]
    \centering
    \includegraphics[width=0.49\textwidth]{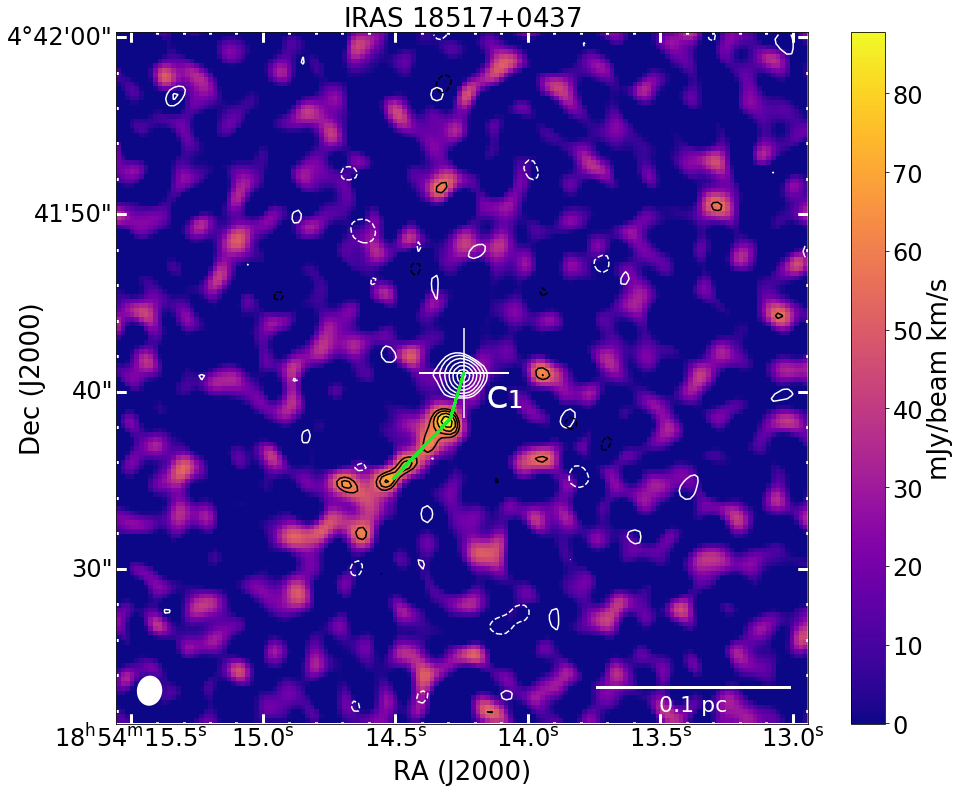}
    \includegraphics[width=0.5\textwidth]{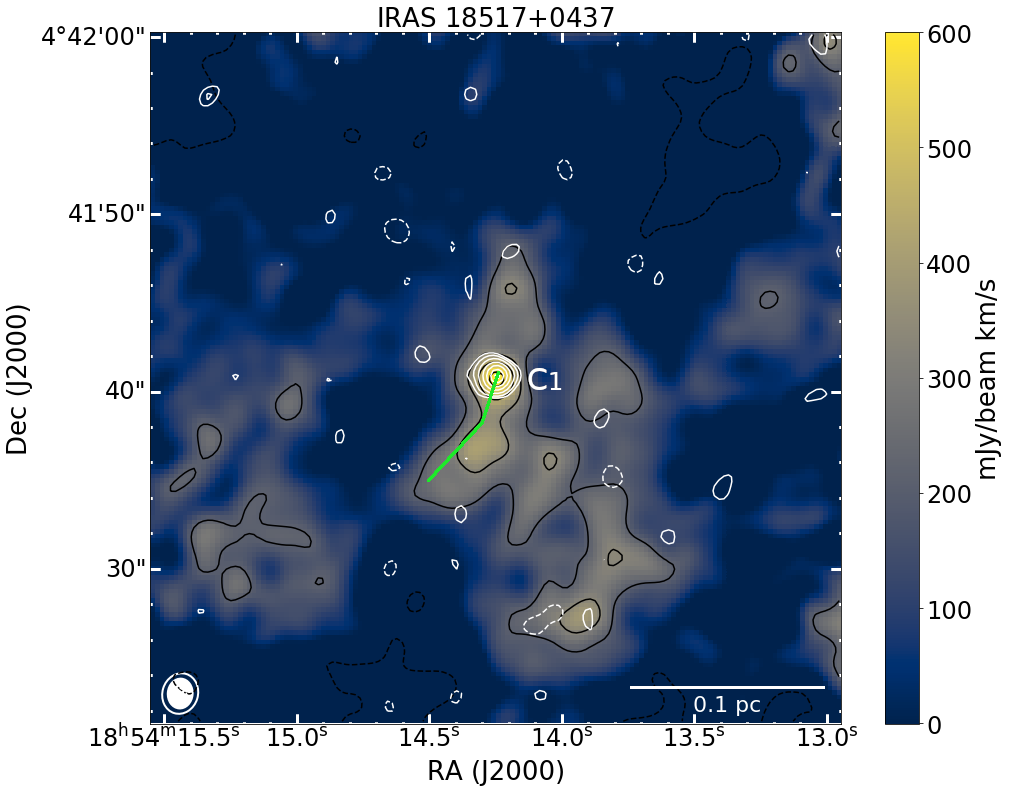}
    \caption{\textit{Left:} SiO(1$-$0) integrated intensity in 18517 in color and black contours. The latter represent $[-3, 3, 3.5, 4, 4.5]\,\times\sigma_{SiO}$ levels, with $\sigma_{SiO} = 18.1\,$\mjy~\kms. The white cross shows the position of the radio continuum source and the white ellipse in the bottom left represents the synthesized beam size. White contours show the 7~mm continuum emission are the same as in the left panel of Fig.~\ref{app:18517-cont}.
    \textit{Right}: CS(1$-$0) integrated intensity in 18517 in color and black contours. Black contours represent $-$35\%, 35\%, 55\%, 75\%, and 95\% of the peak CS(1$-$0) emission, with the peak emission and rms of the map being 600 and 147~mJy~beam$^{-1}\,$\kms, respectively. The white contours are the same as in the left panel, while the outlined and filled white ellipses in the bottom left represent the CS and 7~mm continuum synthesized beam sizes, respectively.
    In both panels the field of view is the same, and the green lines were drawn to guide the reader on the possible flow directions.
    }
    \label{app:18517-mol}
\end{figure}

\begin{figure}[H]
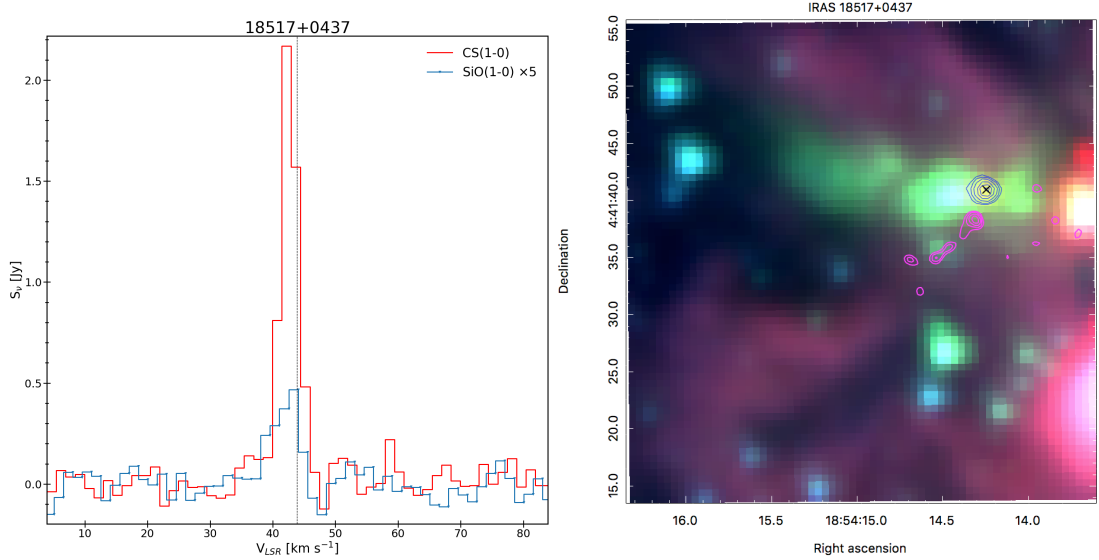

    \centering
    \includegraphics[width=0.4\textwidth]{spectra-18517-wCS.png}
    \includegraphics[width=0.4\textwidth]{MIR-18517-IR.png}
    \caption{
    \textit{Left}: spectra of the SiO ($\times5$) and CS emission in 18517 in blue and red, respectively. The spectra were obtained integrating over all the emission that appears to be connected to C$_1$.
    The vertical dotted line marks the systemic velocity.
    \textit{Right:} Spitzer IRAC color composite, showing the 3.6 (blue), 4.5 (green), and 8.0~$\mu$m (red) emission. The blue and magenta contour levels show the 7~mm continuum emission and SiO integrated intensity (respectively) and are the same as in the left panel of Fig.~\ref{app:18517-mol}.
    The black $\times$ marks the position of the radio continuum source. 
    }
    \label{app:18517-IR}
\end{figure}

\newpage

\subsection{IRAS 18553$+$0414}
\renewcommand{\thefigure}{\Alph{section}\arabic{subsection}.\arabic{figure}}
\setcounter{figure}{0}
We detected one 7~mm source, C$_1$, shown in the left panel of Figure~\ref{app:18553-cont}. The radio continuum and 7~mm peaks coincide in position, and C$_1$ appears to be slightly elongated to the West. In the right panel we show the SED of the core, where the 7~mm flux raises above the ionized emission from \cite{Rosero16}, indicating it is dominated by thermal dust.
Both SiO(1$-$0) and CS(1$-$0) emission were detected, and their integrated intensity maps are presented in the top panels of Figure~\ref{app:18553-mol}. 
We identified two SiO knots to the East of the core, with sizes $\gtrsim0.1\,$pc. There is no clear SiO counterpart to the West of the core. The CS emission, on the other hand, presents a more continuous morphology. Its peak emission coincides with the core center, and it appears to extend in the North-East direction. There is some emission to the East and West of C$_1$ as well. 
In the bottom panel of Figure~\ref{app:18553-mol}, we show the spectra of the SiO ($\times5$) and CS line emission in blue and red, respectively. These were obtained integrating over both SiO knots and all the CS emission enclosed in the 35\% contour level from the top panels.
The bulk of the SiO emission is found toward blue-shifted velocities, while the CS line presents a more symmetric morphology with respect to the systemic velocity.

\cite{Sridharan02} detected CO(2$-$1) line wings in this region using the IRAM 30~m telescope. The authors also report the detection of a water maser in this region, confirmed using VLA observations by \cite{Beuther02c}. In addition, \cite{Rodriguez-Garza17} detected a 44 GHz methanol maser close to C$_1$, and Sanchez-Tovar, E. et al. (submitted) detected methanol 25 GHz emission.

\begin{figure}[H]
    \centering
    \includegraphics[width=0.53\textwidth]{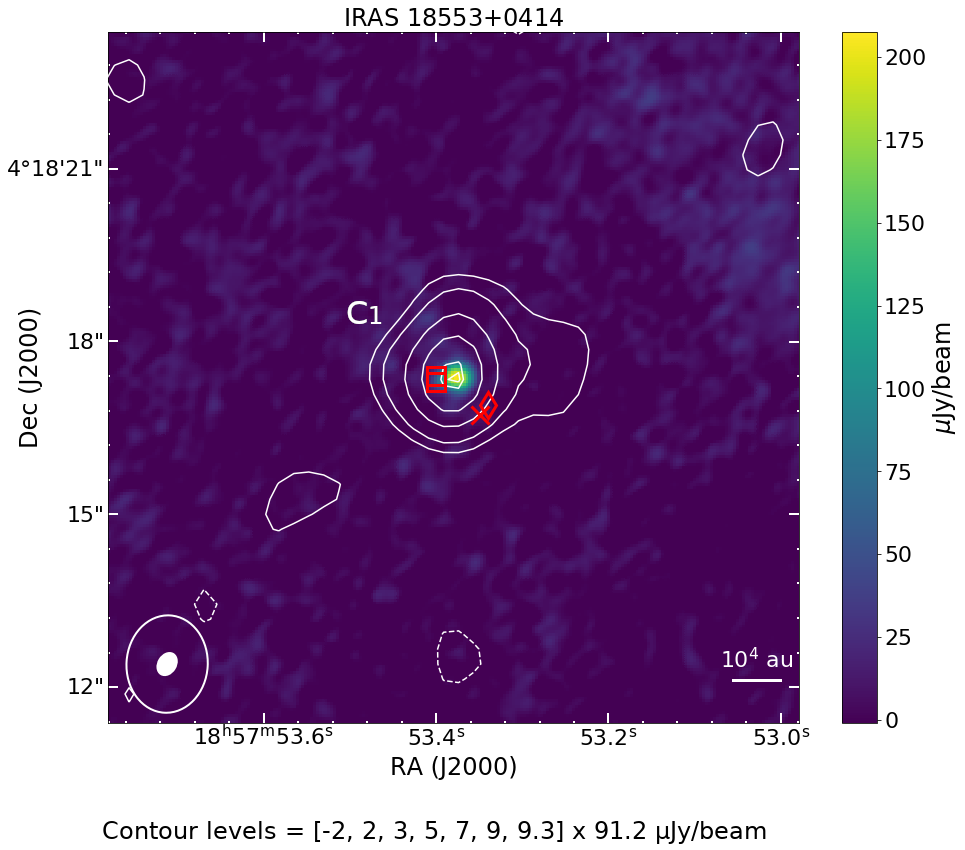}
    \includegraphics[width=0.45\textwidth]{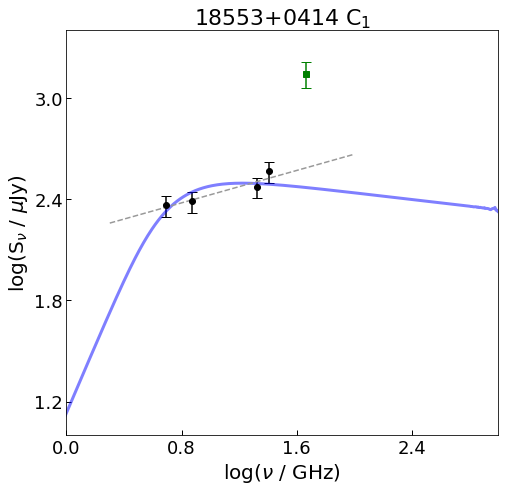}
    \caption{Same as Fig.~\ref{app:18345-cont} but for 18553. The red square, diamond, and $\times$ mark the position of the 22 GHz water maser, 44~GH methanol maser, and 25~GHz methanol emission from \cite{Beuther02c}, \cite{Rodriguez-Garza17}, and Sanchez-Tovar, E. et al. (submitted), respectively.
    }
    \label{app:18553-cont}
\end{figure}

\begin{figure}[H]
    \centering
    \includegraphics[width=0.49\textwidth]{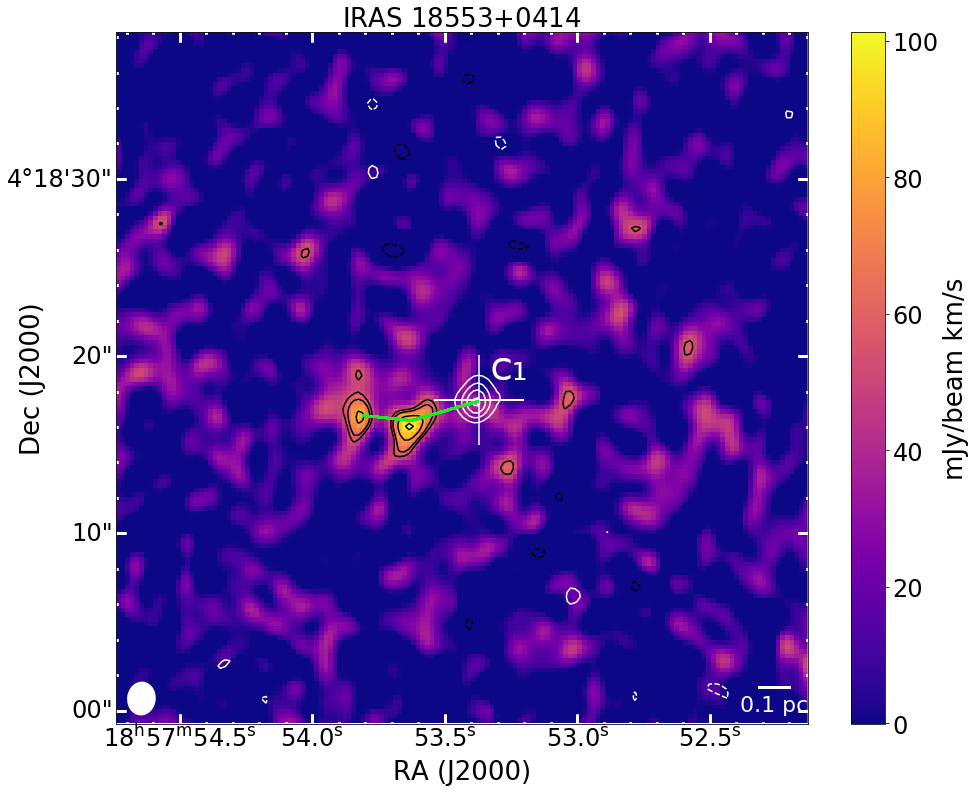}
    \includegraphics[width=0.5\textwidth]{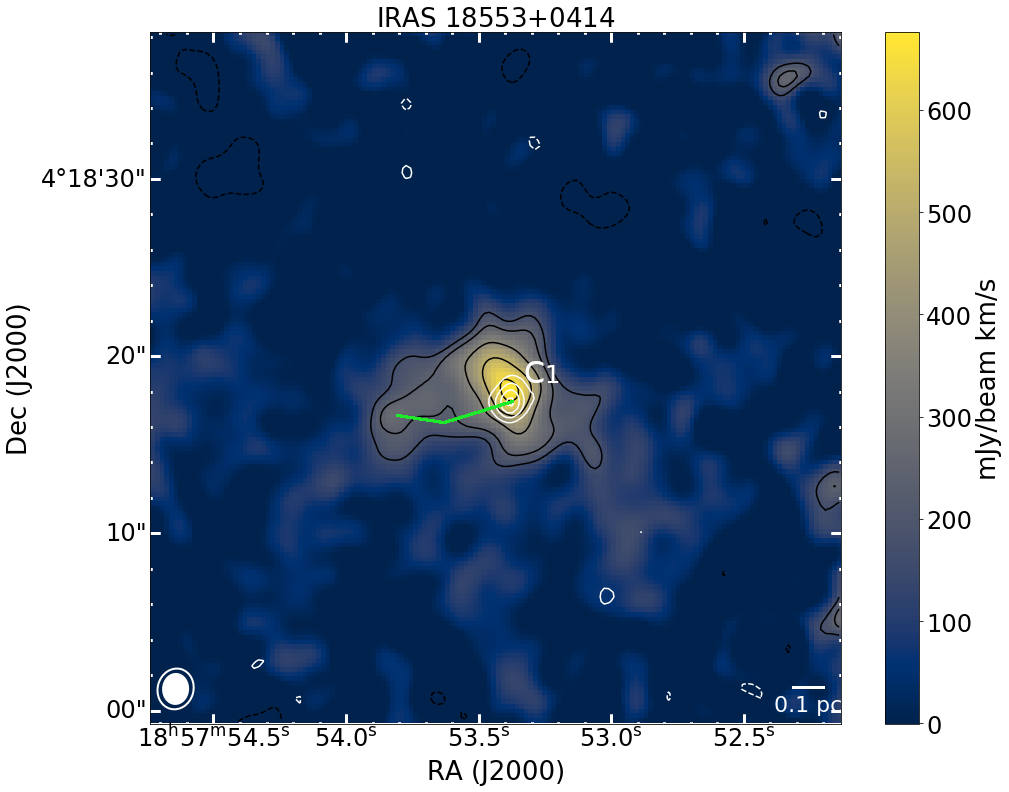}\\
    \includegraphics[width=0.42\textwidth]{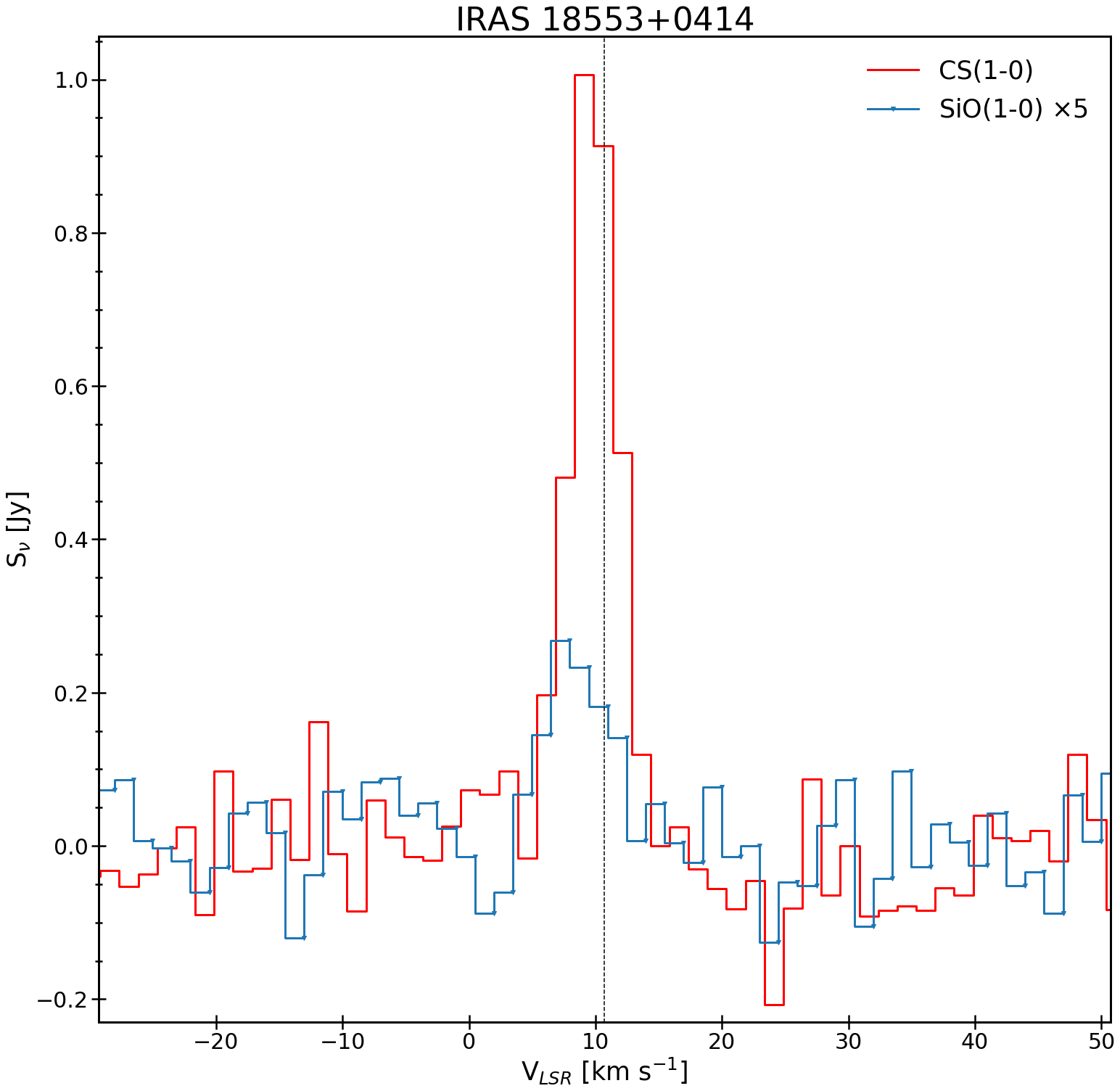}
    \caption{\textit{Top left:} SiO(1$-$0) integrated intensity in 18553 in color and black contours. Black contours represent [$-$3, 3, 3.5, 4.5, 5.5]$\,\times\sigma_{SiO}$ levels, with $\sigma_{SiO} =18 \,$\mjy~\kms. White contours show the 7~mm continuum emission, and represent [$-3$, 3, 5, 7, 9, 9.3]$\,\times\sigma_{7mm}$ levels, with $\sigma_{7mm}=91.2\mu$Jy~beam$^{-1}$.
    The white cross marks the position of the radio continuum source and the white ellipse in the bottom left represents the synthesized beam sizes.
    \textit{Top right}: CS(1$-$0) integrated intensity in color and black contours. The latter represent $-$25\%, 25\%, 35\%, 55\%, 75\%, and 95\% of the peak CS(1$-$0) emission, with the peak emission and rms of the map being 676 and 107~mJy~beam$^{-1}\,$\kms, respectively. White contours show the 7~mm continuum emission are the same as in the left panel. The outlined and filled white ellipses in the bottom left represent the CS and 7~mm continuum synthesized beam size, respectively.
    In both top panels the field of view is the same, and the green lines were drawn to guide the reader on the possible flow direction.
    \textit{Bottom:} spectra of the SiO ($\times5$) and CS $J=1-0$ line emission in blue and red, respectively. The former was obtained integrating over both SiO knots, while the latter represents all the emission surrounding C$_1$ and enclosed in the 35\% contour level.
    The vertical dotted line marks the systemic velocity.
    }
    \label{app:18553-mol}
\end{figure}

\newpage

\subsection{IRAS 19012$+$0536}
\label{sec:app-19012}
\renewcommand{\thefigure}{\Alph{section}\arabic{subsection}.\arabic{figure}}
\setcounter{figure}{0}
We detected two 7~mm sources in this region, C$_1$ and C$2$, shown in Figure~\ref{app:19012-cont}. They both have a radio continuum counterpart.
C$_1$, also referred to as G39.389$-$0.143, presents an extended morphology at both 7~mm and 1.3/6~cm. In Figure~\ref{app:19012-IR}, we show a Spitzer IRAC color composite image of this region, where C$_1$ appears bright and extended at 8~$\mu$m. 
Its structure and size indicate that this is likely an Ultra-Compact (UC) H~II region. Following the formulae in \cite{Kurtz94} and the tabulation in \cite{Panagia73}, we find that the ionizing star is consistent with a ZAMS B3 type star.

C$_2$ is associated with the jet candidate, which is located toward the center of the 7~mm core. C$_2$ appears to be slightly elongated to the East, and its SED is shown in Figure~\ref{app:19012-cont} as well. The 7~mm flux is consistent with the power-law fit, which indicates that the emission at this wavelength is ionized, i.e., no dust emission was detected toward this core.
We detected both SiO(1$-$0) and CS(1$-$0) emission emanating from C$_2$, as shown in the top panels of Figure~\ref{app:19012-mol}. 
The SiO emission is highly asymmetric, collimated, and elongated in the North-East/South-West direction. 
The CS emission peak is coincident with the center of C$_2$, and it also seems to be oriented in the North-East/South-West direction. In the bottom panel of Figure~\ref{app:19012-mol}, we show the spectra of the SiO ($\times5$) and CS line emission in 19012, which are both quite symmetric and slightly blueshifted from the systemic velocity. An extended green object (EGO) associated with C$_2$ is observed in a direction almost perpendicular to the outflow axis (see Figure~\ref{app:19012-IR}).

\cite{Beuther02} detected a North-East/South-West CO(2$-$1) outflow in this region using the IRAM 30~m telescope, and \cite{Lu14} observed extended ammonia emission oriented in the East/West direction and encompassing both cores. \cite{Beuther02c}, using the ATCA and the VLA, detected a water maser toward C$_2$. Later, \cite{Tan20} reported the detection of 6.7 GHz methanol maser emission with the Arecibo telescope, and Sanchez-Tovar, E. et al. (submitted) found 25 GHz methanol emission close to C$_2$ using wide band VLA data.

\begin{figure}[b]
    \centering
    \includegraphics[width=0.53\textwidth]{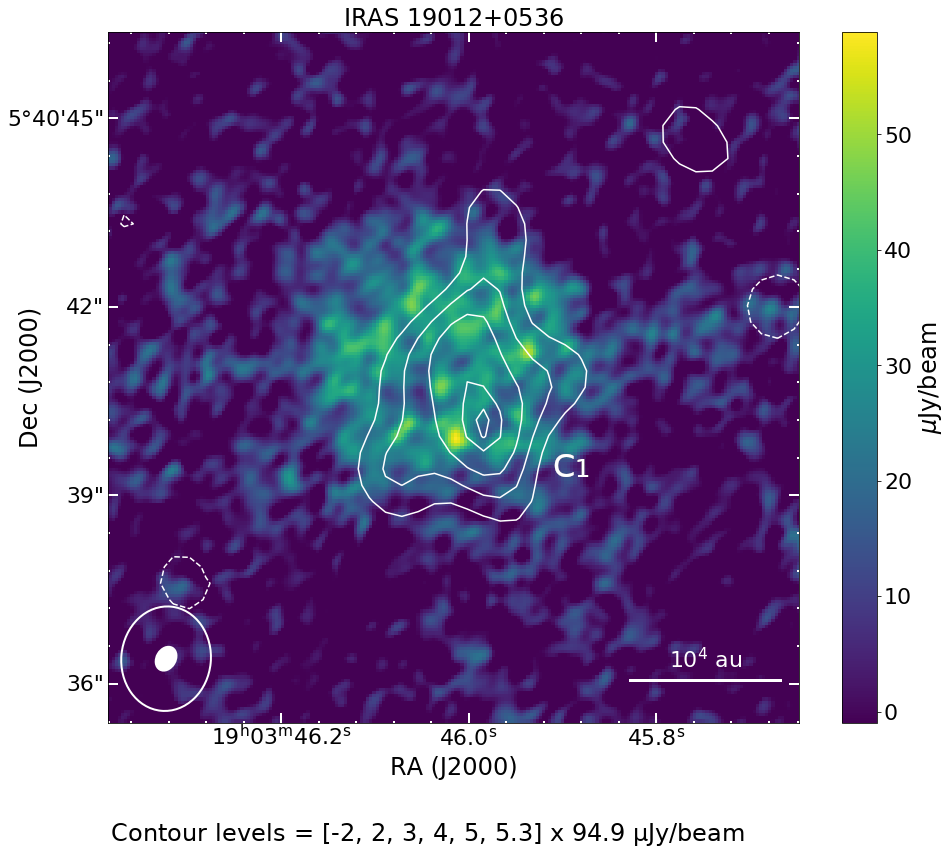}
    \includegraphics[width=0.43\textwidth]{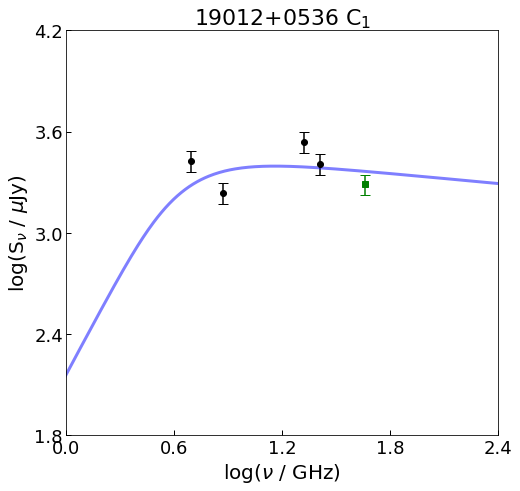}
    \caption{Same as the left panel of Fig.~\ref{app:18345-cont}, but for 19012.
    The red square and $\times$ mark the position of the 22~GHz water maser and 25 GHz methanol emission from \cite{Beuther02c} and Sanchez-Tovar, E. et al. (submitted), respectively. Note the different color scales and contour levels used in each panel.
    }
    \label{app:19012-cont}
\end{figure}

\begin{figure}[H]
\ContinuedFloat
    \centering
    \includegraphics[width=0.53\textwidth]{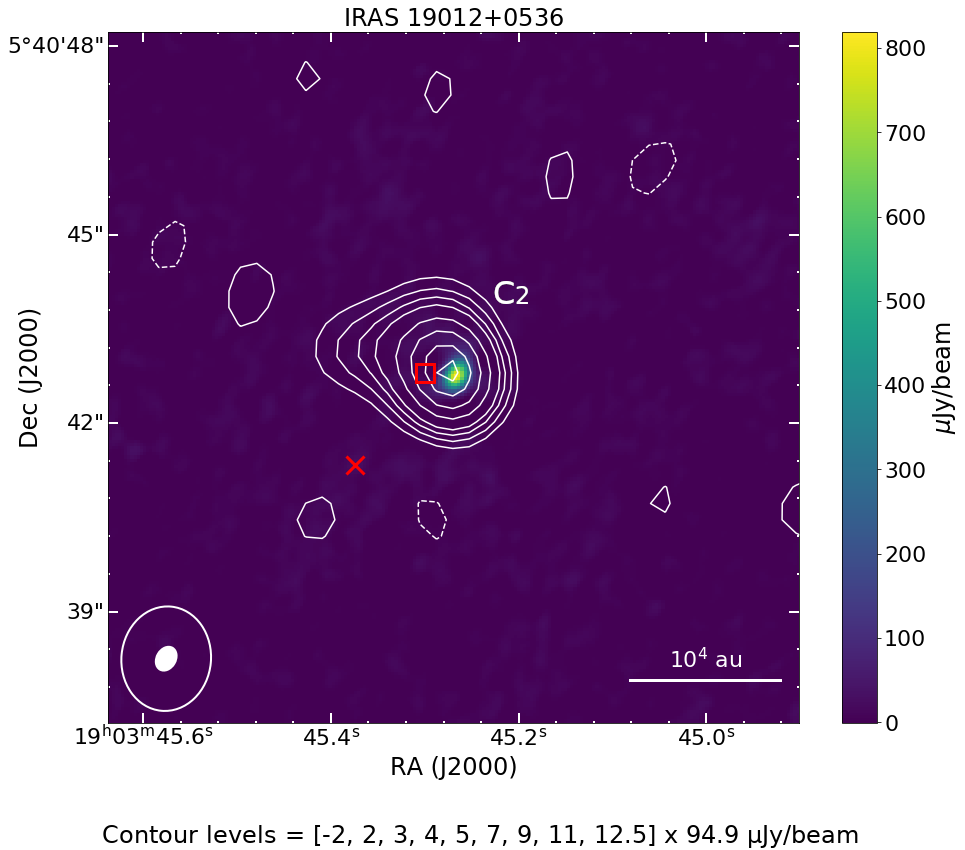}
    \includegraphics[width=0.43\textwidth]{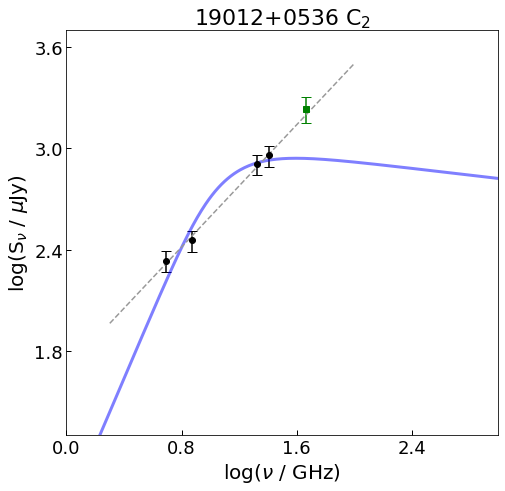}\\
    \caption{\textit{Continued.}
    }
    \label{app:19012-cont}
\end{figure}

\begin{figure}[H]
    \centering
    \includegraphics[width=.5\textwidth]{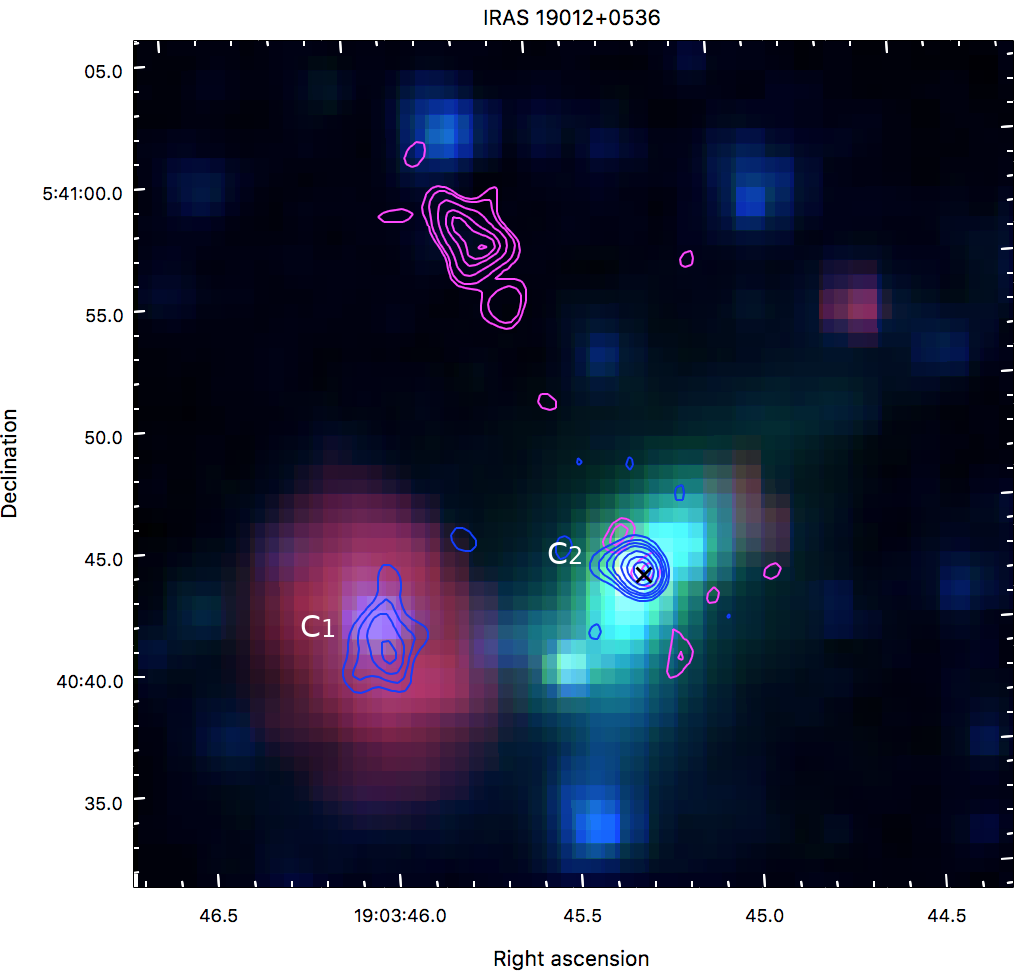}
    \caption{Spitzer IRAC color composite, showing the 3.6 (blue), 4.5 (green), and 8.0~$\mu$m (red) emission. Blue contours show the 7~mm continuum emission, and represent [2, 3, 4, 5, 7, 9, 11, 12.5]$\,\times\sigma_{7mm}$ levels, with $\sigma_{7mm}=94.9\,$\mjy. 
    The magenta contours show the SiO integrated intensity and represent $[2.5, 3, 4, 5, 5.5,6]\,\times\sigma_{SiO}$ levels, with $\sigma_{SiO}=32.4\,$\mjy~\kms. 
    The black $\times$ marks the position of the radio continuum source. 
    }
    \label{app:19012-IR}
\end{figure}

\begin{figure}[H]
    \centering
    \includegraphics[width=0.49\textwidth]{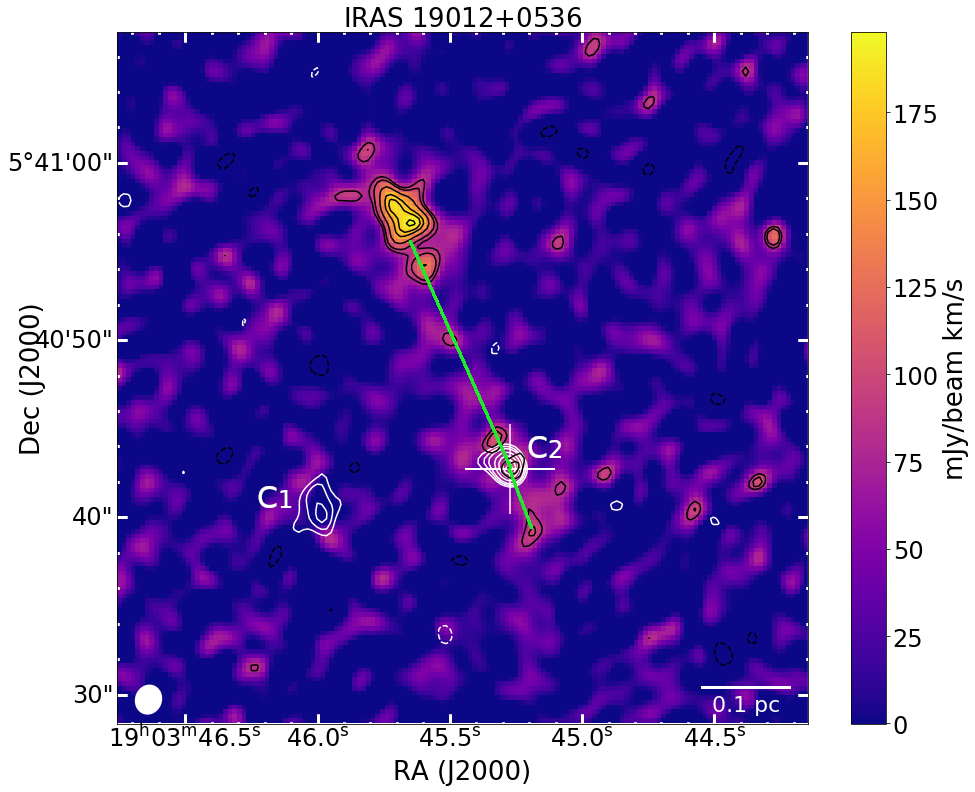}
    \includegraphics[width=0.5\textwidth]{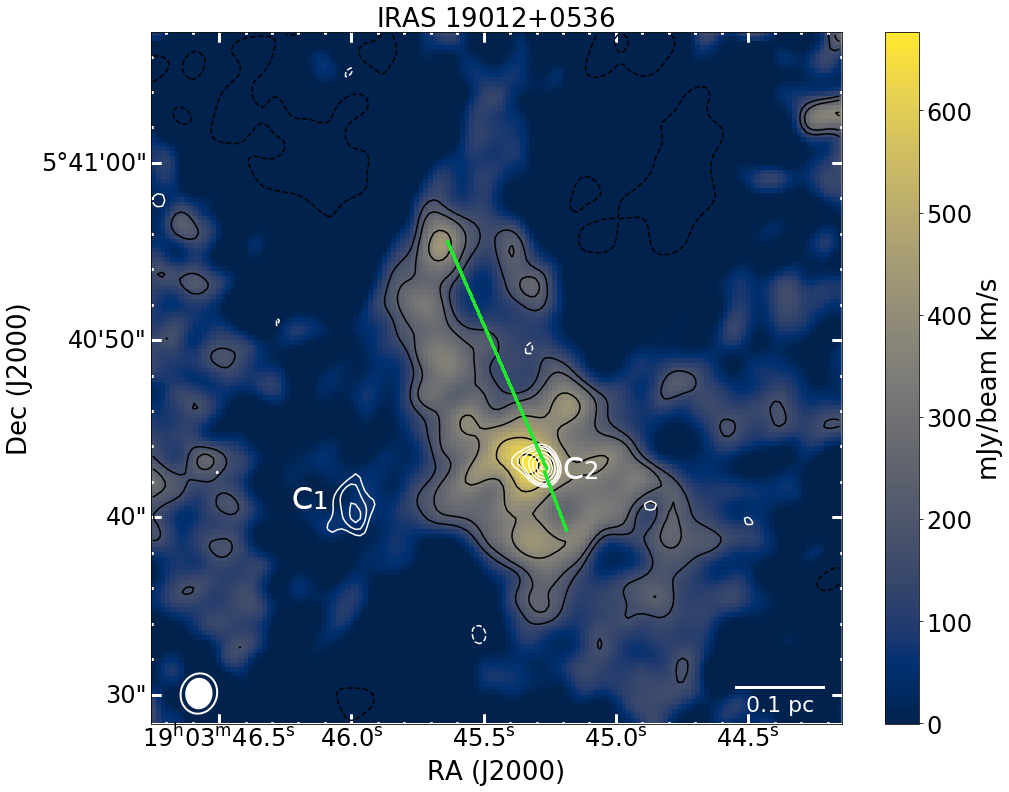}\\
    \includegraphics[width=0.4\textwidth]{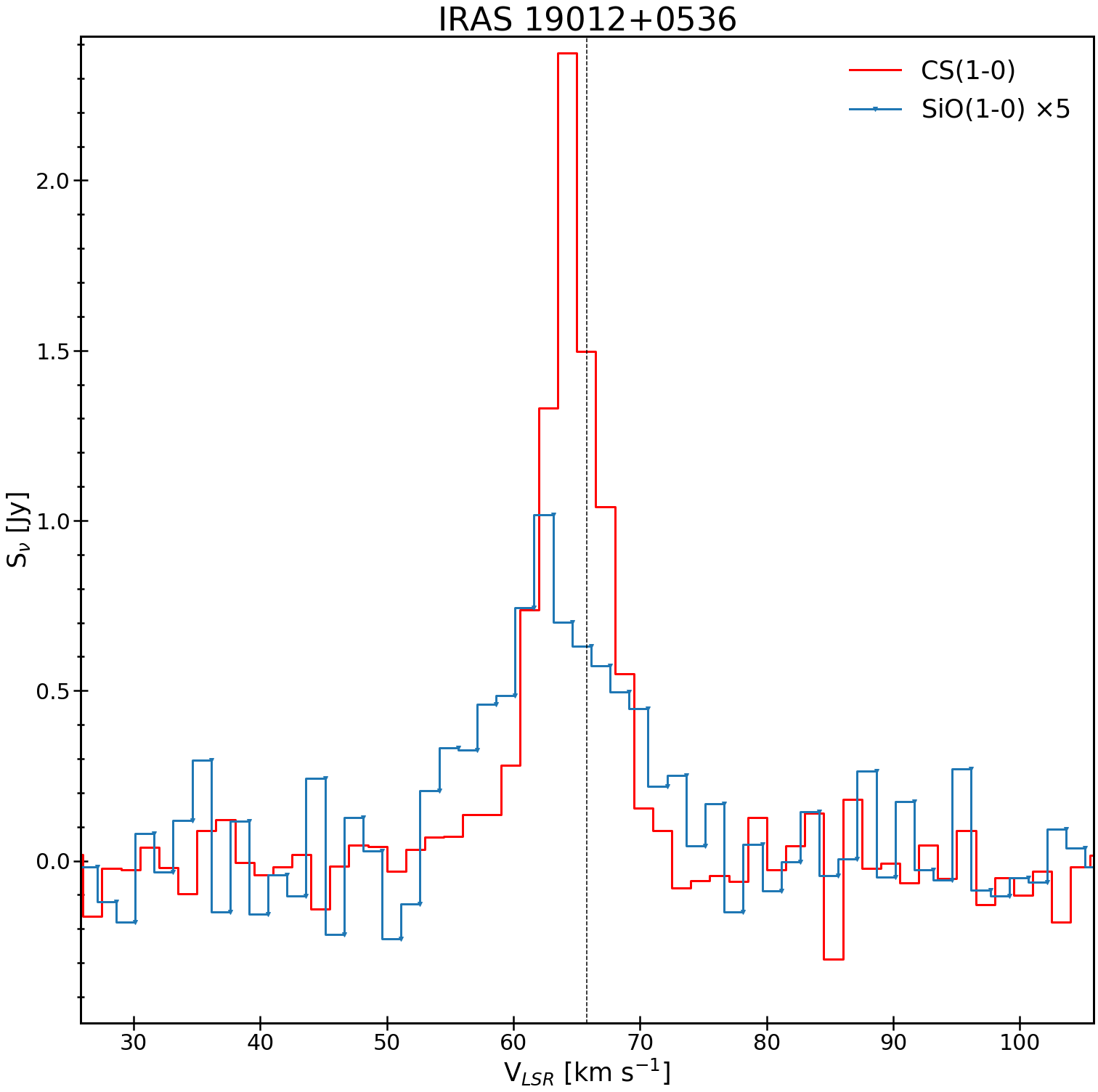}
    \caption{\textit{Top left:} SiO(1$-$0) integrated intensity map in color and black contours. The latter represent [$-$2.5, 2.5, 3, 4, 5, 5.5, 6]$\,\times\sigma_{SiO}$ levels, with $\sigma_{SiO}=32.4\,$mJy~beam$^{-1}\,$\kms. White contours show the 7~mm continuum emission, and are the same as in the right panel of Figure~\ref{app:19012-cont}.
    The white cross shows the position of the radio continuum source and the white ellipse in the bottom left represents the synthesized beam size, respectively.
    \textit{Top left:} CS(1$-$0) integrated intensity in color and black contours. The latter represent $-$35\%, 35\%, 55\%, 75\%, and 95\% of the peak CS(1$-$0) emission, with the peak emission and rms of the map being 676 and 166~mJy~beam$^{-1}\,$\kms, respectively. White contours show the 7~mm continuum emission and are the same as in the left panel.
    The outlined and filled white ellipses in the bottom left represent the CS and 7~mm continuum synthesized beam sizes, respectively.
    In both top panels the field of view is the same, and the green lines were drawn to guide the reader on the possible flow directions.
    \textit{Bottom:} spectra of the SiO ($\times5$) and CS line emission in 19012 in blue and red, respectively. These were obtained integrating over all the emission that appears to be connected with C$_2$.
    The vertical dotted line marks the systemic velocity.}
    \label{app:19012-mol}
\end{figure}

\newpage

\subsection{IRAS 19266$+$1745}
\label{19266}
\renewcommand{\thefigure}{\Alph{section}\arabic{subsection}.\arabic{figure}}
\setcounter{figure}{0}
We detected six individual cores in this region, shown in Figure~\ref{app:19266-cont}. Cores C$_1$, C$_2$, C$_3$, and C$_5$ have a radio continuum counterpart. The radio continuum source is essentially coincident with the $7\,$mm peak in the cases of C$_1$ and C$_3$, and it is slightly offset in the cases of C$_2$ and C$_5$.
As shown in Figure~\ref{app:19266-SED}, the radio continuum emission associated with C$_3$ has a flat spectrum ($\alpha=-0.1\,\pm0.1)$, while C$_1$ and C$_2$ have spectral indices of $-0.8\, (\pm0.1)$ and $-0.9\, (\pm0.1)$, respectively, which indicates non-thermal emission.
Similarly, C$_5$ presents a decreasing spectrum ($\alpha=-1.6\,\pm1.6)$ and a slightly extended structure at 1.3/6~cm.
A well-known source of non-thermal emission in star forming regions are the active magnetospheres of T-Tauri stars. As discussed in \cite{Rosero2019}, the probability of detecting a low-mass young star at a distance greater than 2~kpc is extremely low considering the sensitivity of their observations. The distance of IRAS 19266 is more than 4$\times$ greater than that limit, hence it is unlikely
that the cm continuum emission is caused by such a scenario.
Another source of non-thermal radiation is synchrotron emission, 
which has been detected in a small number of high-mass YSOs along the jet axis (e.g., IRAS 16547$-$4247, \citealt{Garay03}; IRAS 18162$-$2048 \citealt{RK17,Vig18}; G035.02$+$0.35, \citealt{Sanna19}). The non-thermal nature of the emission has only been conclusively associated with synchrotron radiation in the case of IRAS 18162$-$2048 (HH80/81), through the study of polarized emission \citep{Carrasco-Gonzalez10}, although this is inferred in all the other cases.

SiO(1$-$0) emission was not detected in the region. Each radio continuum source has a 7~mm counterpart, which indicates that they are individual objects and discarding the possibility of them being jet knots emanating from the same driving source. 
We detected CS(1$-$0) emission, shown in the left panel of Figure~\ref{app:19266-mol}. The bulk of the CS emission is located between the cores C$_2$ and C$_3$, and it is quite extended. 
Additionally, \cite{Tan20} detected 6.7~GHz methanol maser emission in their single-dish observations, and \cite{Lu14} found ammonia emission extended in the North-South direction.

\begin{figure}[b]
    \centering
    \includegraphics[width=0.49\textwidth]{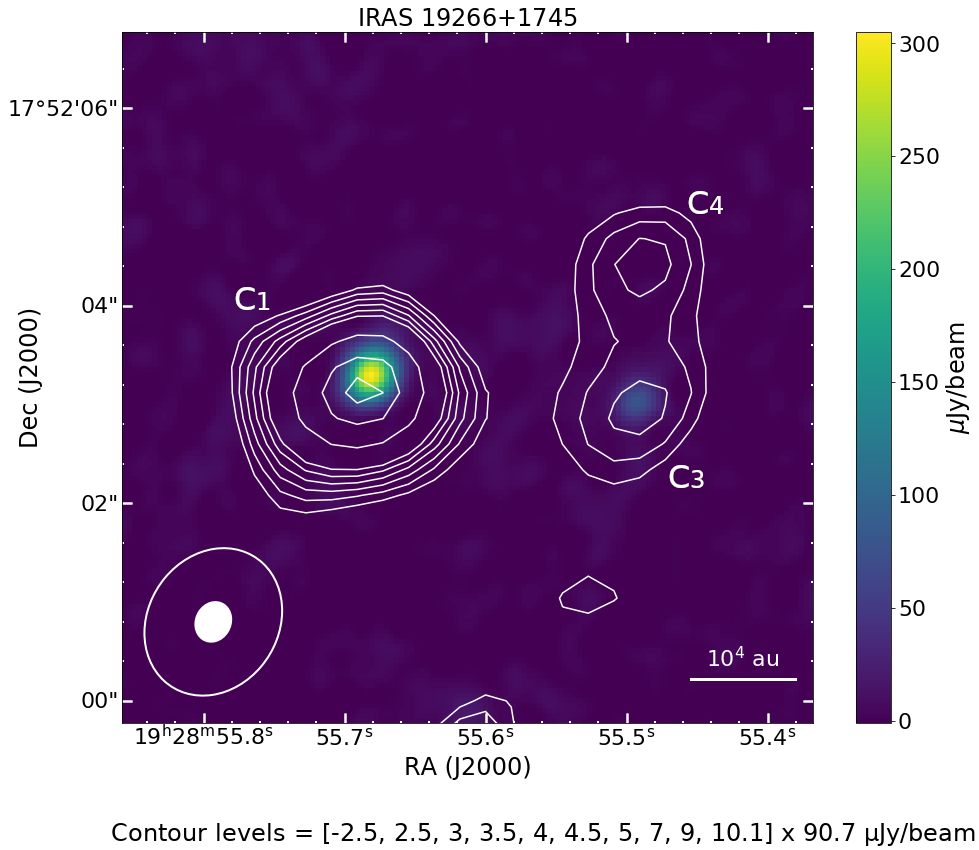}
    \includegraphics[width=0.49\textwidth]{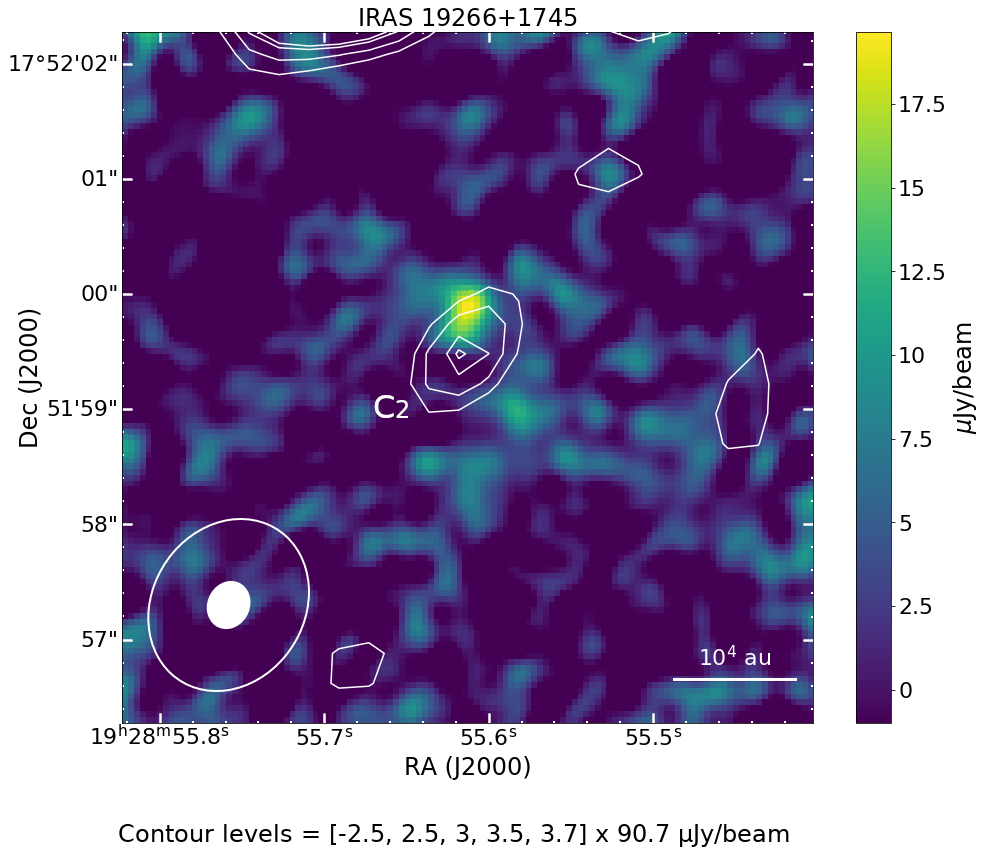}
    \caption{Same as in the left panel of Figure~\ref{app:18345-cont}, but for 19266. Note the different color scales and contour levels used in each panel. 
    }
    \label{app:19266-cont}
\end{figure}

\begin{figure}[H]
\ContinuedFloat
    \centering
    \includegraphics[width=0.5\textwidth]{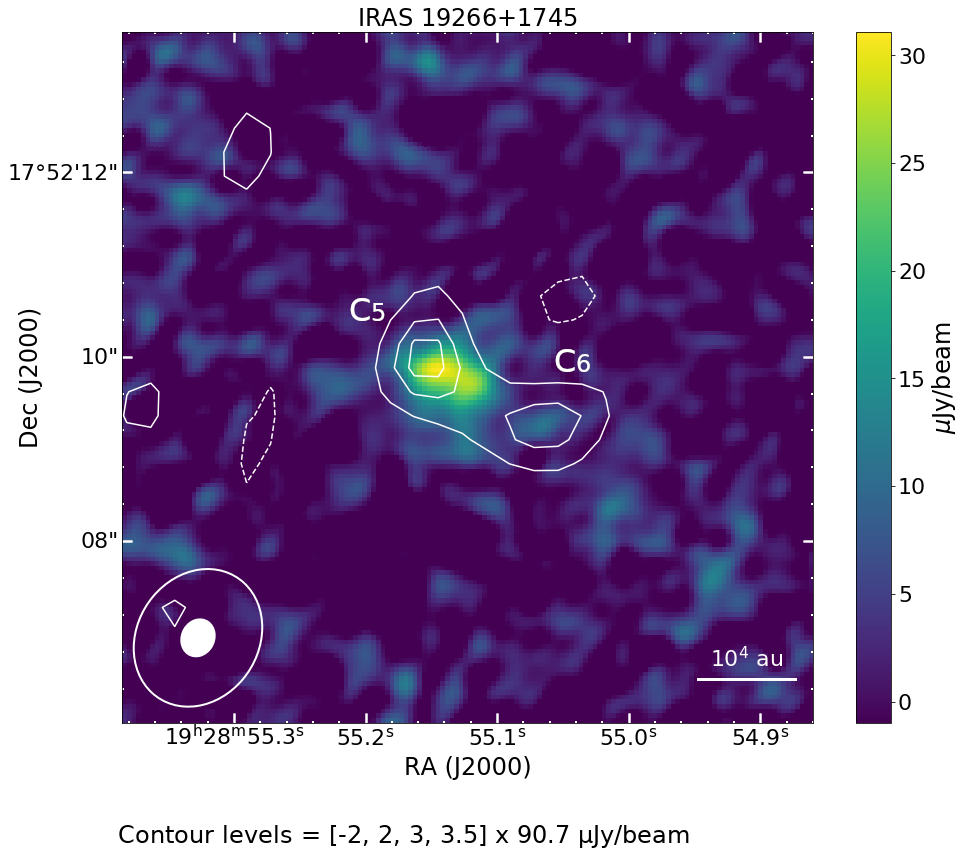}
    \caption{\textit{Continued.}
    }
    \label{app:19266-cont}
\end{figure}

\begin{figure}[H]
    \centering
    \includegraphics[width=0.4\textwidth]{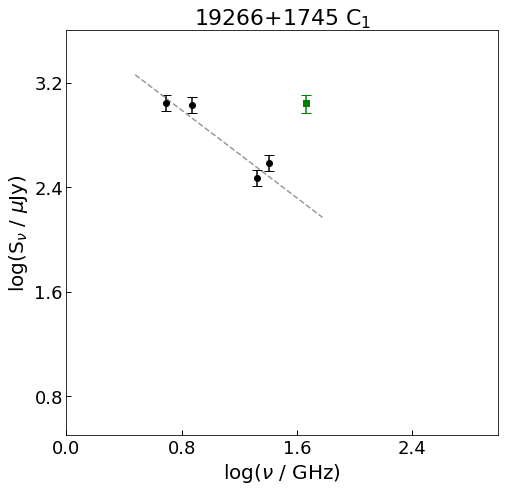}
    \includegraphics[width=0.4\textwidth]{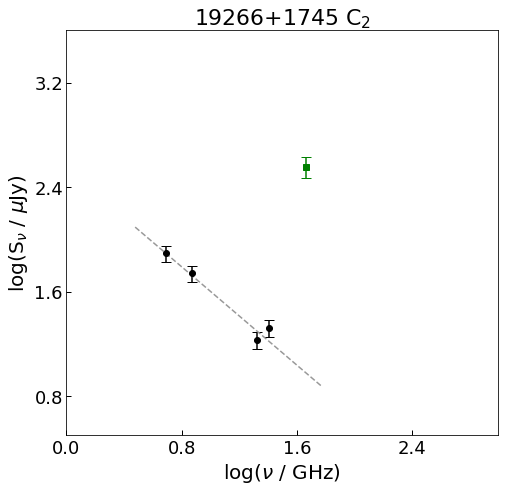}
    \caption{SED of the cores in 19266 with centimeter counterpart. The black circles and green square represent the 1.3/6~cm data from \cite{Rosero16,Rosero2019} and the 7~mm continuum emission, respectively. The gray dashed line shows the power-law fit applied to the radio continuum emission.
    }
    \label{app:19266-SED}
\end{figure}

\begin{figure}[H]
\ContinuedFloat
    \centering
    \includegraphics[width=0.4\textwidth]{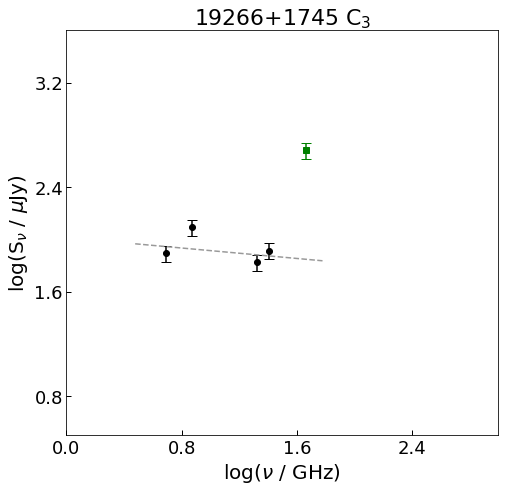}
    \includegraphics[width=0.4\textwidth]{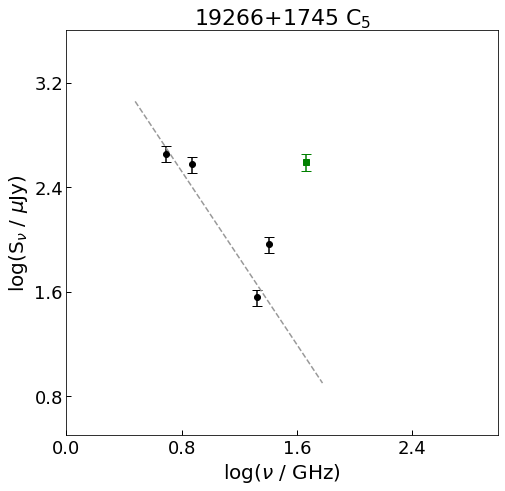}
    \caption{\textit{Continued.}
    }
    \label{app:19266-SED}
\end{figure}

\begin{figure}[H]
    \centering
    \includegraphics[width=0.55\textwidth]{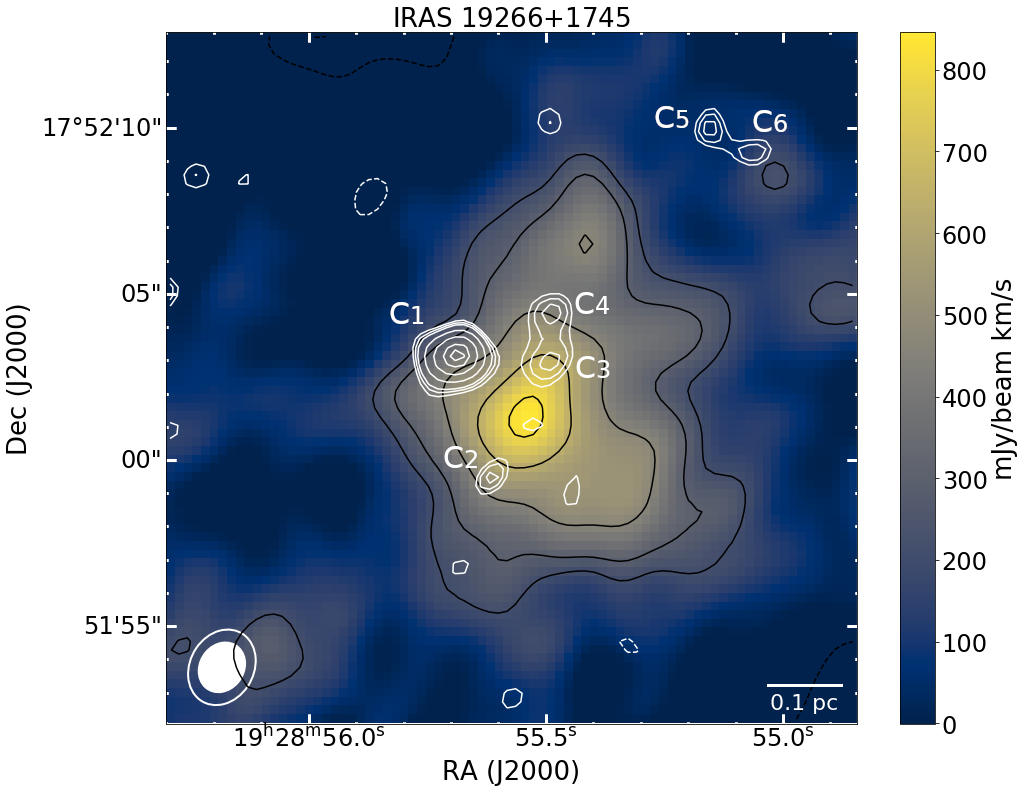}
    \includegraphics[width=0.42\textwidth]{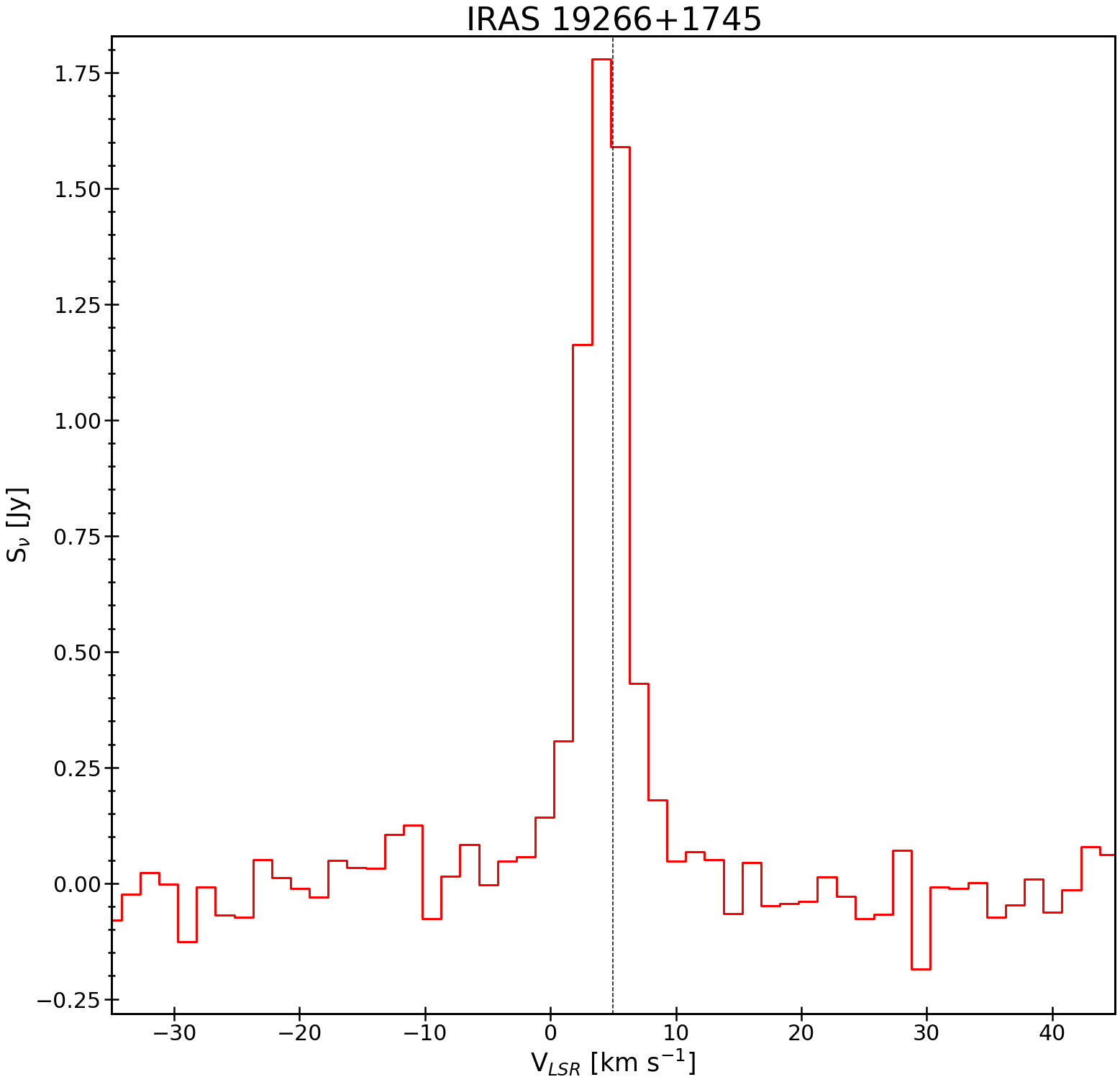}
    \caption{\textit{Left:} CS(1$-$0) integrated intensity in color and black contours. The latter represent $-$15\%, 25\%, 35\%, 55\%, 75\%, and 95\% of the peak CS(1$-$0) emission, with the peak emission and rms of the CS moment-0 map being 846 and 181~mJy~beam$^{-1}\,$\kms, respectively. 
    The white contours show the 7~mm continuum emission and represent [$-$3, 2.5, 3, 3.5, 5, 7, 9, 10]$\,\times\sigma_{7mm}$, with $\sigma_{7mm}=90.7\mu$Jy~beam$^{-1}$.
    The outlined and filled white ellipses in the bottom left represent the CS and 7~mm continuum synthesized beam size, respectively.
    \textit{Right:} spectrum of the CS line emission in 19266, obtained integrating over all the emission enclosed in the 35\% contour level from the left panel. The vertical dotted line marks the systemic velocity.
    }
    \label{app:19266-mol}
\end{figure}

\newpage

\subsection{G53.11$+$00.05 mm2}
\renewcommand{\thefigure}{\Alph{section}\arabic{subsection}.\arabic{figure}}
\setcounter{figure}{0}

\begin{figure}[H]
    \centering
    \includegraphics[width=0.53\textwidth]{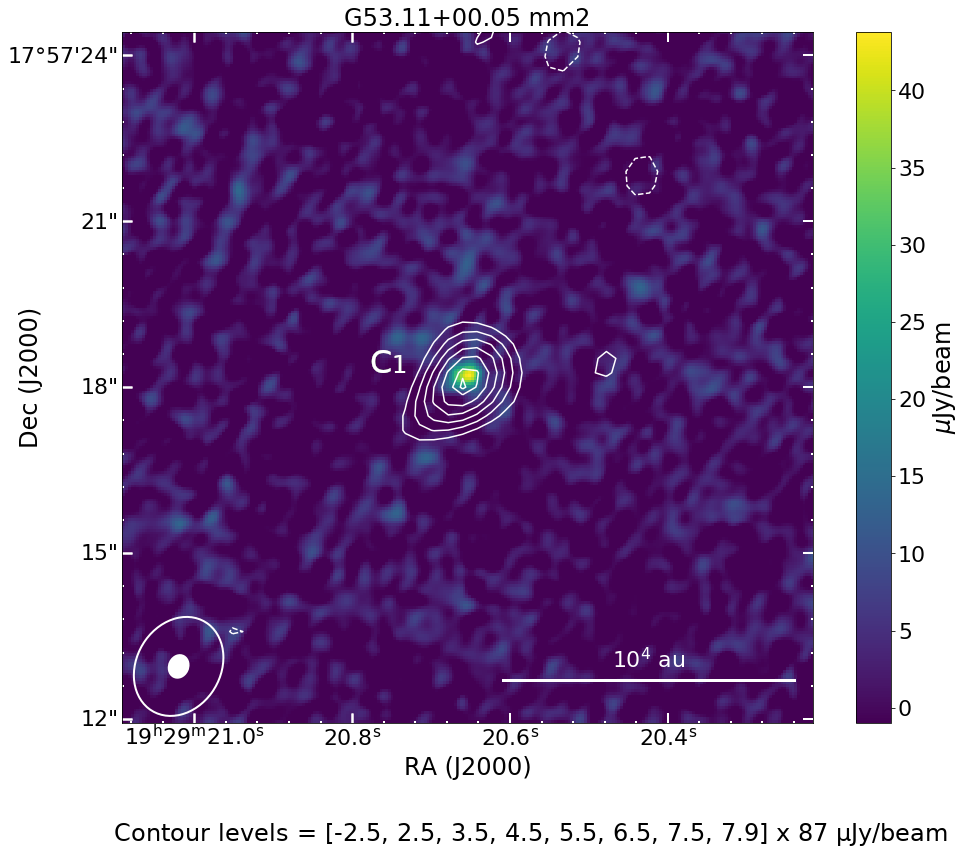}
    \includegraphics[width=0.45\textwidth]{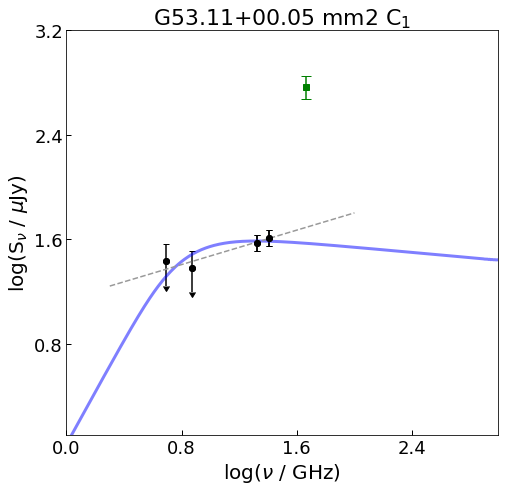}
    \caption{Same as Fig.~\ref{app:18345-cont}, but for G53.11.
    }
    \label{app:G53-cont}
\end{figure}

\begin{figure}[H]
    \centering
    \includegraphics[width=0.485\textwidth]{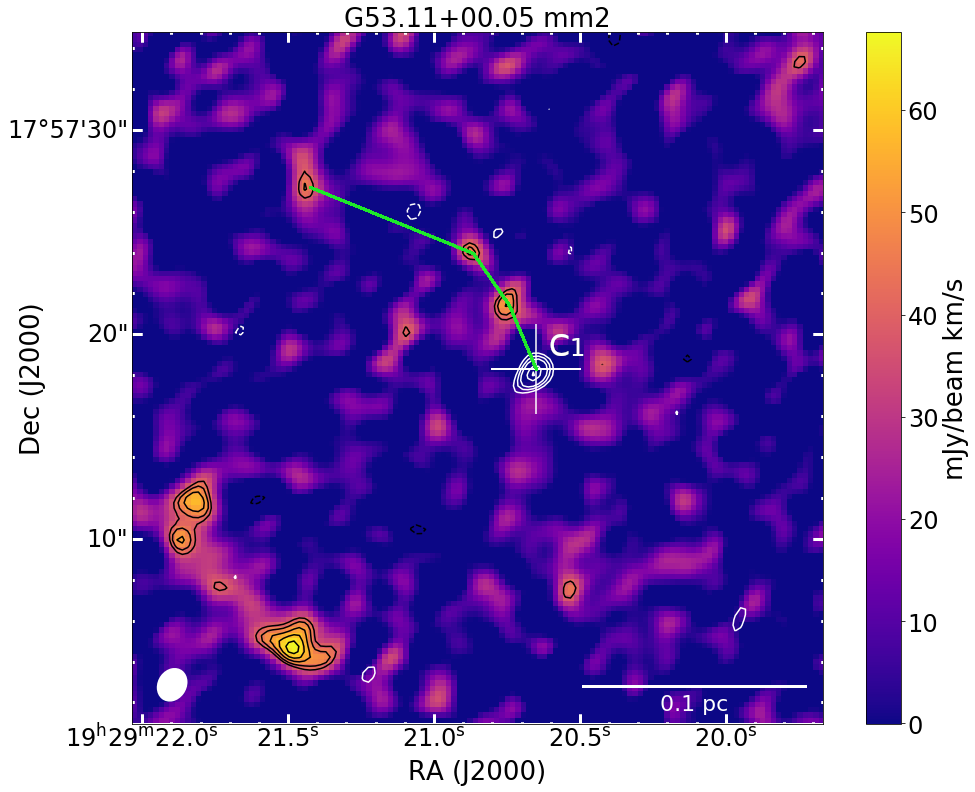}
    \includegraphics[width=0.50\textwidth]{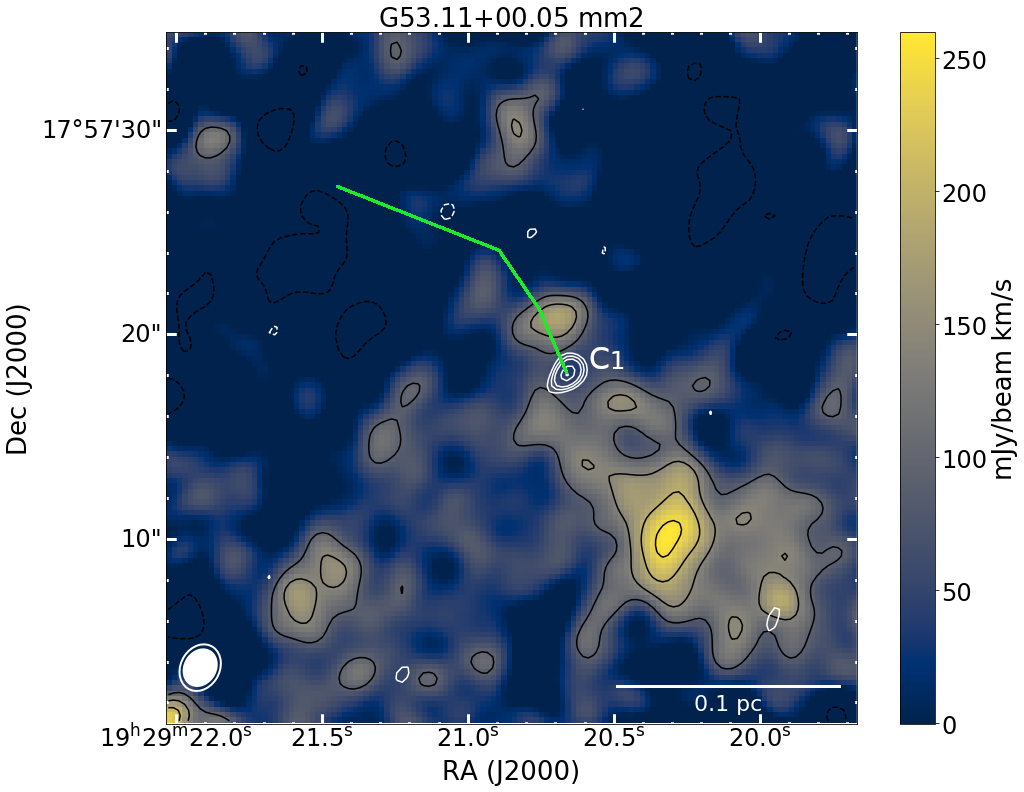}
    \includegraphics[width=0.42\textwidth]{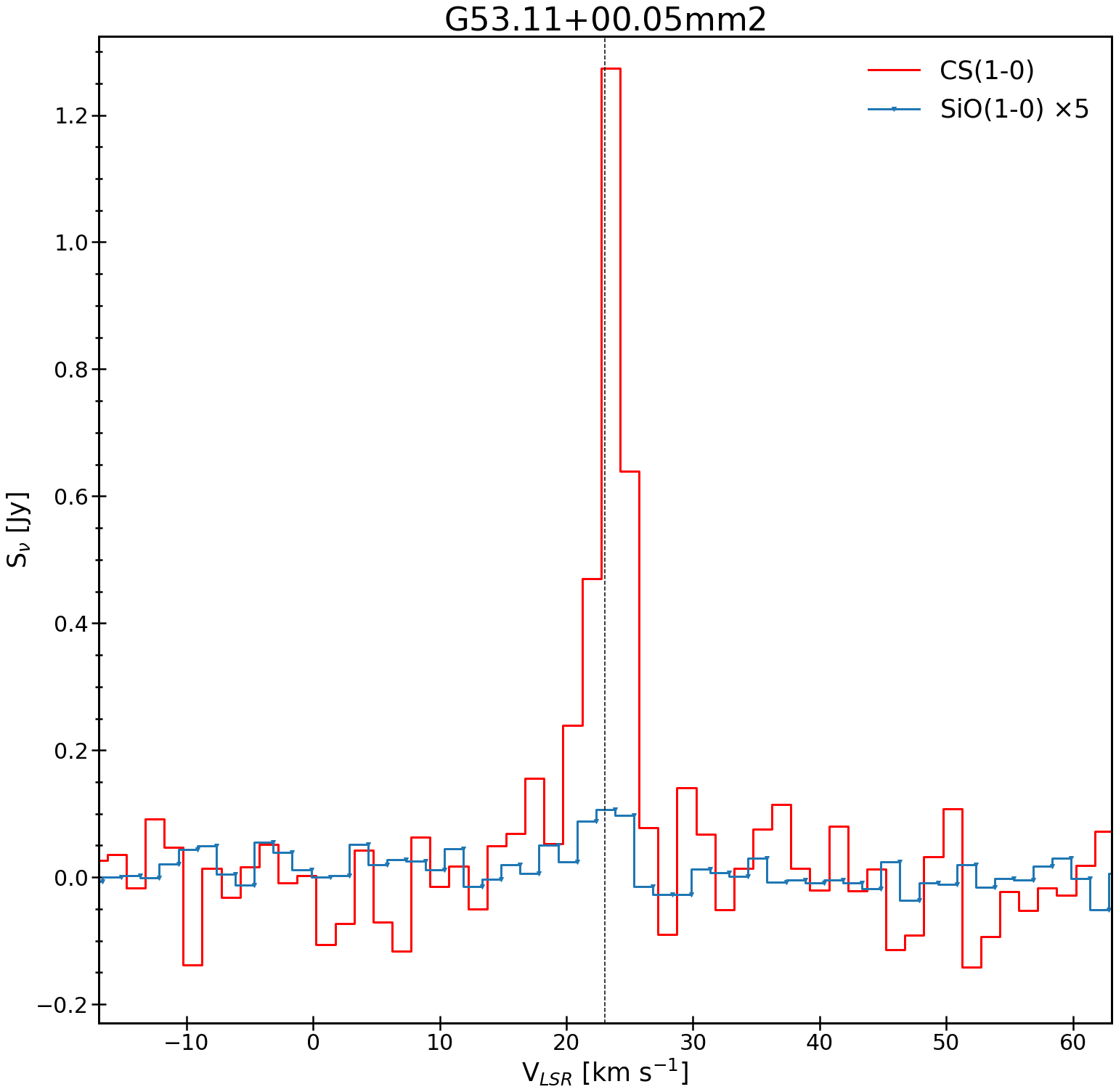}
    \caption{\textit{Top left:} SiO(1$-$0) integrated intensity map in color and contours. The latter represent [$-$3, 3, 3.5, 4, 4.5, 5]$\,\times\sigma_{SiO}$ levels, with $\sigma_{SiO}=12.9\,$mJy~beam$^{-1}\,$\kms. White contours show the 7~mm continuum emission and are the same as in the left panel of Figure~\ref{app:G53-cont}.
    The white cross shows the position of the radio continuum source and the white ellipse in the bottom left represents the synthesized beam size. 
    \textit{Top right:} CS(1$-$0) integrated intensity in color and black contours. The latter represent $-$25\%, 35\%, 55\%, 75\%, and 95\% of the peak CS(1$-$0) emission, with the peak emission and rms of the map being 260 and 69~mJy~beam$^{-1}\,$\kms, respectively. White contours are the same as in the left panel, while the outlined and filled white ellipses in the bottom left represent the CS and 7~mm continuum synthesized beam sizes, respectively. 
    In both top panels the field of view is the same, and the green lines were drawn to guide the reader on the possible flow directions.
    \textit{Bottom:} spectra of the SiO ($\times5$) and CS line emission associated with C$_1$ in blue and red, respectively. 
    These were obtained integrating over the three SiO knots to the North-East of C$_1$ and over all the CS emission to the South-West of the core enclosed by the 25\% contour level, as shown in the top right panel. 
    The vertical dotted line marks the systemic velocity.
    }
    \label{app:G53-mol}
\end{figure}

We detected one 7~mm continuum source, C$_1$, shown in the left panel of Figure~\ref{app:G53-cont}. The radio continuum source is located near the center of the core. In the right panel, we show the SED of C$_1$. The flux rises at 7~mm, indicating that it is dominated by thermal emission from dust. 
We detected SiO(1$-$0) emission to the South-East and to the North-East of the core (see top left panel of Figure~\ref{app:G53-mol}). The former is clumpy and it appears to be oriented in a projected direction perpendicular to what we would expect to be the jet axis, while the latter is composed of three individual knots. No counterpart to either structure is observed, and both cover a similar velocity span.
It is possible that both SiO structures are tracing a precessing jet close to our line of sight, and their spatial separation is solely a projection effect. However, our data are insufficient to confirm or discard this scenario, as we find no clear evidence to connect the South-East arc-shaped emission to the radio continuum source.
Thus, we will consider only the SiO knots to the North-East of C$_1$ as associated with the jet candidate.
Additionally, we detected CS(1$-$0) emission to the South-West of C$_1$, shown in the top right panel of Figure~\ref{app:G53-mol}. The CS emission is rather elongated and it peaks at almost 0.1~pc from the $7\,$mm core.
The spectra of the SiO $(\times5)$ and CS line emission in G53.11 are presented in the bottom panel of Figure~\ref{app:G53-mol}. These were obtained integrating over the three SiO knots to the North-East of C$_1$ and over all the CS emission to the South-West of the core enclosed by the 25\% contour level.

This source is one of the earliest in our sample, characterized as a CMC by \cite{Rosero16}. C$_1$ is located in an infrared dark cloud (IRDC), and \cite{Butler09} detected several YSOs in its surroundings. 
A previous study by \cite{Cosentino20} using the IRAM 30~m telescope found SiO(2$-$1) emission in this region, which they associated with mass outflows rather than a cloud-cloud collision.

\newpage

\subsection{G53.25$+$00.04 mm2}
\renewcommand{\thefigure}{\Alph{section}\arabic{subsection}.\arabic{figure}}
\setcounter{figure}{0}
This source was detected by \cite{Rosero16} at both 1.3 and 6~cm, and classified it as a CMC-IR. \cite{Rathborne2010} reported the detection of a $\sim25\,$\msun\ submm dust core consistent in position with the radio continuum emission. However, we did not detect 7~mm continuum, nor molecular emission in this region. 
If the non-detection is due to a lack of sensitivity, it would imply that the intensity of the 7~mm continuum, SiO(1$-$0), and CS(1$-$0) line emission are lower than our 3$\sigma$ detection limits, which are 0.25, 13.5, and 31.8~\mjy, respectively.

\newpage

\subsection{IRAS 20293$+$3952}
\renewcommand{\thefigure}{\Alph{section}\arabic{subsection}.\arabic{figure}}
\setcounter{figure}{0}
We detected four 7~mm sources in this region, shown in Figure~\ref{app:20293-cont}. 
C$_1$ is associated with the jet candidate, and its SED is presented in the top right panel of Figure~\ref{app:20293-cont}. C$_3$ and C$_4$ also have a radio continuum counterpart. 
The 1.3~cm source is slightly offset from the core peak in the cases of C$_1$ and C$_4$, although we note that the offset is within the 7~mm beam size and both are relatively weak detections ($\sim5\sigma$).
The unresolved radio source 20293 B from \cite{Rosero16,Rosero2019} is also offset from the C$_3$ core peak. The association of the 7~mm emission with the H~II region 20293 C is unclear, hence we will not analyze it further. 
We detected a highly collimated, monopolar SiO(1$-$0) outflow emanating from C$_1$, shown in the top left panel of Figure~\ref{app:20293-mol}. We also detected CS(1$-$0) emission consistent with the SiO structure (top right panel), i.e., it is oriented in the South-East direction appears to be monopolar, and their emission peaks seem to coincide in position as well. The SiO and CS spectra are presented in the bottom panel. The SiO is detected only at velocities blue-shifted from the systemic velocity. On the other hand, the CS spectrum appears to have 2 peaks, one at the systemic velocity, and one coincident with the SiO peak. Both peaks are still observed when integrating only over the CS emission coincident with the SiO flow, which indicates the presence of two velocity components.

\cite{Beuther04} detected several CO(2$-$1) structures in this region, and associated two of those with C$_1$: a bipolar outflow in the North-East/South-West direction, and a monopolar outflow to the South-East. 
They also detected SiO(2$-$1) outflows consistent with the CO structures, and report the observation of H$_2$ emission that seems to be coincident with the SiO(1$-$0) flow. 
Additionally, \cite{Palau07b} detected methanol and ammonia thermal emission in this region. Water and methanol maser emission were detected by \cite{Beuther02c} and \cite{Rodriguez-Garza17}, respectively.

\begin{figure}[H]
    \centering
    \includegraphics[width=0.53\textwidth]{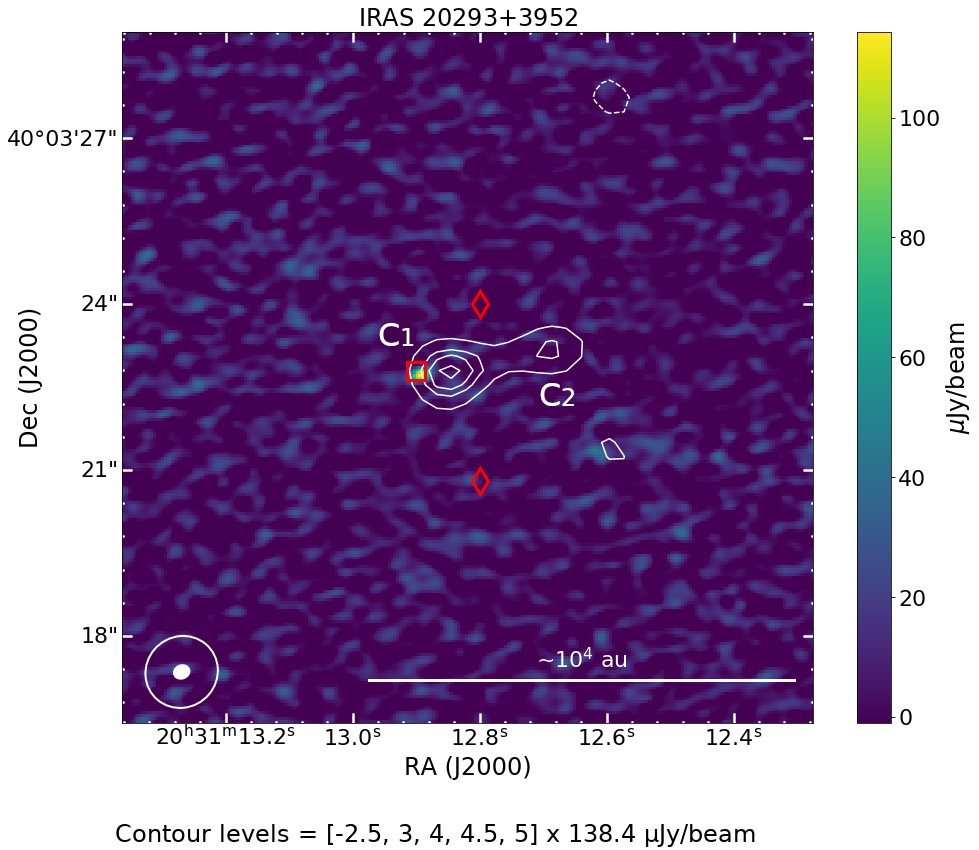}
    \includegraphics[width=0.45\textwidth]{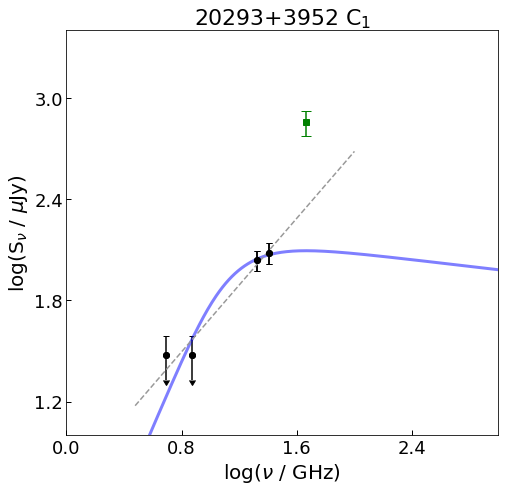}
    \caption{Same as Figure~\ref{app:18345-cont} but for 20293. The red square and diamonds mark the position of the 22 GHz water and 44 GHz methanol masers from \cite{Beuther02c} and \cite{Rodriguez-Garza17}, respectively. The SED of C$_3$ was created using the radio continuum flux of the ionized source 20293~C from \cite{Rosero16}, which contains the embedded component 20293~B. The scale bar marks $\sim10^4\,$au because of the distance uncertainty of the region, which represents $10^4$ and $\sim15,000\,$au. 
    Note the different color scales and contours used in each continuum image. 
    }
    \label{app:20293-cont}
\end{figure}

\begin{figure}[H]
\ContinuedFloat
    \centering
    \includegraphics[width=0.53\textwidth]{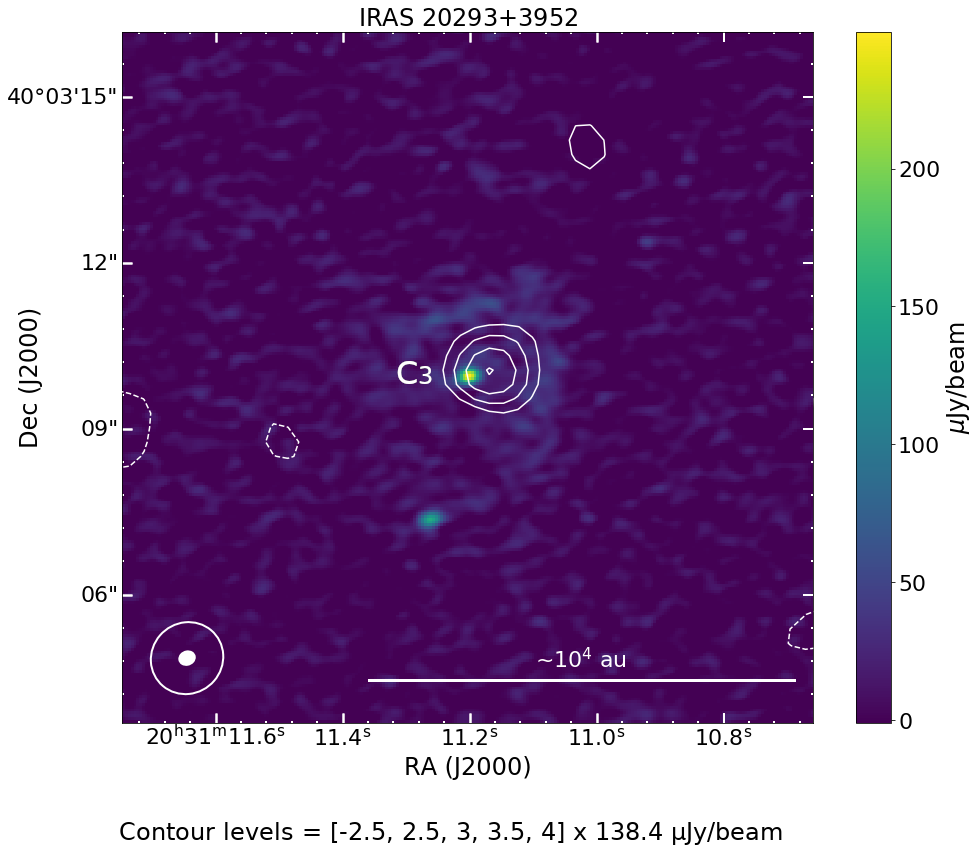}
    \includegraphics[width=0.45\textwidth]{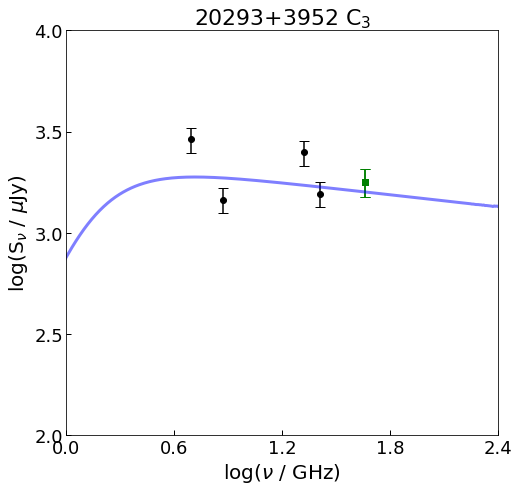}\\
    \includegraphics[width=0.53\textwidth]{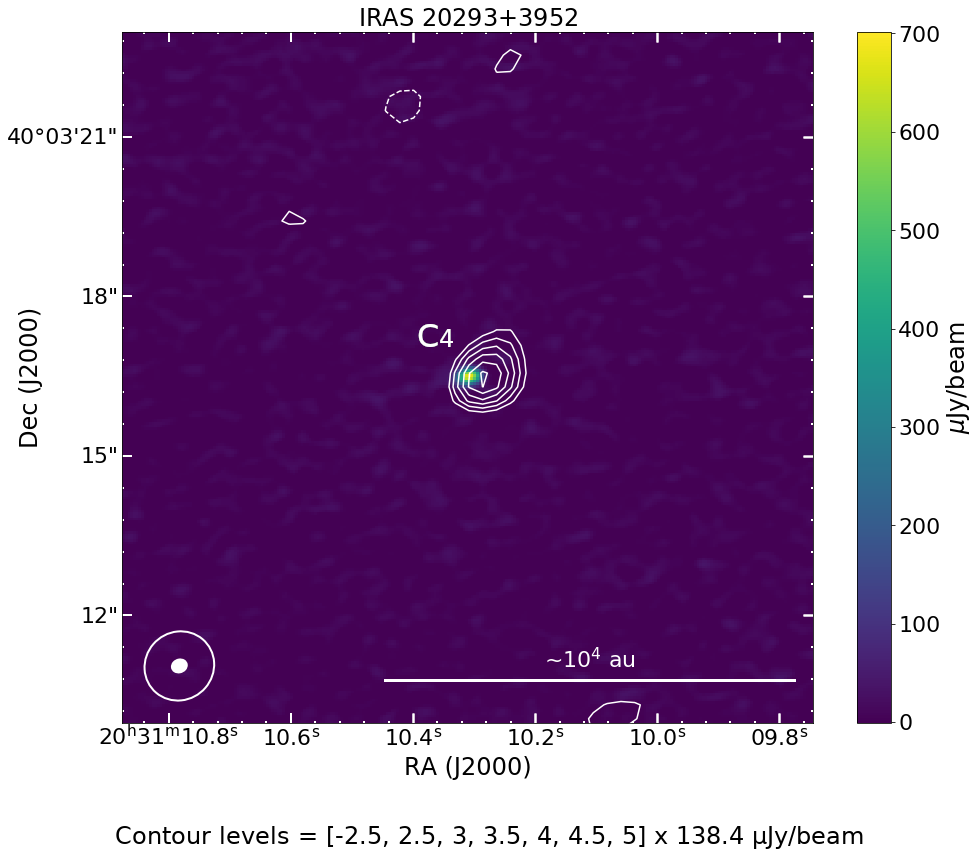}
    \includegraphics[width=0.45\textwidth]{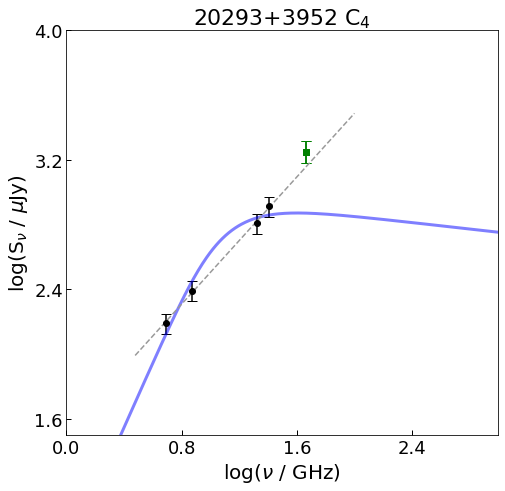}\\
    \caption{\textit{Continued.}
    }
    \label{app:20293-cont}
\end{figure}

\begin{figure}[H]
    \centering
    \includegraphics[width=0.49\textwidth]{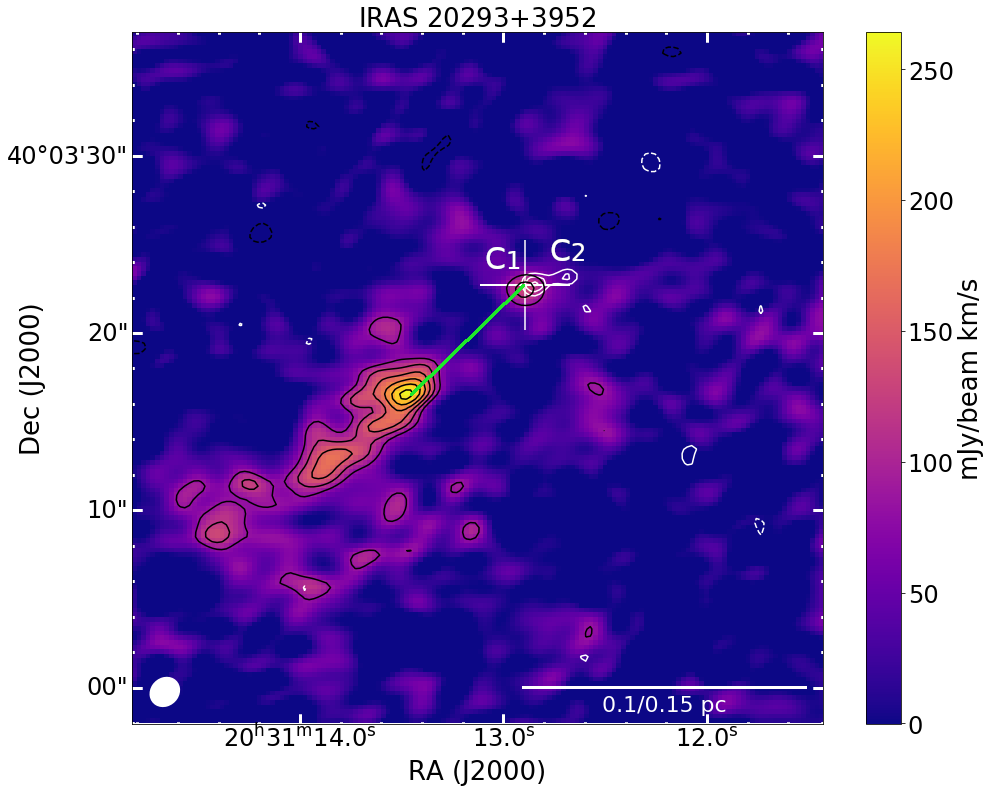}
    \includegraphics[width=0.5\textwidth]{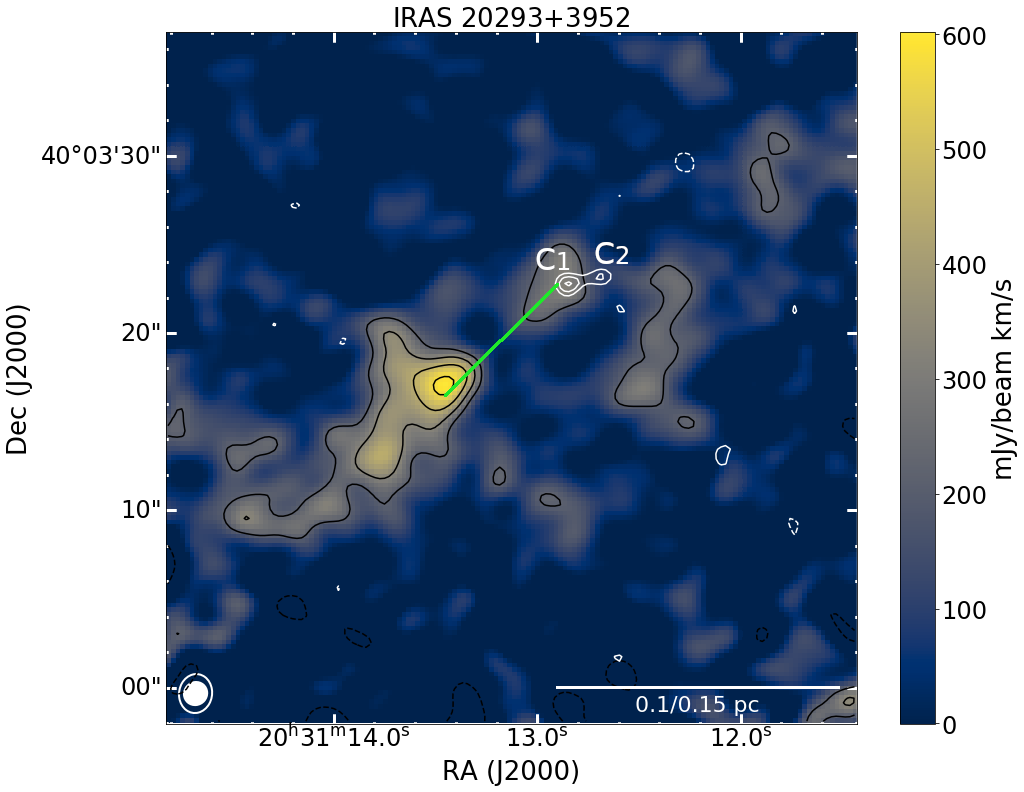}\\
    \includegraphics[width=0.38\textwidth]{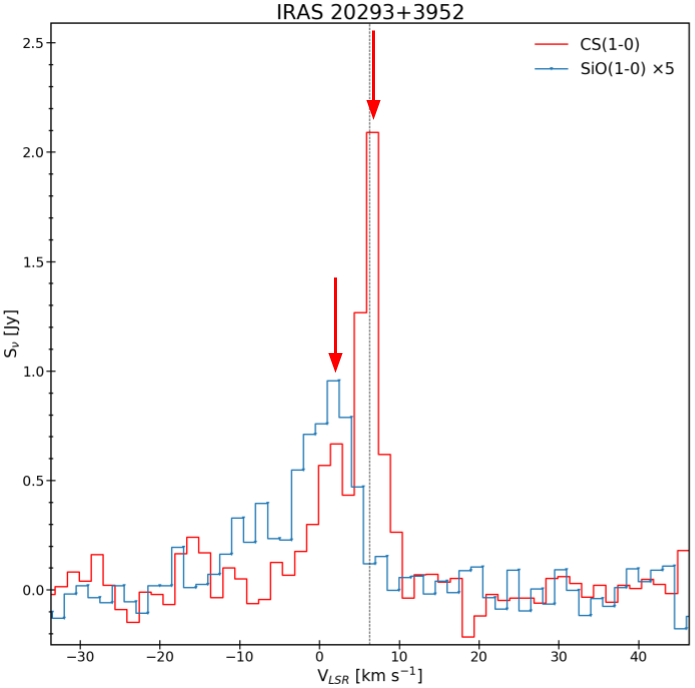}
    \caption{\textit{Top left:} SiO(1$-$0) integrated intensity in 20293 in color and black contours. The latter represent [$-$2.5, 2.5, 3.5, 4.5, 5.5, 6.5, 7.5] $\times\sigma_{SiO}$ levels, with $\sigma_{SiO}=33.5\,$mJy~beam$^{-1}\,$\kms. 
    The white contours show the 7~mm continuum emission and are the same as in the first panel of Figure~\ref{app:20293-cont}.
    The white cross shows the position of the radio continuum source and the white ellipse in the bottom left represents the synthesized beam size.  
    \textit{Top right:} CS(1$-$0) integrated intensity in color and black contours. The latter represent $-$35\%, 35\%, 55\%, 75\%, and 95\% of the peak CS(1$-$0) emission, with the peak emission and rms of the map being 600 and 62~mJy~beam$^{-1}\,$\kms, respectively. White contours represent the 7~mm continuum emission and are the same as in the left panel.
    The outlined and filled white ellipses in the bottom left represent the CS and 7~mm continuum synthesized beam sizes, respectively. 
    In both top panels the field of view is the same, and the green lines were drawn to guide the reader on the possible flow directions.
    \textit{Bottom:} spectra of the SiO ($\times5$) and CS $J=1-0$ emission in 20293 in blue and red, respectively. These were obtained integrating over all the SiO emission to the South-East of the core and over all the CS emission enclosed by the 35\% contour level shown in the top right panel.
    The vertical dotted line marks the systemic velocity. The red arrows were drawn to show the two CS peaks.
    }
    \label{app:20293-mol}
\end{figure}

\newpage

\subsection{IRAS 20343$+$4129}
\renewcommand{\thefigure}{\Alph{section}\arabic{subsection}.\arabic{figure}}
\setcounter{figure}{0}
We detected five 7~mm continuum sources, which are presented in Figure~\ref{app:20343-cont}.
C$_1$ and C$_3$ have a radio continuum counterpart. 
C$_1$ seems to be associated with the jet candidate, although the ionized source is offset to the North of the core peak, while C$_3$ is associated with an HC~H~II region \citep[see Table 1 of][]{Rosero2019}. In the right hand top panel of Figure~\ref{app:20343-cont} we present the SED of C$_1$.

SiO(1$-$0) emission was not detected in this region. We found a rather extended CS(1$-$0) structure, shown in the left panel of Figure~\ref{app:20343-mol}. The 7~mm cores C$_1$, C$_2$, C$_4$, and C$_5$ are located near the center of the CS structure. 

\cite{Beuther02} detected a bipolar, North-South CO(2$-$1) outflow in this region using the IRAM 30~m telescope. Later, \cite{Palau07} found no indication of a North-South structure in their SMA ($\theta\sim3$\arcs) data, but they detected an East-West CO(2$-$1) flow emanating from a position consistent with C$_1$. This outflow aligns well with the $2.122\,\mu$m H$_2$ emission observed by \cite{Kumar02}. In a subsequent study, \cite{Fontani12} detected ammonia thermal emission and several dense gas tracers in this region as well.

\begin{figure}[H]
    \centering
    \includegraphics[width=0.5\textwidth]{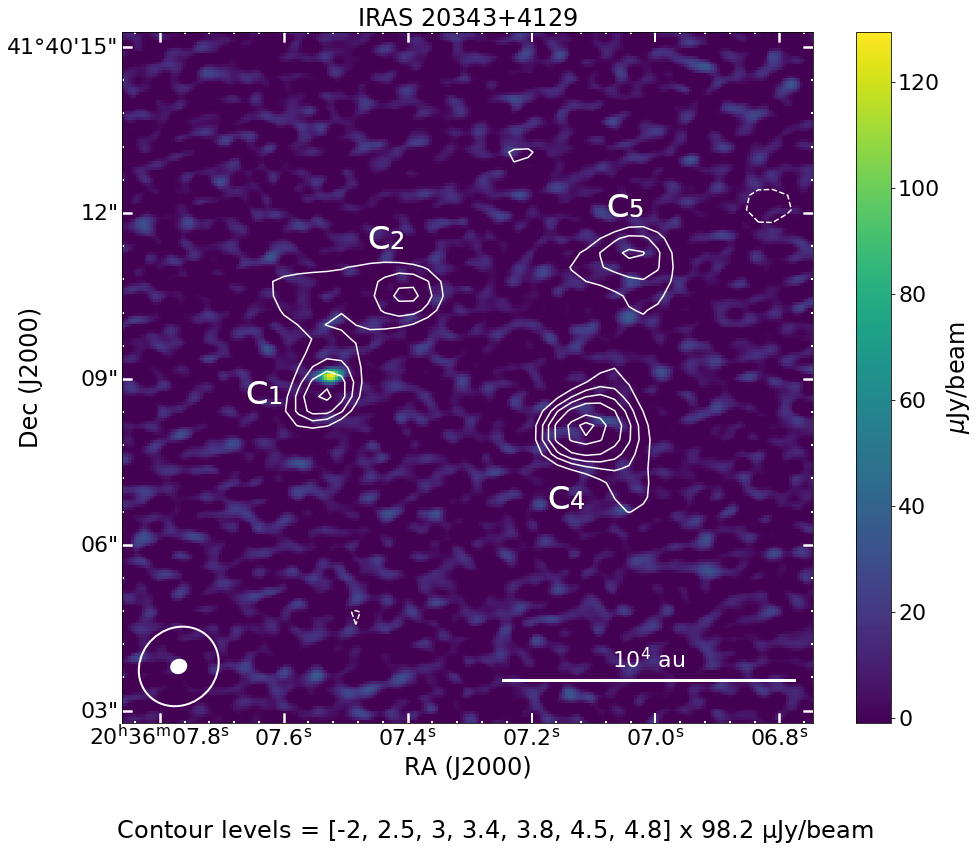}
    \includegraphics[width=0.43\textwidth]{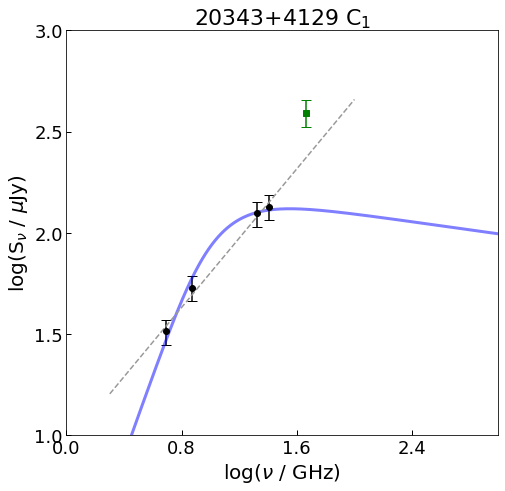}\\
    \includegraphics[width=0.5\textwidth]{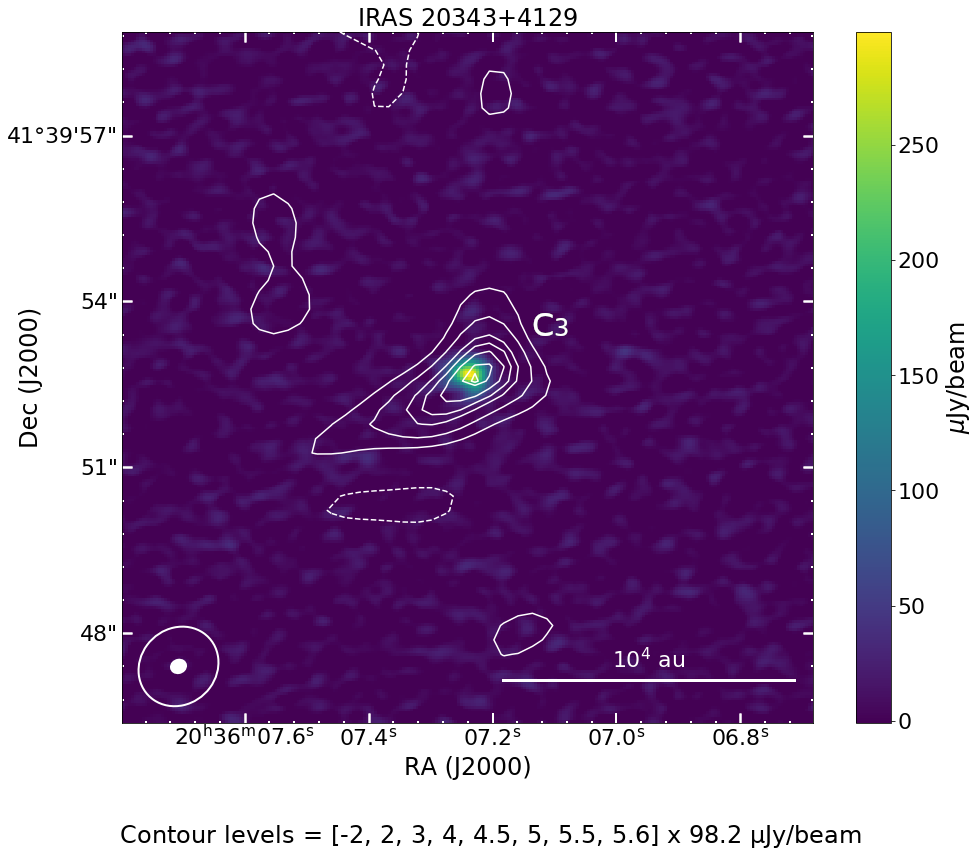}
    \includegraphics[width=0.45\textwidth]{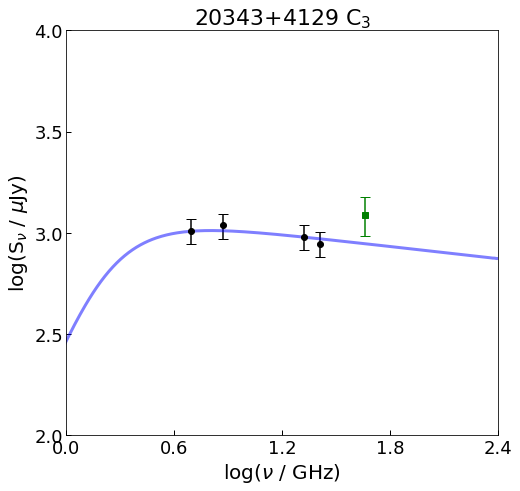}
    \caption{Same as \ref{app:18345-cont} but for 20343. Note the different color scale and contour levels used in the continuum images.
    }
    \label{app:20343-cont}
\end{figure}

\begin{figure}[H]
    \centering
    \includegraphics[width=0.55\textwidth]{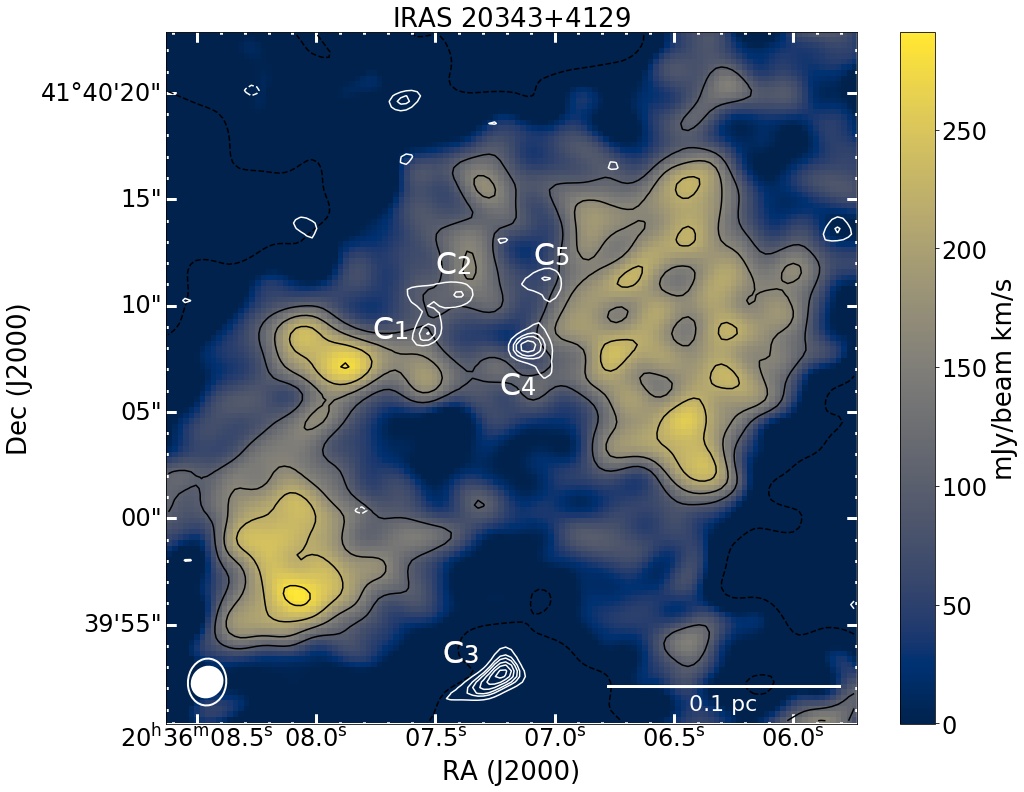}
    \includegraphics[width=0.38\textwidth]{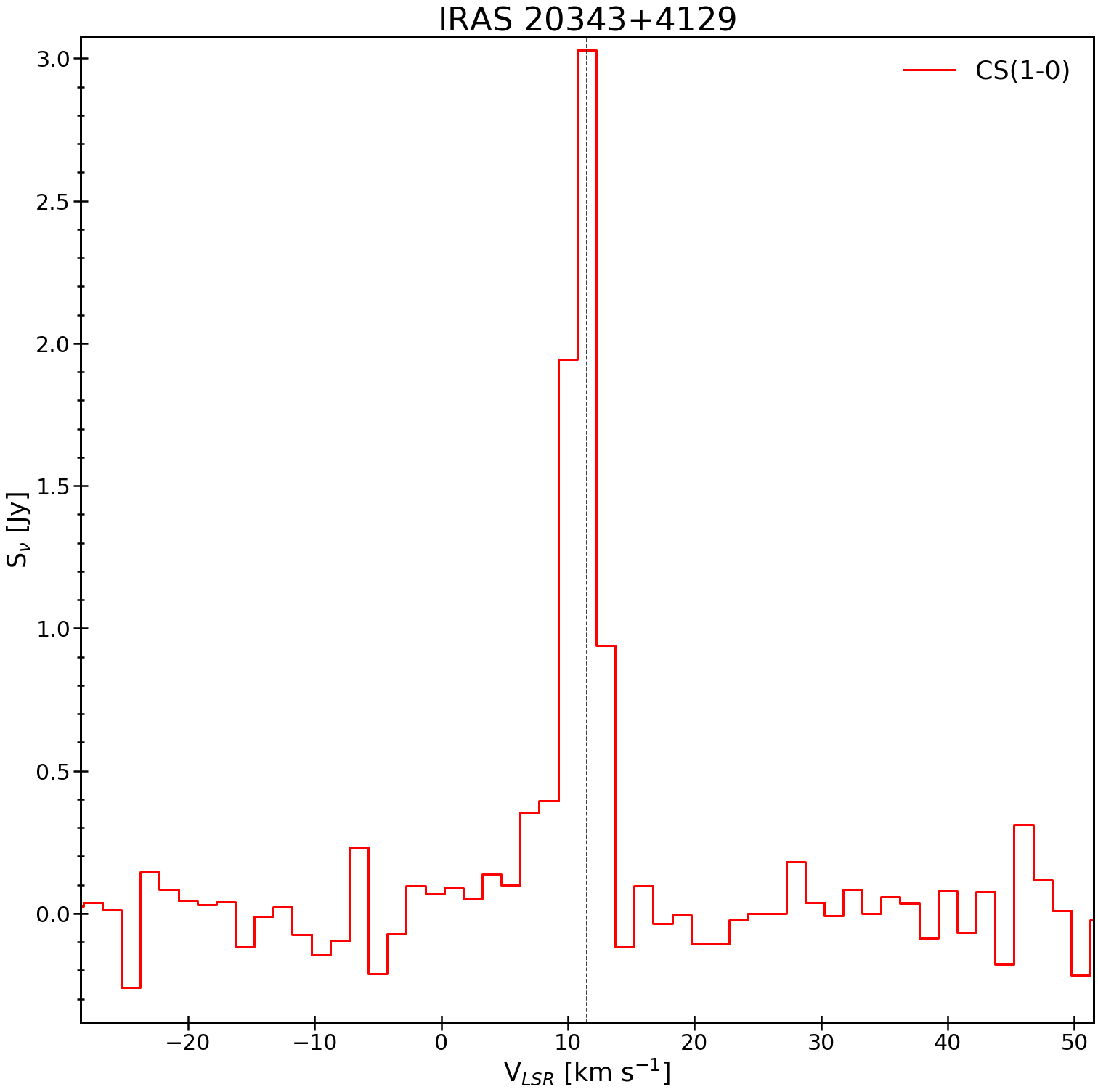}
    \caption{\textit{Left: }CS(1$-$0) integrated intensity map in color and black contours. The latter represent $-$25\%, 35\%, 55\%, 75\%, and 95\% of the peak CS(1$-$0) emission, with the peak emission and rms of the map being 291 and 113~mJy~beam$^{-1}\,$\kms, respectively. White contours show the 7~mm continuum emission and represent [$-$2.5, 2.5, 3, 4, 4.5, 5, 5.5]$\,\times\sigma_{7mm}$, with $\sigma_{7mm}=98.2\,\mu$\jy.
    The outlined and filled white ellipses in the bottom left represent the CS and 7~mm continuum synthesized beam size, respectively. 
    \textit{Right:} spectrum of the CS(1$-$0) line emission in 20343, obtained integrating over all the emission enclosed by the 35\% contour level from the left panel. The vertical dotted line marks the systemic velocity.
    }
    \label{app:20343-mol}
\end{figure}

\end{appendix}
\end{document}